\documentclass[]{aa}

\usepackage[varg]{txfonts}
\usepackage{subcaption}
\usepackage[normalem]{ulem}
\usepackage{placeins}

\newcommand{\gaia}{{\emph{Gaia}}}
\newcommand{\kepler}{\emph{Kepler}}
\newcommand{\tess}{{\emph{TESS}}}
\newcommand{\ktwo}{{\emph{K2}}}
\newcommand{\simbad}{{\textsc{Simbad}}}
\newcommand{\epic}{{\textsc{EPIC}}}
\newcommand{\tmass}{{\textsc{2mass}}}
\newcommand{\gsc}{{\textsc{GSC}}}
\newcommand{\usno}{{\textsc{USNO}}}
\RequirePackage[]{xcolor}
\definecolor{AIPblue}{RGB}{  0, 85,160}

\usepackage[colorlinks]{hyperref}
\hypersetup{urlcolor=AIPblue,linkcolor=AIPblue,citecolor=AIPblue}

\begin{document}

\title{New insights into the rotational evolution of near-solar age stars from the open cluster M\,67%
    \thanks{
        Table\,\ref{tab_results_large} and the corresponding light curves are available in electronic form at the CDS via anonymous ftp to cdsarc.u-strasbg.fr (130.79.128.5) or via \href{http://cdsweb.u-strasbg.fr/cgi-bin/qcat?J/A+A/672/A159}{http://cdsweb.u-strasbg.fr/cgi-bin/qcat?J/A+A/672/A159}
        }%
    }

\subtitle{}

\author{D. Gruner\inst{1,2}, S. A. Barnes\inst{1,3}, J. Weingrill\inst{1}}

\institute{
    Leibniz-Institute for Astrophysics Potsdam (AIP), An der Sternwarte 16, 14482, Potsdam, Germany
        \and 
    Institut f\"ur Physik und Astronomie, Universit\"at Potsdam, Karl-Liebknecht-Str. 24/25, 14476 Potsdam, Germany
        \and
    Space Science Institute, Boulder, CO, USA
    }

\date{Received 19 January 2023 / Accepted 3 March 2023}

\abstract 
  {Gyrochronology allows the derivation of ages for cool main sequence stars from their observed rotation periods and masses, or a suitable proxy of the latter. It is increasingly well explored for FGK stars, but requires further measurements for older ages and K\,--\,M-type stars. }
  {Recent work has shown that the behavior of stellar spindown differs significantly from prior expectations for late-type stars. We study the 4\,Gyr-old benchmark open cluster M\,67 to explore this behavior further.}
  {We combined a \gaia{} DR3 sample with the \kepler{} \ktwo{} superstamp of Campaign~5 around M\,67 and created new light curves from aperture photometry. 
  The light curves are subjected to an extensive correction process to remove instrumental systematics and trending, followed by period analysis to measure stellar rotation.
  }
  {We identify periodic signals in 136 light curves, 47 of which are from the rotation of effectively single main-sequence stars that span from early-G to mid-M type.
  These results connect well to prior work on M\,67 and extend it to much later spectral types.
  } 
  {We find that the rotation periods of single stars of age 4\,Gyr define a tight relationship with color, ranging from spectral types F through M.
  The corresponding surface of rotation period against age and mass is therefore well-defined to an older age than was previously known. 
  However, the deviations from prior expectations of the stellar spindown behavior are even more pronounced at 4\,Gyr.
  The binary cluster members do not follow the single star relationship. The majority are widely scattered below the single star sequence. Consequently, they do not seem to be suitable for gyrochronology at present.
  }

\titlerunning{A\&A 672, A159 (2023)}

\authorrunning{Gruner, D., et al.}


\maketitle

\section{Introduction} \label{sec_intro}

    A key issue in cool star science is to understand the extent to which the rotation rate of a star can be used to infer its age. While empirical data are available in a series of young open clusters, they are lacking for cooler spectral types in older clusters. Here we confirm related prior results and extend the empirical knowledge of this extent to K- and M-type stars of age 4\,Gyr by presenting measured rotation periods in the open cluster M\,67.

    The fact that the rotation rate of a star depends on its age was originally proposed by \cite{1972ApJ...171..565S} following $v \sin i$ measurements for stars of solar mass in open clusters. Older stars were found to rotate slower than younger ones,  that means the stars spin down as they age, approximately as a power law. This relationship was later extended to other late-type stars by \cite{2003ApJ...586..464B} who showed the existence of a mass-dependence of the spindown rate and suggested that the age of a star can be inferred from its measured rotation period and mass (or a suitable proxy of the latter, such as color). This marks the onset of what is now known as gyrochronology.

    Using stellar rotation for age estimates of late-type main sequence stars opens up new possibilities in age dating for stars whose classical parameters change only very little over the course of billions of years (e.g., temperature and luminosity), which can only be used reliably in young stars (e.g., Li abundance), or which change as long-term averages but vary significantly on short timescales (e.g., activity). However, to be able to use gyrochronology reliably, it needs to be first established if and how stellar rotation changes systematically with age for stars of different masses. Now, if the rotation periods of stars change systematically with age, then regardless of the details of the dependencies on stellar mass and age, appropriate measurements of open cluster stars of known ages can be used to set up a series of "mileposts" at suitable ages to derive stellar ages from their measured rotation periods. This has led to the exploration of accessible open clusters of differing ages to construct the sequence of spindown empirically as a function of mass and age. Open clusters are the preferred means to explore the spindown as they provide sets of stars of different masses with well-established ages and metallicities.

    Fortunately, the relationship has so far been found to be single-valued as a function of mass at least for clusters older than the Hyades ($\sim$600\,Myr). Unfortunately, only a small number of such clusters are readily available. Outstanding among them is M\,67; with its sun-like age of 4\,Gyr, it is the oldest cluster explored in this regard today. However, only two distinct groups have been measured reliably to date:  F- and G-type stars by \citet{ApJ...823..2016.16B} and mid to late M-type stars by \citet{2022ApJ...938..118D}. Here we present new measurements for stars spanning from early-G to early-M types, overlapping with the prior work and bridging the gap between them. Crucially, we find that the rotation periods of single cluster members in M\,67 continue to define a single-valued relationship with stellar mass at this age. Consequently, we claim that rotation is usable as an age indicator even though its mass and age dependencies are likely more complex than was thought earlier. 

    The systematic exploration of rotation in stellar open clusters using timeseries photometry can be traced back to \citet[][see also \citet{1987A&AS...67..483V}]{1982Msngr..28...15V} in the Pleiades ($\sim$125\,Myr) and \citet{1987ApJ...321..459R} in the Hyades. However, $v \sin i$ measurements were typical at that time and, despite the ambiguity introduced by the unknown inclination $i$, showed traces of the patterns we observe today \citep[e.g.,][]{1984ApJ...280..202S,1993ApJS...85..315S,1998A&A...335..183Q}. Subsequently, CCD photometry from ground-based facilities and, later, the onset of space-based CCD photometry (mostly thanks to the \kepler{} mission) has allowed large-scale studies of various readily available open clusters. Photometric rotation periods,  that means measuring the periodic brightness variations induced by activity features traversing the stellar disk, remove the ambiguity of unknown inclination.

    Numerous ground-based, and more recently, space-based photometric studies have together constructed a set of mileposts with open clusters, including the Pleiades \citep[125\,Myr, \cite{1987A&AS...67..483V}, revisited by][]{2016AJ....152..113R}, the Hyades \citep[650\,Myr, \cite{1987ApJ...321..459R}, revisited by][]{2019ApJ...879..100D}, Praesepe \citep[700\,Myr,][]{2011ApJ...740..110A}, NGC\,6811 \citep[1\,Gyr,][]{2011ApJ...733L...9M,2019ApJ...879...49C}, NGC\,6819 \citep[2.5\,Gyr,][]{2015Natur.517..589M}, and Ruprecht\,147 \citep[2.7\,Gyr,][]{2020A&A...644A..16G,2020ApJ...904..140C}. \cite{2020A&A...641A..51F} have also recently shown that the distribution of stellar rotation periods in the 125\,Myr-old clusters NGC\,2516, Pleiades, M\,35 \citep{2009ApJ...695..679M}, M\,50 \citep{2009MNRAS.392.1456I}, and Blanco\,1 \citep{2014ApJ...782...29C} are indistinguishable. For additional work, and especially for rotation in the wider context of stellar activity, readers may refer to chapter 5 in \cite{2021isma.book.....B}.
    
    Recent work has advanced to large field star samples \citep[e.g.,][]{2020A&A...635A..43R,2022ApJ...933..114D,2022arXiv220605500D} which provide a much more extensive parameter space than the clusters alone, especially with respect to metallicity but lack the definitive nature of clusters to allow their usage for calibration. Furthermore, especially concerning slower rotating stars, the detection and identification of accurate stellar rotation periods is somewhat problematical. As \citet[][see also \cite{2018ApJ...863..190B}]{2020AN....341..513T} have pointed out, a significant fraction ($\approx10$\,\%) of the widely used \cite{2014ApJS..211...24M} sample is likely  listed with only half the actual period due to double-dipping. Field stars require more rigorous and extensive vetting than cluster stars, which have uniform metallicity, age, and distances among them. If not, unaccounted non-solar metallicities, binarity, or an evolutionary state even marginally past the main sequence are prone to lead to flawed conclusions. Given those shortcomings, open clusters remain the primary calibrators for the evolution of stellar rotation, and provide the main motivation for this work.

    M\,67, also known as NGC\,2682, has long been of great interest because it is one of the oldest open clusters known, has an age near that of the sun and is located close enough to allow extensive study. In Table\,\ref{tab_m67_overview}, we give an overview of some of the fundamental parameters of M\,67. Apart from the work on stellar rotation in M\,67 by \cite{ApJ...823..2016.16B} and  \cite{2022ApJ...938..118D} there have been the studies of \cite{2016MNRAS.459.1060G,2016MNRAS.463.3513G} and \cite{2018PhDT........63E} whose findings challenge not only the prior work on M\,67 but also those for other clusters and one bedrock principle of gyrochronology: the single-valued nature of the spindown relation. We also investigate and make some sense of their findings below.

 \begin{table}[ht!]
     \centering
     \caption{Basic properties of M67 \label{tab_m67_overview}}
     \begin{tabular}{lccl}
     \hline\hline
        Parameter                  &                                  & Value       & Ref. \\
     \hline
                      Ra           & [h\,min\,s]                      &  08 51 18   & 1 \\
                     Dec           & [$^\circ$\,$\arcmin$\,$\arcsec$] & +11 48 00   & 1 \\
        $\mu_\text{Ra}$            & [mas/yr]                         & -10.97      & 2 \\
        $\mu_\text{Dec}$           & [mas/yr]                         &  -2.94      & 2 \\
        $v_\text{rad}$             & [km/s]                           &  33.92      & 3 \\
        Age $t$                    & [Gyr]                            &   4.0       & 4 \\
        Parallax $\varpi$          & [mas]                            &   1.1325    & 1 \\
        Distance $d$               & [pc]                             & 883         & from parallax \\
        Distance mod.              & [mag]                            & 9.6         & from distance \\
        $E(B-V)$                   & [mag]                            &   0.04      & 5 \\
        $A_V$                      & [mag]                            &   0.124     & from $E(B-V)$ \\
        $\left[\text{Fe/H}\right]$ &                                  & 0.028       & 6 \\
     \hline
     \end{tabular}
     \tablebib{
        (1)~\cite{2020A&A...633A..99C};
        (2)~\cite{2018A&A...616A..10G};
        (3)~\cite{2019A&A...623A..80C};
        (4)~\cite{2010A&A...513A..50B}:
        (5)~\cite{2007AJ....133..370T};
        (6)~\cite{2019MNRAS.490.1821C}
    }
 \end{table}

    This paper is structured in the following way. In Sect.\,\ref{sec_data}, we describe the observations that this work is based on, as well as the stellar content of the \kepler{} \ktwo{} superstamp. This section includes a membership analysis based on GDR3 data which is detailed in Appendix\,\ref{appendix_cluster_membership}. Section\,\ref{sec_lightcurves} addresses the creation of light curves from the \ktwo{} data. The section gives a short overview of the steps carried out while Appendix\,\ref{appendix_correction_main} describes all the details regarding the difficulties of \ktwo{} data and our ways of overcoming them.  The light curves thus constructed are subjected to a period analysis which is described in Sect.\,\ref{sec_analysis}.  The results of the analysis are presented in Sect.\,\ref{sec_results}. There, we construct a Color-period diagram (CPD) for M\,67 after scrutinizing each and every star in the sample. We then go on to discuss our findings and their implications in the contexts of prior work on open cluster and the modeling of rotational evolution in Sect.\,\ref{sec_discussion} before we finish with some conclusions in Sect.\,\ref{sec_conclusion}.

\section{Observational data} \label{sec_data}

    In this section we describe the photometric data we use throughout this work: the \kepler{} \ktwo{} Campaign\,05 superstamp. We obtain the stellar content of the field of view from the \gaia{} catalogs and carry out a membership analysis. The cluster sample constructed is then used to reconfirm the age of M\,67 and estimate reddening parameters in different colors.

\subsection{The K2 superstamp}

    Our goal is to explore the cluster center of M\,67 to its fullest. Therefore, we ignore the few \kepler{} \ktwo{} light curves supplied by the mission itself, and construct our own based on a new, more extensive target list. This allows us to use the richness of the cluster center without being bound by the preselection of targets in the \emph{K2} program. We show in Appendix\,\ref{appendix_validation} the meager extent of the original \ktwo{} program, especially when it comes to the superstamp. However, this also means that we now have to solve by ourselves the problems addressed by the mission itself when it provided the \ktwo{} light curves.

    After the conclusion of its primary mission, the \kepler{} telescope started its \ktwo{} program. Between 2014 and 2019, the telescope was pointed at 19 different fields along the ecliptic for approximately 80\,d each (denoted as campaigns, C\#\# hereafter). Prior to and during this long observation run several parts of the telescope ceased to function. However, \kepler{} was able to continue its mission until it ultimately ran out of coolant in 2019, long after its originally designated lifetime. Some of the defects make \ktwo{} data problematic to use if not properly addressed.
    
\begin{figure}[ht!]
    \centering
    \includegraphics[width=\linewidth]{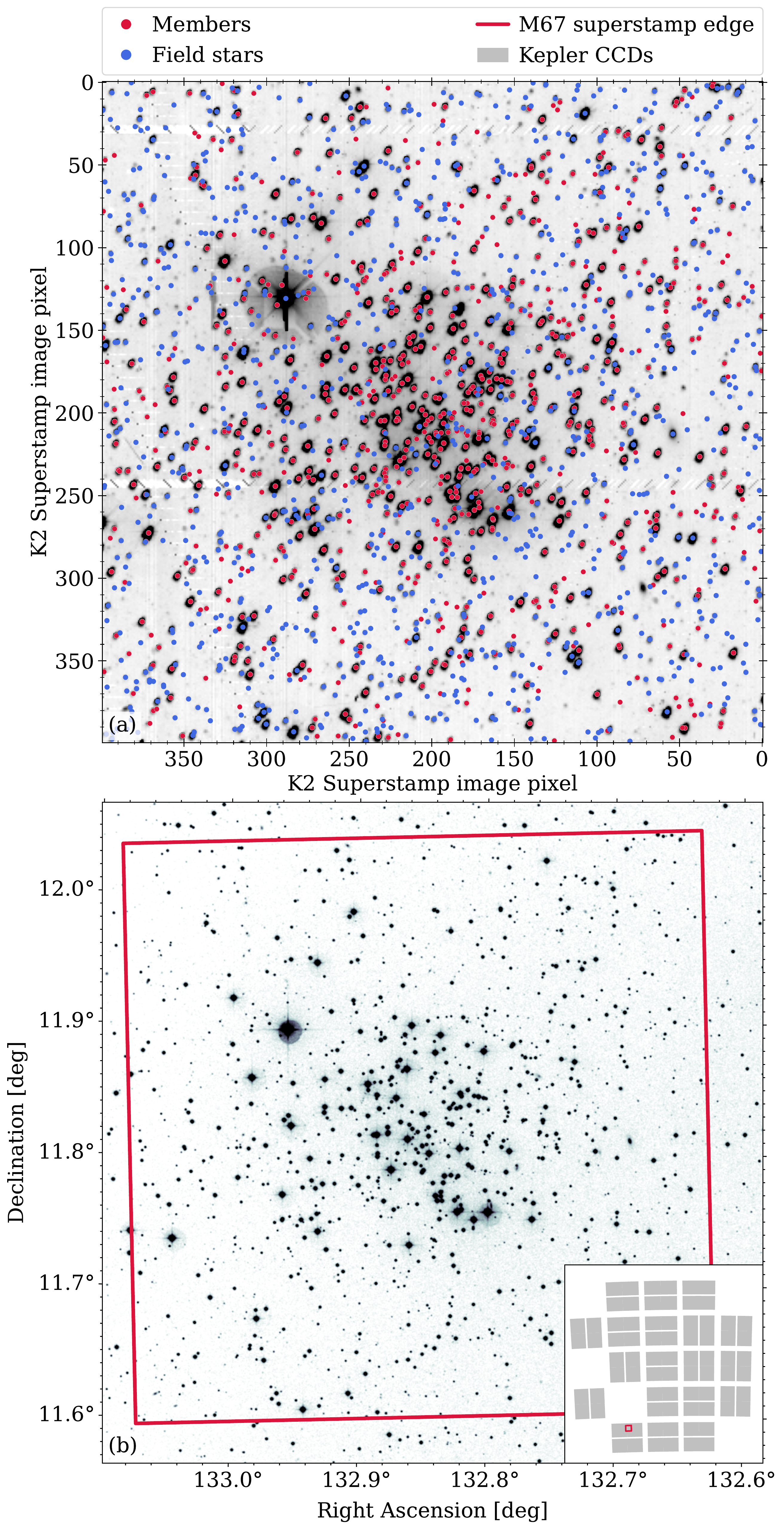}
    \caption{
        Overview of the \ktwo{} C05 superstamp coverage of M67. The upper panel shows one superstamp image on a logarithmic gray-scale. Stars recognized by GDR3 are overplotted, color-coded according to our membership evaluation (see Sect.\,\ref{sec_membership} and Appendix\,\ref{appendix_cluster_membership} for details). The lower panel shows a DSS2 (red channel) image of the same region. The extent of the superstamp is indicated in red. The small inset in the corner shows the location of the superstamp in the Kepler FOV C05. Both panels are shown such that their orientations are as similar as possible (see coordinates in panel b).
    }
    \label{fig_superstamp_overview}
\end{figure}

    M\,67 was in the field of view during three of those campaigns, namely C05, C16, and C18. Special attention was paid to the cluster during C05 and data for an extended region around the cluster center were collected. This \emph{superstamp} covers the central part of the cluster (cf. Fig.\,\ref{fig_superstamp_overview}) from April to July in 2015 and is the basis for the present work. The coverage during C16 (Dec.\,2017 to Feb.\,2018) and C18 (May\,2018 to Jul.\,2018) may provide additional opportunities for future explorations. 
    
    During C05, \kepler{} obtained 3663 images in its slow-cadence mode. These span a baseline of approximately 70\,d. Because of data transmission limitations, \kepler{} did not transmit full frame images, but only certain preselected regions. These are stored and made available as target pixel files (TPF). For C05, TPFs were selected such that they cover a continuous region around the center of M\,67. \cite{2018RNAAS...2Q..25C} used those TPFs to create a continous region, the \emph{superstamp}, around the cluster, measuring 400 by 400 pixels in size ($\approx0.5^\circ\times0.5^\circ$). The coverage of the cluster region in the superstamp and its location in the \kepler{} FOV are illustrated in Fig.\,\ref{fig_superstamp_overview}. Additionally, they created a new, time dependent astrometric calibration for each individual image, overwriting the default, time independent \ktwo{} pipeline results to account for the jitter of the targets. We note that this astrometric solution turns out to be both very good and useful, and it is of fundamental importance for this work. We note further that \cite{2018RNAAS...2Q..25C} omitted a small number of cadences, resulting in only 3620 superstamp images for M\,67.

\subsection{Stellar content of the superstamp}\label{sec_membership}

    The superstamp covers a region of about $0.25$\,deg$^2$, aligned with the cluster center. The most consistent and extensive catalogs for any random region of the sky are from the \gaia{} mission and thus we use \gaia{} DR3 \citep[][GDR3 hereafter]{2022yCat.1355....0G} as a foundation for our work. We obtain a subset of GDR3 data for the region covered by the superstamp. This subset contains about 2000 stars, but the \ktwo{} mission only provided light curves for 96 of these. Due to the motion of the telescope, individual targets on the edges may shift in and out of the field of view. Here, we did not limit the stars by brightness or any other parameter. However, we note that, as expected, the parameter coverage of the individual stars varies widely in the sample, generally becoming more sparse toward the faint end. We note further that the spatial resolution of \gaia{} greatly exceeds that of \kepler{}\footnote{\kepler{} had 4\,\arcsec pixels.}, while on the other hand \gaia{} omits certain significantly brighter sources\footnote{e.g., those with problematic astrometric solutions, i.e., very active stars or binaries.}. This requires us to carefully assess which star's light we are analyzing.

\begin{figure}
    \centering
    \includegraphics[width=\linewidth]{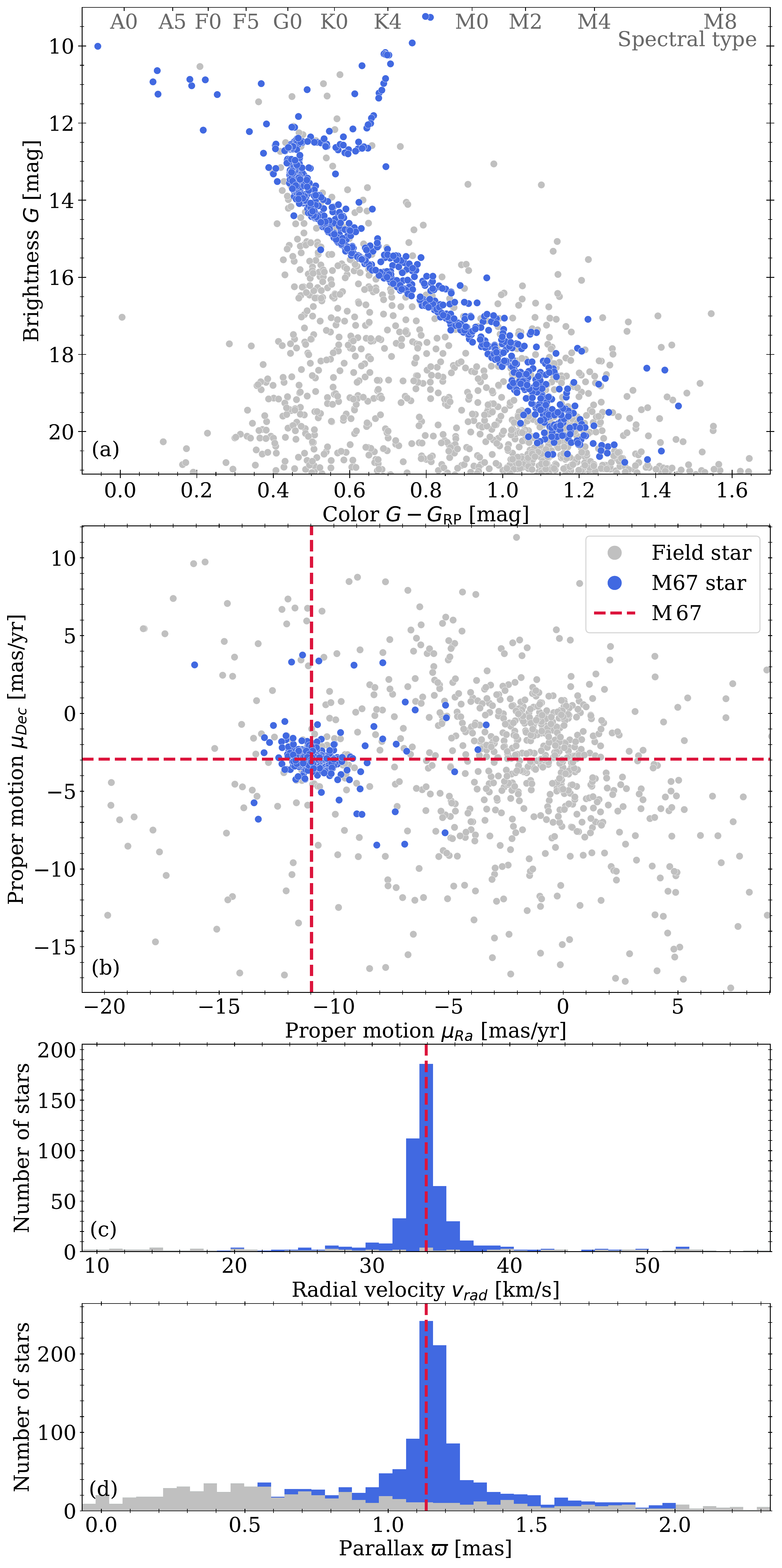}
    \caption{
        Overview of GDR3 stars in the field of view and their M\,67 membership status (following Sect.\,\ref{sec_membership} and Appendix \ref{appendix_cluster_membership}). In all panels, blue stars denote \emph{members} while non-members are gray. Panel (a) shows a color-magnitude diagram in \gaia{} $G-G_\mathrm{RP}$ color, panel (b) the proper motions, panel (c) a histogram of radial velocities, and panel (d) a histogram of the stellar parallax. Dashed red lines indicate the position of M\,67.
        }
    \label{fig_sample_overview}
\end{figure}

    M\,67 is obvious in the field-of-view (FOV) sample in both photometry and astrometry (see Fig.\,\ref{fig_sample_overview}). In a color-dependent sense, the brightest stars in the field belong to M\,67, with the exception of a few very bright foreground stars. We complement the GDR3 sample with data from the \emph{Two-micron All Sky Survey} \citep[\tmass{},][]{2003yCat.2246....0C,2006AJ....131.1163S}, \emph{US Naval Observatory B catalog} \citep[\usno{},][]{2003AJ....125..984M}, \emph{Guide Star Catalog 2.4.2} \citep[\gsc{},][]{2008AJ....136..735L}, and the \emph{Ecliptic Plane Input Catalog} \citep[\epic{},][via their \tmass{} IDs, where available]{2017yCat.4034....0H} and the \simbad{} database\footnote{\url{simbad.u-strasbg.fr/simbad}}. This crossmatch provides us with multiband photometry and radial velocity information for a large number of stars. Most of the radial velocity measurements are from groundbased surveys by \cite{2015AJ....150...97G,2021AJ....161..190G} and \cite{2018AJ....156..142D}. We note that these surveys prioritized cluster stars and, as such, the availability of radial velocities is strongly biased toward member stars and is brightness limited, with the cutoff being approximately $G=17$\,mag. We supplement those radial velocities with additional ones provided in GDR3. Based on this, admittedly heterogeneously available information, we designate each star in the field either as a \emph{member}, or as \emph{field star}. Details about this designation are described in Appendix\,\ref{appendix_cluster_membership}. Figure\,\ref{fig_m67_membership} depicts the resulting cluster sample which amounts to 971 \emph{members} and 1042 \emph{field stars}. Around 80\,\% of the \emph{members} can be assumed to be main sequence stars, spanning a brightness range from $G=13$\,mag down to $G=21$\,mag.

    We also include a designation for stars displaying signs of binarity. Those are derived either from the extensive information present in the \simbad{} archive or from a star's position in the CMD indicating a photometric binary. The former leads to a star being assigned binary status if it is listed in \simbad{} as an eclipsing binary (\verb+EB*+), spectroscopic binary (\verb+SB*+), or cataclysmic variable (\verb+RSCVn+ or \verb+CataclyV+). Their designations are based on various catalogs and studies, too numerous to be listed here, do not include all actual binaries in the sample, or could misidentify a single star as a binary. However, judging from the overall picture that emerges, we find this designation to be relatively reliable and consistent. An identification of binaries is crucial because multiplicity can lead to a number of issues, for example, tidal interactions or mass exchange may change the stellar angular momentum and as such disqualify a star for our purposes.  Those stars need to be discarded from our final sample.

    Henceforth, we will use $G-G_\mathrm{RP}$ as the color parameter for the stars. This has the advantage that it is essentially independent of \gaia{} $G_\mathrm{BP}$, which is either very uncertain or unavailable for the faint and red stars. The disadvantage of this choice is the unavailability of a measured reddening parameter in this color. We will address this issue below. Plots in other colors (such as $B-V$, $V-K$, or $G_\mathrm{BP}-G_\mathrm{RP}$) that are potentially useful to the community are provided either parallel to the main plot or as supplementary plots in Appendix\,\ref{appendix_supp_figures}. We will not use the individual parallaxes of the stars to obtain absolute magnitudes. Stellar photometry is much better constrained and more widely available. As such, for a subsequent isochrone fit to the cluster, we will apply the cluster parallax (and reddening, cf. Table.\,\ref{tab_m67_overview}) to the isochrone to match the cluster data instead.

\subsection{Color magnitude diagram for M\,67}\label{sec_cluster_reddening}

\begin{figure}
    \centering
    \includegraphics[width=\linewidth]{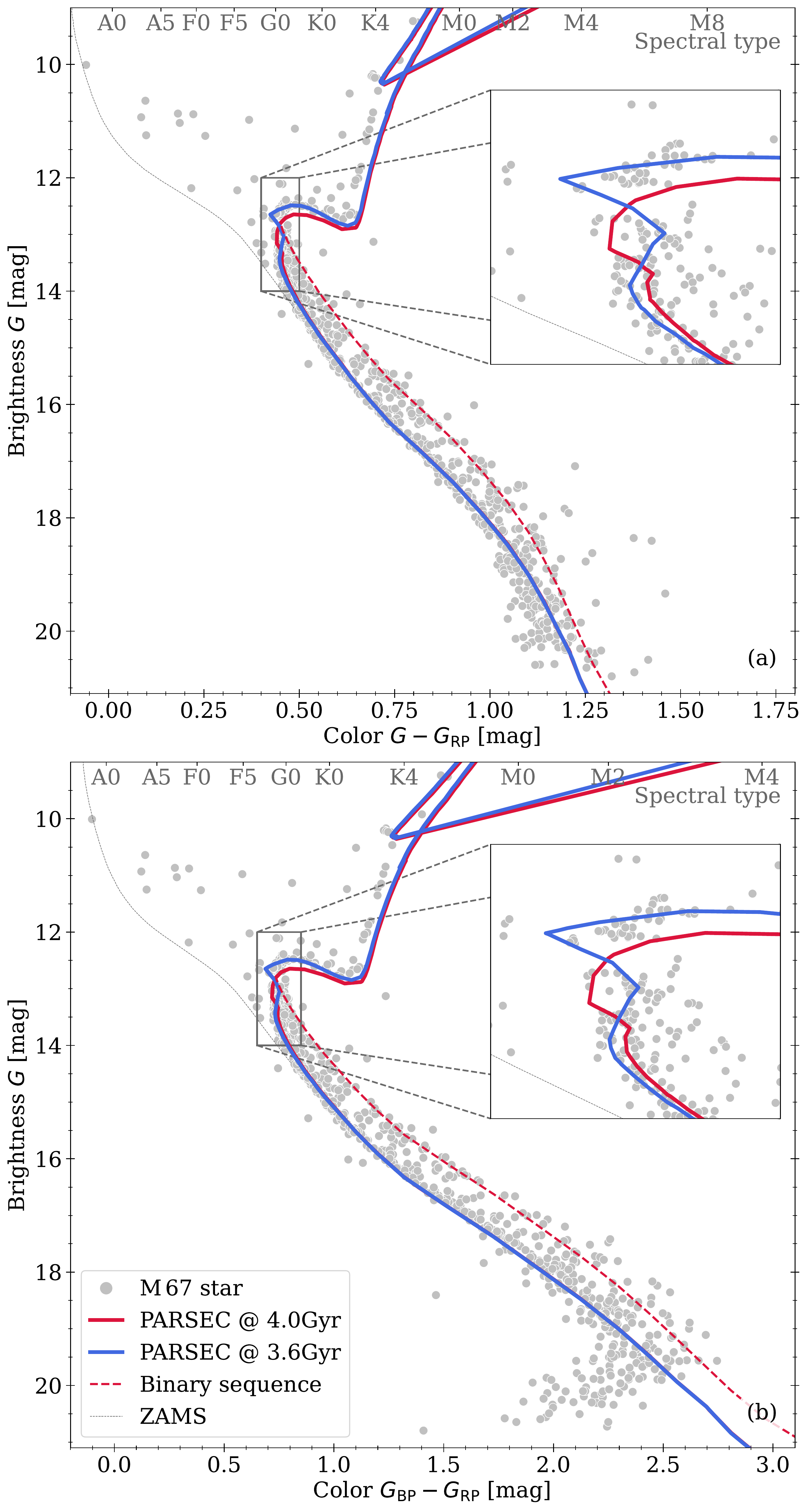}
    \caption{
        Isochrone fit to the identified cluster members in two different \gaia{} colors. The {\tt PARSEC} isochrone $G$-band magnitude is adjusted for  the parallax $\varpi$ and extinction $A_G$ using Eq.\,\eqref{eq_m67_gaia_extinction}. The {\tt PARSEC} $G_\mathrm{BP}-G_\mathrm{RP}$  color is reddened by $E(G_\mathrm{BP}-G_\mathrm{RP})$ from Eq.\,\eqref{eq_m67_gaia_reddening}. The dashed line indicates the nominal position of the equal-mass binary sequence (main sequence stars only). An alternative version using $B-V$ and $V-K$ is shown in Fig\,\ref{fig_m67_johnson_cmd}. 
    }
    \label{fig_m67_membership}
\end{figure}

    To reconfirm the age of the cluster and to determine the $E(G-G_\mathrm{RP})$ reddening and extinction $A_G$, we proceed with an isochrone fit to the cluster sample (cf. Fig.\,\ref{fig_m67_membership}). We opt for an empirical determination of $E(G-G_\mathrm{RP})$ for M\,67 based on the offset observed between a distance-corrected isochrone and the observed cluster sample. We work with the isochrones from the Padova and Trieste Stellar Evolutionary Code \cite[{\tt PARSEC}\footnote{\url{stev.oapd.inaf.it/cgi-bin/cmd}}][]{2012MNRAS.427..127B,2014MNRAS.444.2525C,2015MNRAS.452.1068C}. In principle, the isochrone fit has four free parameters: the cluster metallicity, cluster age, reddening, and extinction. However, we can constrain some of them. The cluster metallicity has been measured repeatedly in the past, with values ranging from $\left[\text{Fe/H}\right]=-0.01$ to $\left[\text{Fe/H}\right]=0.03$. Thus M\,67 is essentially of solar metalicity. We adopt $\left[\text{Fe/H}\right]=0.03$ \citep{2016A&A...585A.150N} and note that changes within the range described above do not change the result below in a significant way.

    Following the method described in \cite{2020A&A...644A..16G}, we use the coefficients obtained by \cite{2018MNRAS.479L.102C} for the relation between the extinction in \gaia{} $G_\mathrm{BP}-G_\mathrm{RP}$ and Johnson $B-V$ colors. We adopt $E(B-V)=0.04$\,mag\footnote{$A_V=0.124$\,mag} \citep{2007AJ....133..370T} and calculate
    \begin{equation}
        E(G_\mathrm{BP}-G_\mathrm{RP}) = 1.339 \cdot E(B-V) = 0.054\,\text{mag} \label{eq_m67_gaia_reddening}
    \end{equation}
    and combine it with the relation that we already used for Ruprecht\,147 \citep{2020A&A...644A..16G} for extinction
    \begin{equation}
        A_G = 2.0 \cdot E(G_\mathrm{BP}-G_\mathrm{RP}) = 0.107\,\text{mag.} \label{eq_m67_gaia_extinction}
    \end{equation}
    With $A_G$ constrained, we adjust $E(G-G_\mathrm{RP})$ to best reproduce the observed cluster sequence and find
    \begin{equation}\label{eq_m67_grp}
        E(G-G_\mathrm{RP})=0.03\pm0.005\,\text{mag}.
    \end{equation}
    This fit uses the the distance modulus listed in Table\,\ref{tab_m67_overview}. As a byproduct of this calculation, we have obtained $E(G_\mathrm{BP}-G_\mathrm{RP})$. We note that this value also produces a consistent fit between the isochrone and the cluster in $G_\mathrm{BP}-G_\mathrm{RP}$ (see panel (b) of Fig.\,\ref{fig_m67_membership}). Below, we also provide $V-K$ as a \gaia{}-independent color that omits the issues of blue bands. For the corresponding reddening, we use the prescription by \cite{1990ApJ...357..113M} to obtain
    \begin{equation}
        E(V-K) = \frac{A_V}{1.1} = \frac{3.1\cdot E(B-V)}{1.1} = 0.11\,\text{mag.}
    \end{equation}

    M\,67 is generally believed to be about 4.0\,Gyr old, somewhat younger than the sun. In the process of fitting isochrones to the cluster sequence, we notice that the isochrones suggest a slightly younger age for the cluster, namely $t=3.6\,$Gyr. This is mostly caused by the shape of the turn-off point and the subgiant sequence (see zoomed plots in Fig.\,\ref{fig_m67_membership}). However, a 4\,Gyr {\tt PARSEC} isochrone still provides a reasonably good fit to the cluster stars.After all, isochrone fitting is not the purpose of this paper. We note that in no color do the isochrones reproduce the faint red end of the main sequence completely satisfactorily. We note further that the observed red giant branch is slightly bluer than the isochrone prediction. This seems to point to an underlying problem in the isochrones (or the input physics) rather than the photometry as it is also visible in other, non-\gaia{} colors (cf. Fig.\,\ref{fig_m67_johnson_cmd}).

\section{Light curves and rotation periods from the \ktwo{} superstamp} \label{sec_lightcurves}

    In this section we describe the creation and correction of stellar light curves from the \ktwo{} superstamp. We lay out briefly the problems inherent to \ktwo{} data and our approach to dealing with them. A more thorough explanation of all the technical details is provided in Appendix \ref{appendix_correction_main}.

    During the \ktwo{} mission, some vital parts of the \kepler{} telescope became dysfunctional. The loss of certain parts of the detector (see missing CCDs in panel (b) inset in Fig.\,\ref{fig_superstamp_overview}) does not affect us. However, the well known pointing problems of the telescope do. Essentially, the telescope was in a constant state of drift throughout the observations, causing the stars to move slowly across the detector. This drift was periodically corrected (i.e., at $\approx6$\,h intervals) by firing the telescope's thrusters \citep{2016PASP..128g5002V}. Consequently, stars move across the detector during the run. The movement is small for individual exposures and as a result there are no noticeable star trails on the images. Between the thruster firings, however, a star moves up to two pixels ($\lesssim8$\,\arcsec) across the detector. There are also significant sensitivity variations between the pixels and within the individual pixel. This means that due to the changed positions of an otherwise constant star on the detector, the recorded flux varies from image to image. Those changes, while systematic, are unique for each pixel and its environment. As such, there is no general way to correct for those systematics in a simple, wholesale manner. However, those variations are rather fast and there have been approaches to correct those systematic effects \cite[e.g., {\sc k2sc} and {\sc everest} programs by][respectively]{2016MNRAS.459.2408A,2016AJ....152..100L} with varying levels of success and usability of the resulting corrected light curves for various purposes. 

    Unfortunately, none of these prior light curves are very well suited for our purpose. The main reason is their availability for only a limited number of targets. {\sc k2sc} and {\sc everest} (and essentially all other works in this regard) operate on the \epic{} catalog and the sample of stars \emph{observed} by \kepler\footnote{see \emph{Kflag} column in the \epic{} catalog}. The list of observed stars originates from the original proposals that shaped the \kepler{} \ktwo{} mission. However, for the superstamp there are only 96 individually designated targets with light curves whereas the field contains more than 2000 stellar sources. Therefore, we create our own light curves, directly based on the superstamp FFI and a list of sources obtained from \gaia{} DR3 rather than the EPIC catalog. This means that we have to perform the photometry and corrections thereof from scratch. Below, we lay out the principal ideas and steps that we employ to create the light curves we subsequently investigate for rotation signals.

\subsection{Light curve extraction and correction}

    We have constructed a procedure aiming to extract light curves and to remove the artificial variations introduced by the effects mentioned above. The plan is to create an empirical model that captures the systematic position dependence of the flux in order to remove it. The position dependence turns out to be of a very complex nature and requires special attention to the individual stars. As such, it is rather labor intensive and requires iteration between different steps until the best result is achieved. Ultimately, we are able to create light curves that are sufficiently free of systematics for a large number of stars.

    In Appendix\,\ref{appendix_correction_main}, we provide a detailed explanation of the extraction and correction process applied to the data in order to create the light curves. There we include all the technical details and illustrate them on a sample star. Here, we only describe the general steps taken. 

\begin{figure}
    \centering
    \includegraphics[width=\linewidth]{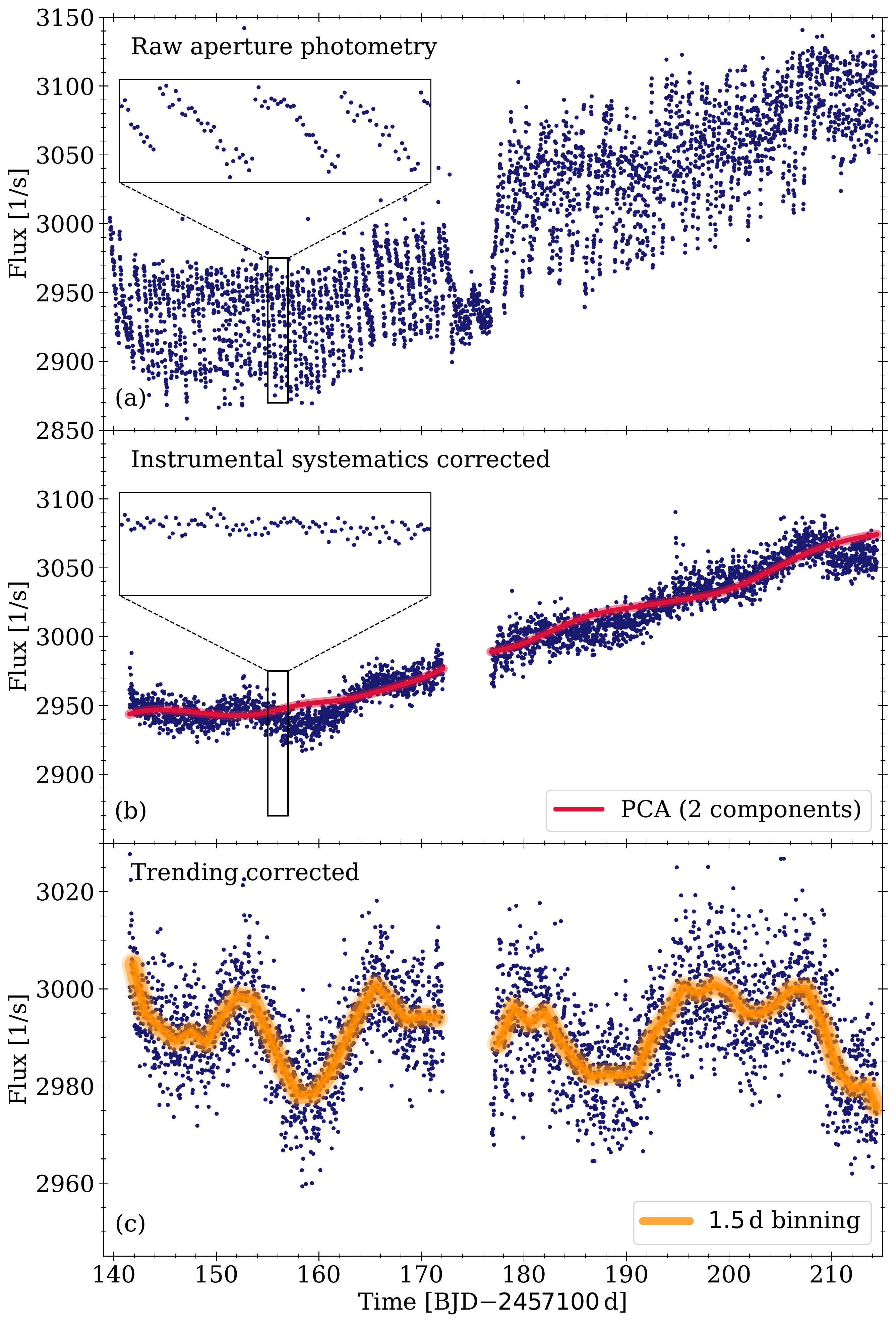}
    \caption{
        Example of the light curve reduction process using \emph{Gaia\,DR3\,604971466769552128}. Panel (a) shows the raw light curve (blue dots), Panel (b) shows an intermediate light curve after the systematics correction (blue dots) together with the PCA correction (red) that is applied to the create the final reduced light curve that is shown in panel (c). The orange line in panel (c) shows the reduced light curve with 1.5\,d binning to highlight the now visible intrinsic stellar variability. The insets in (a) and (b) zoom in on a 2\,d span of the light curve. The drift-induced flux changes and the jumps caused by the realignment of the telescope are readily visible in panel (a), while their disappearance can be seen in panel (b).
    }
    \label{fig_full_lightcurve_process}
\end{figure}

    First, we introduce a naming convention that we will use throughout this work. The \emph{total flux variation} of an extracted light curve for a star is the combination of three things. First there is the \emph{intrinsic astrophysical variability} of the star, which is the variation we are actually seeking. This flux is modified by the motion of the star across the detector in combination with the detailed pixel sensitivity. We will refer to the changes caused by this effect as the \emph{instrumental systematics}. Superimposed on those are quasi-systematic, long-term patterns in the data which are common between similar stars, and which we will refer to as \emph{trending} or \emph{trends}. We illustrate the correction steps for an example star in Fig.\,\ref{fig_full_lightcurve_process}.

    We define an aperture mask for each star on the superstamp and create the raw light curve by simply summing up the enclosed flux. The aperture mask is defined manually and adjusted individually for each star to provide the best (i.e., cleanest, low noise, best systematics removal) light curve possible. Depending on the star being considered, the instrumental systematics appear to introduce flux variations up to a few tenths of a magnitude (cf. panel (a) of Fig.\ref{fig_full_lightcurve_process}). We correlate the flux with positional changes of the star on the detector. The time-dependent position is obtained from the world coordinate system \citep[WCS\footnote{See also \url{fits.gsfc.nasa.gov/fits_wcs.html}},][]{2002A&A...395.1061G} which is part of the superstamp FFI data for each individual image and was created by \cite{2018RNAAS...2Q..25C}. We proceed to model this behavior (flux as a function of position) with a fifth-order polynomial in order to remove it. Here, we also need to handle jumps in the data that do not allow us to process the light curve as a whole but only in chunks as described in Appendix\,\ref{appendix_correction_main}. This is probably the most intricate step and provides us with a set of light curves \emph{largely} free of the instrumental systematics (cf. panel (b) of Fig.\ref{fig_full_lightcurve_process}). 

    The processed light curves still contain \emph{trending} at this point, which we then correct using Principal Component Analysis (PCA). For this, we select a subsample which excludes all stars with obvious intrinsic variability. For any given star that is to be corrected we select all similar stars (in terms of brightness and position) from this subset as the basis for the PCA. This procedure needs such selectivity as there are significant differences between the stars in the trends with respect to those parameters. And with an 80\,d light curve, expected rotation periods of 20\,--\,30\,d and trends of similar length, there is a limited time baseline available for the light curve to express those variabilities. Degeneracy between them is a real issue and it again requires attention to the individual stars to identify. From the individually tailored PCA basis we calculate a correction based on two components and apply it to the light curves. This results in the final light curve product (cf. panel (c) of Fig.\ref{fig_full_lightcurve_process}) that we provide and use in the subsequent period analysis.

    This process is not successful for all stars; faint stars and those in particularly crowded areas cannot be processed satisfactorily. In fact, most light curves still show remnants of systematics and trends. However, those are now typically small compared with the stellar signal itself and as such more of a nuisance than a real stumbling block. We remove all stars for which we cannot obtain light curves of reasonable quality. This includes stars that are located in problematic areas on the detector, that is, those where large systematic effects impact groups or rows of pixels. Some of these regions are visible in the upper panel of Fig.\,\ref{fig_superstamp_overview} (horizontal and vertical structures, e.g., around $y=245$). Our usage of aperture photometry, does not allow us to separate heavily blended sources. Furthermore, the correlation between location and flux cannot be adequately reproduced and removed for every star. This is especially true for stars which show very rapid intrinsic variability (i.e., variability similar to the systematics). We are also forced to remove the first 2\,d of the light curves and a central part (172\,--\,178\,d) as those simply defy any correction attempt with our approach.

    The resulting light curves show a wide variety of signals in addition to the rotation signals we seek. We also identify pulsations and eclipsing binaries (some of them with secondary eclipses). Traces of the instrumental systematics remain for a number of stars; however, those are often minor compared with the observed intrinsic signal. We note that we do not (and cannot) attempt to extract and fine tune each and every star in the field of view. Each target apparently requires individual attention, from the design of the pixelmask to the evaluation of the resulting light curve. Given our science goal, we limit ourselves only to M\,67 members, together with a sufficient number of non-members to provide a good basis for the PCA.

    We also emphasize that the light curves we produce with this method are not intended to rival large scale correction endeavors such as {\textsc everest} or {\textsc k2sc}. Our light curves are purpose-built, with assumptions made in the creation that are invalid for other purposes (e.g., astroseismology). However, we include a comparison between our results and light curves provided in the larger endeavors in Appendix\,\ref{appendix_validation} to validate our results. We will provide all light curves of stars for which a signal was identified as part of the auxiliary data to this publication. Figure\,\ref{fig_lightcurve_sample_1} displays the light curves for those 47 stars that are single M\,67 main-sequence stars and where we were able to identify a periodic rotational signal.

\subsection{Period analysis} \label{sec_analysis}

\begin{figure}
    \centering
    \includegraphics[width=\linewidth]{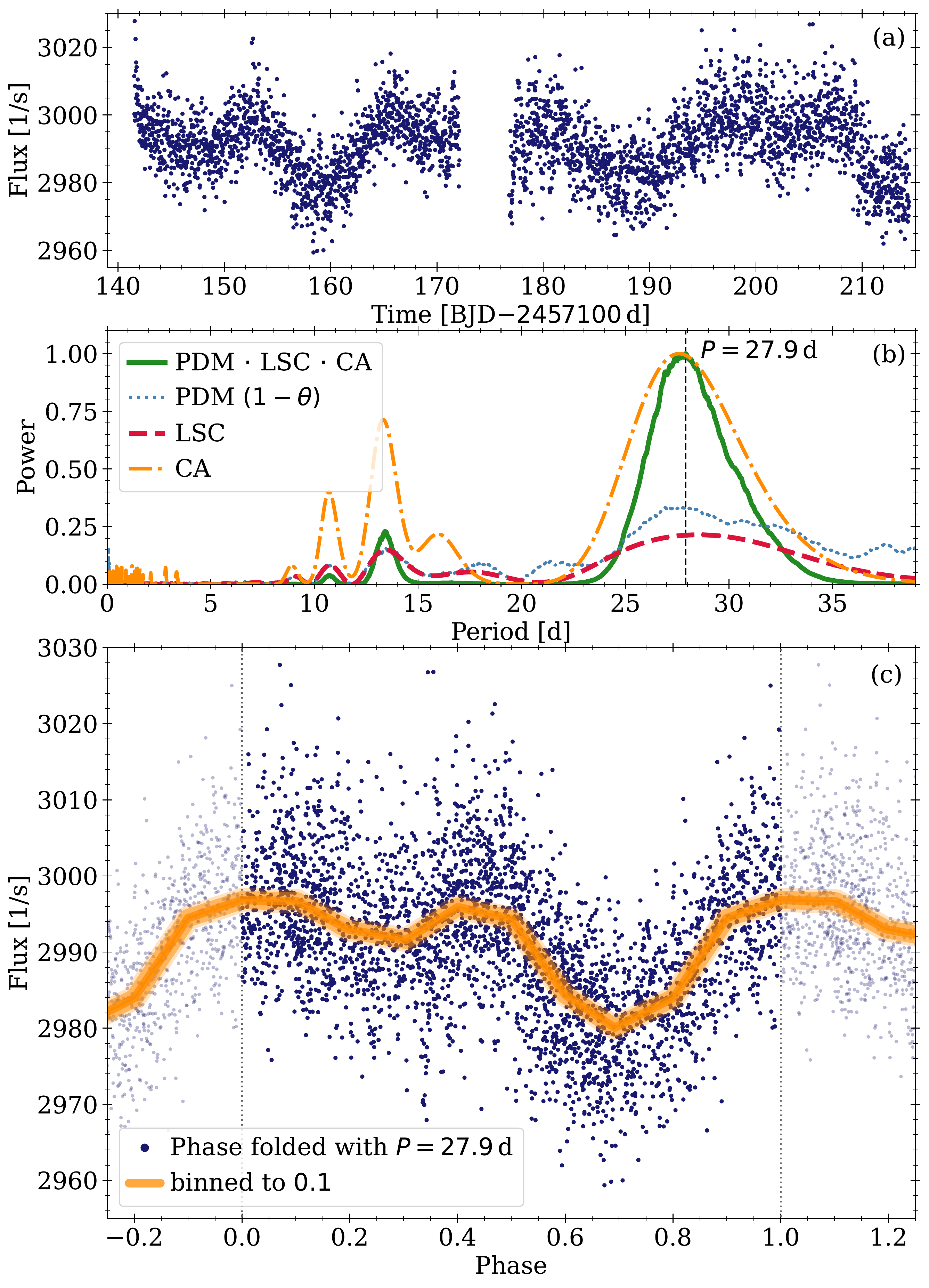}
    \caption{
        Result of period analysis on the light curve for \emph{Gaia\,DR3\,604971466769552128}. Panel (a) shows the light curve produced, one which exhibits a clear periodic signal. The power spectra in panel (b) are obtained using multiple methods as indicated. The Clean (CA) and Combined power spectra are normalized to their maxima and the PDM spectrum is shown as $1-\theta$. Panel (c) shows the phase-folded light curve, folded with $P=27.9$\,d, corresponding to the peak in the combined spectrum. 
    }
    \label{fig_period_analysis}
\end{figure}

    For the period analysis, we continue with a hands-on approach for the individual light curves. Unlike the case of most ground-based data, space-based data are both well-sampled and, for the most part, equally sampled. Owing to that, a long-period signal can usually be identified by eye easily when present (cf. panel (a) in Fig.\,\ref{fig_period_analysis}). Manual inspection also permits the identification of a periodic signal when spot evolution or remnants of systematics or trends are present. Automated algorithms often struggle or fail outright in such cases. However, we subject all light curves to an array of algorithms to verify and quantify a generally obvious signal (panel (b) in the figure). We run period finding algorithms employing a Lomb-Scargle \cite[LSC,][]{1982ApJ...263..835S} periodogram\footnote{Based on the \texttt{astropy} v5.1 implementation \citep{2012cidu.conf...47V,2015ApJ...812...18V}.}, a Clean analysis \cite[CA,][]{1987AJ.....93..968R}, and phase dispersion minimization \citep[PDM,][]{1978ApJ...224..953S}.

    Whenever the results of the individual algorithms are inconsistent, a manually determined period supersedes the algorithms. Typically, the algorithms agree among each other, with the one most prone to failure being LSC, and the most consistently reliable being PDM. This is to be expected as the spot-induced features in the light curve appear to deviate strongly from a sinusoidal shape for the stars we are most interested in. We provide the derived power spectra together with the light curve data in the auxiliary files. Fig.\,\ref{fig_lightcurve_sample_1} includes plots of the phase folded light curves.

    Period errors are derived in much the same way as the periods themselves. The presence of data systematics and spot evolution makes the usage of an automated algorithm for error determination problematical. Therefore, we decided to determine errors manually based on the phase folded light curve. We folded the light curve with different periods and set the error range such that for every period covered a phase folded light curve provides a reasonable result in matching the spot-induced features. The uncertainty arises from a number of factors: (1) the number of periods covered, that is, how many repetitions of the features we observe in the K2 C5 baseline, (2) the general signal-to-noise of the variation, and (3) the amount of spot evolution, together with the remaining systematics. Generally, this results in errors that are in the order of $\pm$5\,--\,10\,\%. This procedure is supported during the comparison between our derived periods and those from the literature (see below in Sect.\,\ref{sec_discussion} and particularly Table\,\ref{tab_period_comparison}.) Furthermore, ignoring the three outliers, 37 of the 44 remaining stars of our sample overlap with the trend line (see Fig.\,\ref{fig_cpd_final} below), suggesting that our uncertainties are on the order of $1.5 \sigma$.

\section{Results}\label{sec_results}

    In this section we present the stars for which periodic signals were identified. We investigate outliers, construct a Color-Period diagram (CPD) for M\,67 of single MS-stars to be used in the subsequent discussion, and present the final CPD for M\,67.

\subsection{The raw Color-Period Diagram}

    We identify 136 stars in the FOV that exhibit periodic signals of which 96 are M\,67 \emph{members} which exhibit signals that can be attributed to stellar rotation and 83 {of which are MS stars. Apart from the few stars that do not have a convective envelope (namely the blue stragglers with spectral types earlier than mid-F), the sample covers spectral types from early-G to mid-M. Figure\,\ref{fig_cpd_raw} gives an overview of the stellar rotation period distribution found. The detected periods range from a few hours up to 38\,d, the latter abutting the detection limit of \ktwo{} C05. A significant fraction of the cluster sample (53 stars, 43 on the MS) are binaries; those are typically both fast rotating and among the bluer stars of the sample.

 \begin{figure}
    \centering
    \includegraphics[width=\linewidth]{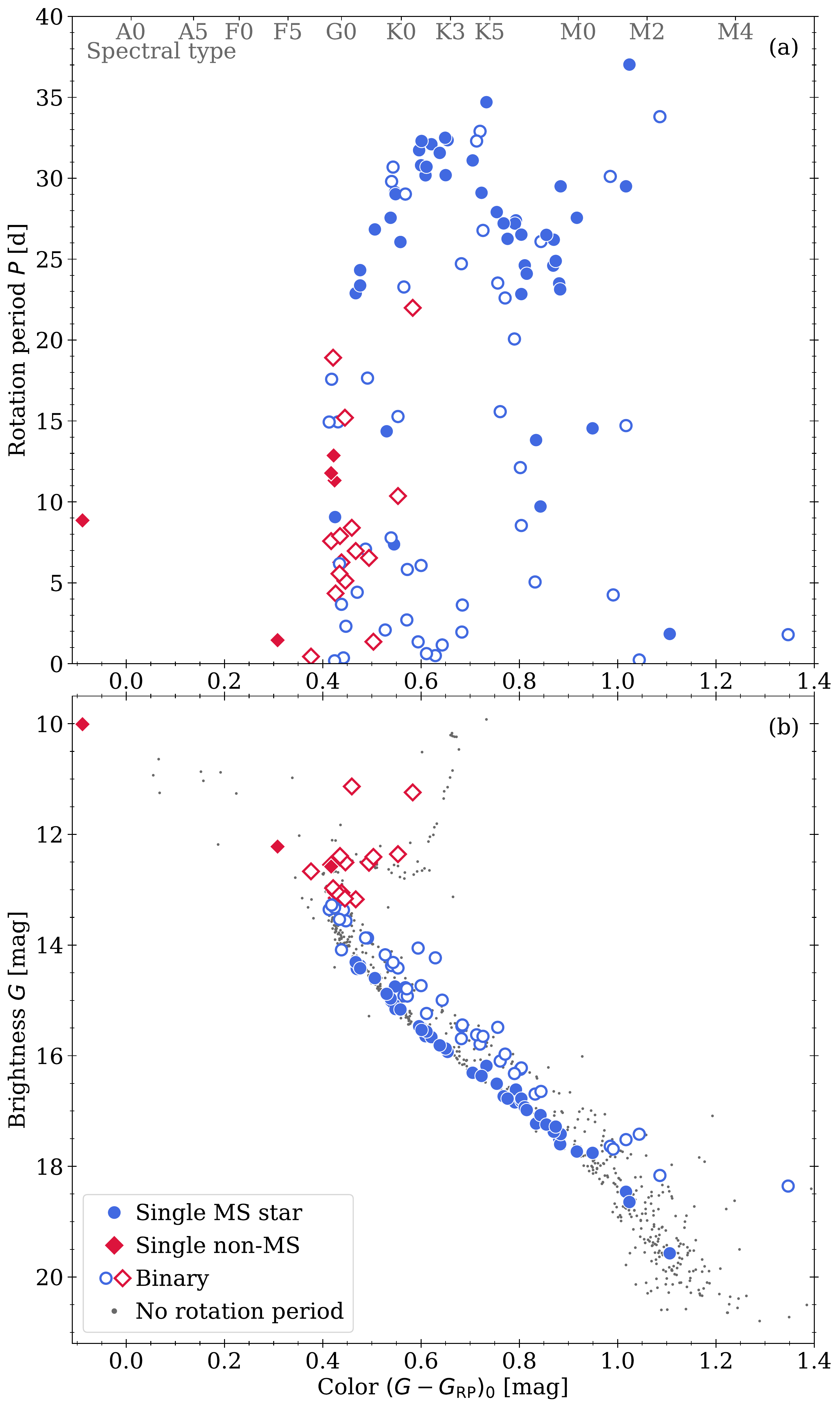}
    \caption{
        Stars for which we have identified rotation periods. Panel (a) shows a color-period diagram (CPD) and panel (b) the corresponding color-magnitude diagram (CMD). Stars belonging to M\,67 according to our membership analysis are shown as blue circles (the main sequence stars) and red squares (post-MS stars). Open symbols (in all of the above) mark those that show signs of binarity. Error bars are suppressed here for visibility reasons.
    }
    \label{fig_cpd_raw}
\end{figure}

    To enable the usage of this sample as a calibration set for gyrochronology, we also remove all stars that detract from this purpose. We define these as any that violate one of the following statements:
    % \begin{enumerate}
        (1) The stars are M\,67 cluster members.
        (2) Their colors are a valid proxy for their mass, which means they are unevolved.
        (3) They are not likely to have experienced angular momentum transfer that has caused \emph{externally induced} changes in their rotation rate, and have spun down over time based on the processes described above.
    % \end{enumerate}

    Adherence to the first statement is simply addressed through our membership analysis. Figure\,\ref{fig_field_star_cpd} in the appendix shows that the field stars (red symbols there) are scattered all across the CPD, corresponding to a wide range of (unknown) ages. Accordingly, we remove them from our sample.

    The second statement is violated by evolved stars. Stellar evolution causes stars to change their colors as well. Changes on the main sequence are small enough to be a nonissue here. However, when stars leave the main sequence they become significant. This means that the color ceases to be a valid proxy for mass for all stars that have left the main sequence and have undergone drastic changes in their colors (typically turning redder toward the red giant branch). We will address those stars separately below in Sect.\,\ref{sec_mass_proxy} with respect to what happens when we forgo color and revert to assess the stars directly via their masses. For now, we omit them from our sample by removing all stars brighter than $G=13.2$\,mag (red symbols in Fig.\,\ref{fig_cpd_raw}).

    The third statement is violated by stars that have experienced angular momentum transfer in the past. Prime candidates for this are stars in close binary systems. Angular momentum is typically transferred from orbit to rotation, causing a spin up (rejuvenation of the rotation rate). Consequently, removal of binaries from studies like ours is performed almost habitually. Here we see the effects of binarity clearly thanks to our sample size. A significant fraction of the cluster sample (53 stars, 43 on the MS) are binaries; and those are both fast rotating and typically among the bluer stars of the sample. Examination of Fig.\,\ref{fig_cpd_raw} shows that whereas the single cluster members are concentrated at the long-period end of the distribution, already suggesting a sequence against color, the binaries are scattered over the entire rotation period range, with almost all having shorter periods than the single stars.

    To our knowledge, this phenomenon has not been observed as starkly before. After an extensive radial velocity and photometric survey of the young open clusters M\,35 ($\sim$150\,Myr) and M\,34 \citep[$\sim$200\,Myr,][]{2009ApJ...695..679M}, \cite{2011ApJ...733..115M} noticed that the rotational distribution of binaries beyond the influence of tides was marginally skewed toward faster rotation. Other work in young open clusters \citep[e.g., in NGC\,2516 and NGC\,3532 by][respectively]{2020A&A...641A..51F,Fr2020} found that while certain (presumably very close) binaries were anomalous, the vast majority of the photometric binaries were rotationally indistinguishable from the single stars. In the far older (2.5\,Gyr) but much sparser cluster Ruprecht\,147 \citet{2020A&A...644A..16G} noted that three of the four binaries therein were off the single star sequence. However, that is too few stars to be able to draw conclusions. In contrast, the situation here is obvious. 34 out of the 43 main sequence stars with signs of binarity exhibit slightly to considerably faster rotation than the single star sequence.

    In fact, the scatter is so great that it seems unlikely that photometric or tighter binaries of this age are suitable for gyrochronology. Their ages will have to be determined by other means. Although we only have indications for photometric binarity for many of our stars (increasingly toward the fainter end), this already appears to be enough to identify a star as unsuitable. We stress that binarity does not mean that a star does not agree with the rotation of a similar single star per se; however, it is obvious in our results that \emph{it is considerably more likely than not} that rotation is affected. Consequently, we remove all stars with signs of binarity from our sample, including the nine binaries whose positions agree with the single star sequence.

\begin{figure}
    \centering
    \includegraphics[width=\linewidth]{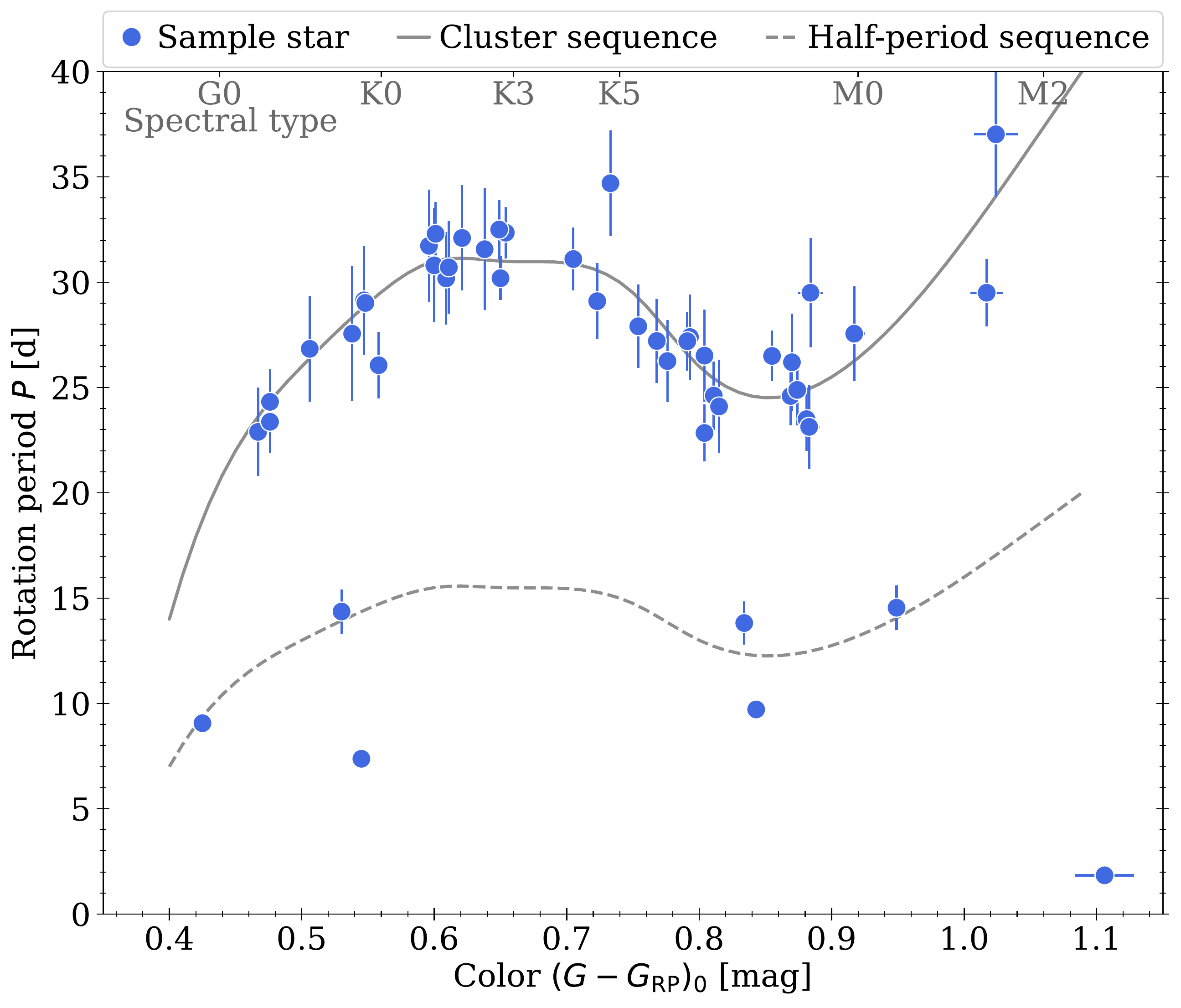}
    \caption{
        Color-period diagram for our sample, now including rotation period uncertainties. Only \emph{member} stars on the main sequence, and with no indications of binarity are displayed. An approximation of the emerging cluster distribution (solid gray line) is overplotted, together with its half-period counterpart (dashed line).
    }
    \label{fig_cpd_reliable}
\end{figure}

    The above pruning leaves us with the 47 stars that are displayed in Fig.\,\ref{fig_cpd_reliable}, enabling a closer look at the emerging period distribution and its features. We find a group of slower rotating stars with rotation periods between 15 and 35\,d, spanning from early-G to early-M. They follow a somewhat sinusoidal shape, with a maximum at $(G-G_\mathrm{RP})_0 \sim 0.65$ and a local minimum at $(G-G_\mathrm{RP})_0 \sim 0.85$. We identify this with the classical \emph{slow rotator} sequence. We have indicated this group with a simple trend line (cf. solid line in the figure). It is created from a simple cubic interpolation to points listed in Table\,\ref{tab_cluster_sequences}. We stress that this line is drawn solely to guide the eye and indicate the cluster sequence. We will employ similar indications in color-period diagrams for other clusters below. Another group of stars apparently follows the same distribution but at half the period. This \emph{half-period} sequence is formed by double dipping stars which feature two spots whose signals in the light curve are indistinguishable and as such appear to have only half the actual period.

\subsection{Double dipping stars}\label{sec_period_doubling}\label{sec_double_spot_stars}

    Stars can exhibit more than one significant star spot (or group). In fact, as \citet[][see also \cite{2020AN....341..513T}]{2018ApJ...863..190B} have shown, it is more likely for stars with longer rotation periods to exhibit more than one spot at a given time. We recall that we use the term \emph{spot} as a simple handle for any coherent activity (combination of) features on the stellar surface that is visible as a significant modulation in our light curves. This explicitly includes faculae, which are assumed to be the dominant flux-altering feature at the age of M\,67 \citep[e.g.][]{2019A&A...621A..21R}. Following the results of \cite{2018ApJ...863..190B}, at the rotation rates we find for M67 it is up to ten times more likely. We indeed find that about half of the stars in our sample exhibit signs of more than one surface feature. This may create an additional problem. If those features are sufficiently similar in shape and close to a phase shift of 0.5, they may become indistinguishable. This means that we are likely to find a number of stars whose determined rotation periods will be half of their actual periods. Sufficient numbers of such stars will form a sequence at half the actual period, a \emph{half-period sequence}. 

\begin{figure}
    \centering
    \includegraphics[width=\linewidth]{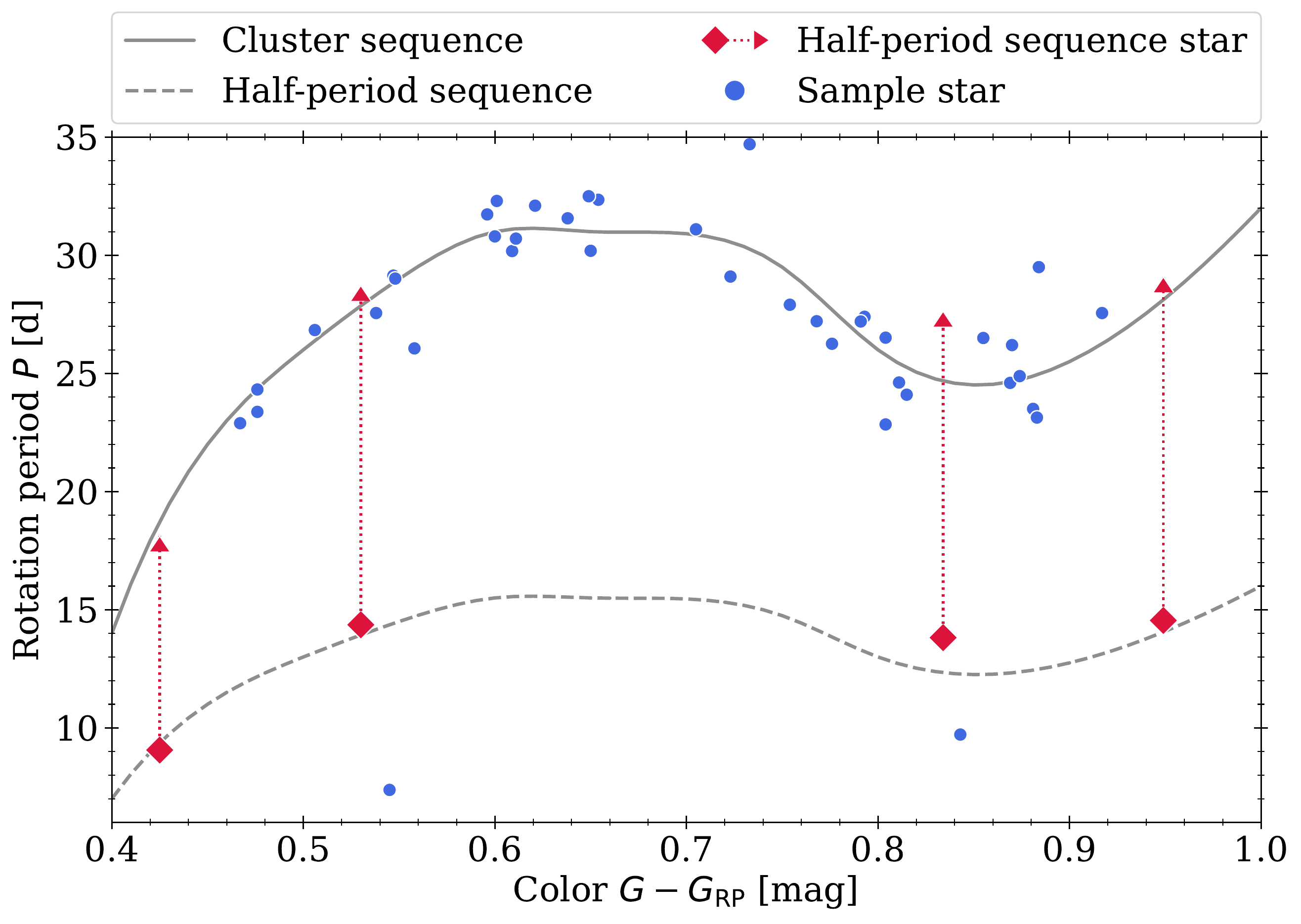}
    \caption{
        CPD for our M\,67 sample emphasizing stars on the \emph{half-period sequence}. The cluster sequence, approximated by an interpolated line, is overplotted in gray. The dashed line shows its half-period counterpart. Stars for which we double the measured periods (red) are connected to their new positions by dashed lines.
    }
    \label{fig_cpd_sample}
\end{figure}

    Indeed, we do observe this behavior (cf. Fig.\,\ref{fig_cpd_sample}). 43 stars define a long-periodic sequence with periods between 15 and 38\,d, while four stars follow the \emph{half-period sequence}. Following \cite{2018ApJ...863..190B} and the distribution of other cluster stars, it is a reasonable assumption that those four stars have rotation periods that are actually twice the identified ones. Therefore, we adopt final rotation periods for those four stars that are twice the measured ones. We note that such light curves are also visible in the prior rotation period work on warmer stars in the M67 work of \cite{ApJ...823..2016.16B}.

\subsection{Color-period diagram for M\,67}\label{sec_final_sample}

    With the half-period sequence stars accounted for and with binaries and evolved stars eliminated, we can construct a color-period diagram for M\,67 that can be compared with other clusters and with spindown models. The final emerging CPD contains 47 stars and is shown in Fig.\,\ref{fig_cpd_final}. Appendix Table\,\ref{tab_results_large} lists these stars.

\begin{figure}
    \centering
    \includegraphics[width=\linewidth]{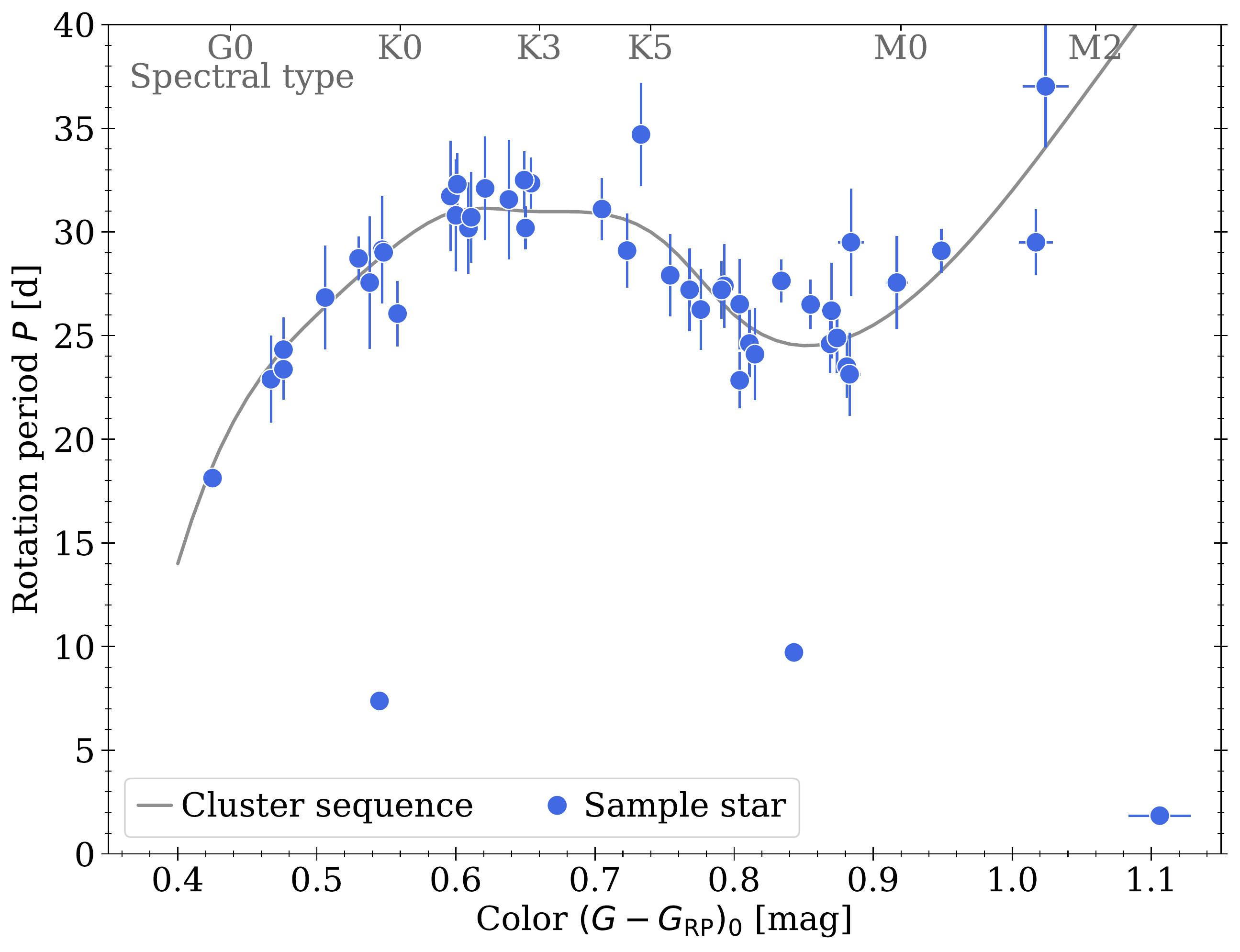}
    \caption{
        Final CPD of M67 based on our study. The corresponding data are displayed in Table\,\ref{tab_results_large}
    }
    \label{fig_cpd_final}
\end{figure}

    The emerging distribution of rotation periods, spanning from early-G to mid-M stars, shows a clear sinusoidal sequence. It rises from around 18\,d periods for G0 stars, the earliest type stars at the age of M\,67 that is still on the MS, to periods around 30\,d for K3 stars. Following the distribution redward, the periods decrease again to around 25\,d at M0 only to rise again to 30\,d and longer for mid-M. This distribution will be compared with prior work and discussed in more detail in the next section, following a discussion of the outliers immediately below.

\subsection{Outliers}

    Our sample contains three fast rotating stars with periods $P\leq10$\,d that deviate strongly from the slow rotator sequence occupied by all other stars. We suspect that the likely origin of this deviation is undiscovered binarity impacting the stellar angular momentum evolution. The stars in question are actually inconspicuous among the sample as regards their \emph{renormalized unit weight error} (\verb+RUWE+) in GDR3. (A larger than average \verb+RUWE+ value can indicate an underlying issue with the astrometric solution, which itself may originate in undiscovered multiplicity.)

    \emph{\object{Gaia DR3 604911204083987584}} ($P=7.4$\,d, $(G-G_\mathrm{RP})_0=0.55$\,mag) sits right within the area of the CPD that is typically occupied by binaries (cf. Fig.\,\ref{fig_cpd_raw}). It does have two very close companions ($0.5$ and $4.3$\,magnitudes fainter) within 5\arcsec. Those cannot be separated during light curve extraction; however, it is clear from an investigation with different pixelmasks that the star itself is the source of the observed variability. \cite{2015AJ....150...97G} actually designate the star as a non-member based on radial velocity work but \gaia{} astrometry indicates a high probability member. Consequently, we retain it in the CPD. 

    \emph{\object{Gaia DR3 604922229264424448}} ($P=9.7$\,d, $(G-G_\mathrm{RP})_0=0.84$\,mag) could potentially be part of the \emph{half-period sequence}. However, closer inspection shows that it lies significantly below, a deviation even more pronounced if one were to double its period. Furthermore, its light curve shows clear signs of continuous spot evolution which would mean both spots evolving identically, if it were a double-dipping star. This seems unlikely.
    
    For those two objects, angular momentum exchange (e.g., in a binary system) appears to be the most likely explanation. The third outlier is the reddest and simultaneously fastest rotating star of our sample \emph{\object{Gaia DR3 604969061592133376}} ($P=1.8$\,d, $(G-G_\mathrm{RP})_0=1.11$\,mag).  It could still be part of the fast rotator sequence, similar to what can be observed in other, younger clusters. It would be the first known star that is part of the fast rotator \citep[denoted \emph{convective} in][]{2003ApJ...586..464B} sequence at a confirmed age older than 1\,Gyr. However, it may also simply be a star with hidden binarity.

\section{Discussion }\label{sec_discussion}

    In this section our results for M\,67 are placed in the larger context of the rotational evolution of stars. First, we compare our results with findings in past studies and investigate deviations and inconsistencies. In the process, we address certain issues with prior work. We then build a sample of combined knowledge for rotation data from M\,67 and compare it with the few old open clusters studied to date and also with the predictions of rotational spindown models.

\subsection{Comparison with prior work on M67}

    There have been three prior studies of M67 using \ktwo{} light curves from C05: \citet[][B16 hereafter]{ApJ...823..2016.16B}, \citet[][with a followup in \cite{2016MNRAS.463.3513G}, G16 hereafter]{2016MNRAS.459.1060G}, and \citet[][see also \cite{2018ApJ...859..167E}, E18 hereafter]{2018PhDT........63E}. Those studies were performed on the \emph{presearch data conditioning} (PDC) light curves provided in the \ktwo{} archive\footnote{\url{archive.stsci.edu/missions-and-data/k2}}. These light curves were created from aperture photometry and include a correction that, to a certain degree, accounts for systematics and trending. However, as pointed out by \cite{ApJ...823..2016.16B}, the light curves are not free of either problem. E18 made additional use of self-extracted light curves from the C05 superstamp and C16 light curves. All studies operated on the low spatial resolution given by the EPIC catalog. For this comparison we match their samples with GDR3. In addition to the work mentioned above, a recent study by \cite{2022ApJ...938..118D} created light curves from 3 years worth of ground-based observation. This latter work is limited to M-dwarfs.
       
    \citet[][20 stars]{ApJ...823..2016.16B} and \citet[][98 stars]{2016MNRAS.459.1060G,2016MNRAS.463.3513G} operated on the light curves provided by the \kepler{} mission itself; only a few of those were from the superstamp region (cf. Fig.\,\ref{fig_lightcurve_distribution}). Thus, the overlap with our sample is small. \citet[][30 stars]{2018PhDT........63E} includes parts of the superstamp, but the overlap is still small. We also have some overlap with \cite{2022ApJ...938..118D}. Table\,\ref{tab_period_comparison} shows the stars that are common between the samples.     The agreement in derived rotation periods is very good, with period differences typically $\lesssim2$\,d. The only star with a significant difference is  EPIC\,211397501\footnote{Gaia\,DR3\,604895948360165888}, for which \cite{2016MNRAS.463.3513G} reported the half-period. (It is identified as a half-period sequence star by us.) In the comparison with \cite{2022ApJ...938..118D} we find that seven rotation periods (including three that were nominally rejected from their sample) are in good agreement with our periods.
    
\begin{table}[ht!]
    \centering
    \caption{
        Overlap between our sample and those from the literature
    }
    \label{tab_period_comparison}
    \begin{tabular}{lccccc}
   \hline\hline
            Gaia DR3 &             $P$ &  $P_\text{B16}$ &  $P_\text{G16}$ &  $P_\text{E18}$ &  $P_\text{D22}$ \\ 
                     &             [d] &             [d] &             [d] &             [d] &             [d] \\ 
   \hline
  598899796057231616 &             --- &            31.8 &            31.7 &             --- &             --- \\ 
  598902716634970240 &             --- &            30.7 &            30.4 &             --- &             --- \\ 
  598903678707639296 &            32.4 &            30.3 &            30.1 &             --- &             --- \\ 
  604895948360165888 &            24.2 &             --- &            12.4 &             --- &             --- \\ 
  604896051439391104 &             --- &            25.1 &            28.3 &             --- &             --- \\ 
  604896498115959296 &            23.5 &             --- &             --- &             --- &            25.9 \\ 
  604896635554924672 &            32.5 &            34.5 &             --- &             --- &             --- \\ 
  604896837417607808 &            12.9 &             --- &            24.9 &             --- &             --- \\ 
  604897631987337600 &             --- &            28.4 &            26.8 &             --- &             --- \\ 
  604897833850019328 &            24.3 &            26.9 &            26.3 &             --- &             --- \\ 
  604900651348634240 &            23.1 &             --- &             --- &             --- &         \ (25.3) \\ 
  604903331408222208 &            29.5 &             --- &             --- &             --- &            31.6 \\ 
  604903438783070208 &            37.0 &             --- &             --- &             --- &          \ (826) \\ 
  604906531159503616 &             4.4 &             --- &             --- &             4.3 &             --- \\ 
  604906840397139584 &            29.1 &             --- &             --- &             --- &         \ (28.2) \\ 
  604917354477131392 &             6.1 &             --- &             --- &             6.4 &             --- \\ 
  604917698074587136 &             7.8 &             --- &             --- &             8.0 &             --- \\ 
  604920549932807296 &            24.6 &             --- &             --- &             --- &          \ (442) \\ 
  604922229264424448 &             9.7 &             --- &             --- &             --- &          \ (9.8) \\ 
  604923848467470976 &            26.2 &             --- &             --- &             --- &            26.0 \\ 
  604930544320911744 &             --- &            18.1 &             --- &            19.8 &             --- \\ 
  604930681760054656 &            27.6 &             --- &             --- &             --- &         \ (63.9) \\ 
  604960235430488960 &             --- &            26.9 &            26.8 &             --- &             --- \\ 
  604969267746267520 &            26.5 &             --- &             --- &             --- &            26.1 \\ 
  604970131035099008 &            24.1 &             --- &             --- &             --- &          \ (237) \\ 
  604973635728426752 &             --- &            24.9 &             --- &            25.3 &             --- \\ 
  604996106997219712 &             --- &            31.1 &            29.2 &             --- &             --- \\ 
   \hline
\end{tabular}

    \tablefoot{Literature sample based on \emph{Kepler} data by \citet[][B16]{ApJ...823..2016.16B}, \citet[][G16]{2016MNRAS.459.1060G,2016MNRAS.463.3513G}, and \citet[][E18]{2018PhDT........63E}, and that of ground-based data from \citet[][D22]{2022ApJ...938..118D}. D22 periods in parentheses were rejected as outliers in their sample.}
\end{table}

\subsubsection{\cite{ApJ...823..2016.16B}}

   Given that instrumental systematics and trending are still present in the data, \cite{ApJ...823..2016.16B} subjected their sample to a PCA to remove trending and, partially, systematics. They identified 20 stars with a rotational signal, shown with red squares in panel (a) of Fig.\,\ref{fig_cpd_barnes_1}, together with our sample. Generally speaking, they subjected each light curve to the same vetting that we performed, and for the common stars, found periods in agreement with ours to within the uncertainties.

   The \cite{ApJ...823..2016.16B} sample forms a sequence spanning from early G stars with rotation rates around 18\,d to mid-K stars with rotation rates around 30\,d, and with a continuous, but non-linear, increase in period in that range. The derived distribution of stars was in very good agreement with the models of the time (that only had the sun as anchor point before) and is still in agreement with the ones today, given that it is used as a calibrator for said models.   The crossmatch with GDR3 reveals that two stars included in the B16 sample are likely photometric binaries (cf. CMD in Fig.\,\ref{fig_cpd_barnes_1} panel (b)). However, they are not deviant in any way in the CPD. An inspection of panel (c) in Fig.\,\ref{fig_cpd_barnes_1} shows that sample stars are all well-constrained cluster members.
   
   Our sample connects well to the B16 sample, overlaps with theirs in the G0 to K3 range of spectral types, and extends it in a continuous fashion to early M\,stars. It is notable that the B16 sample has its reddest stars right at the point where the range of monotonically increasing rotation with color ends (at K4).
   
\begin{figure}
    \centering
    \includegraphics[width=\linewidth]{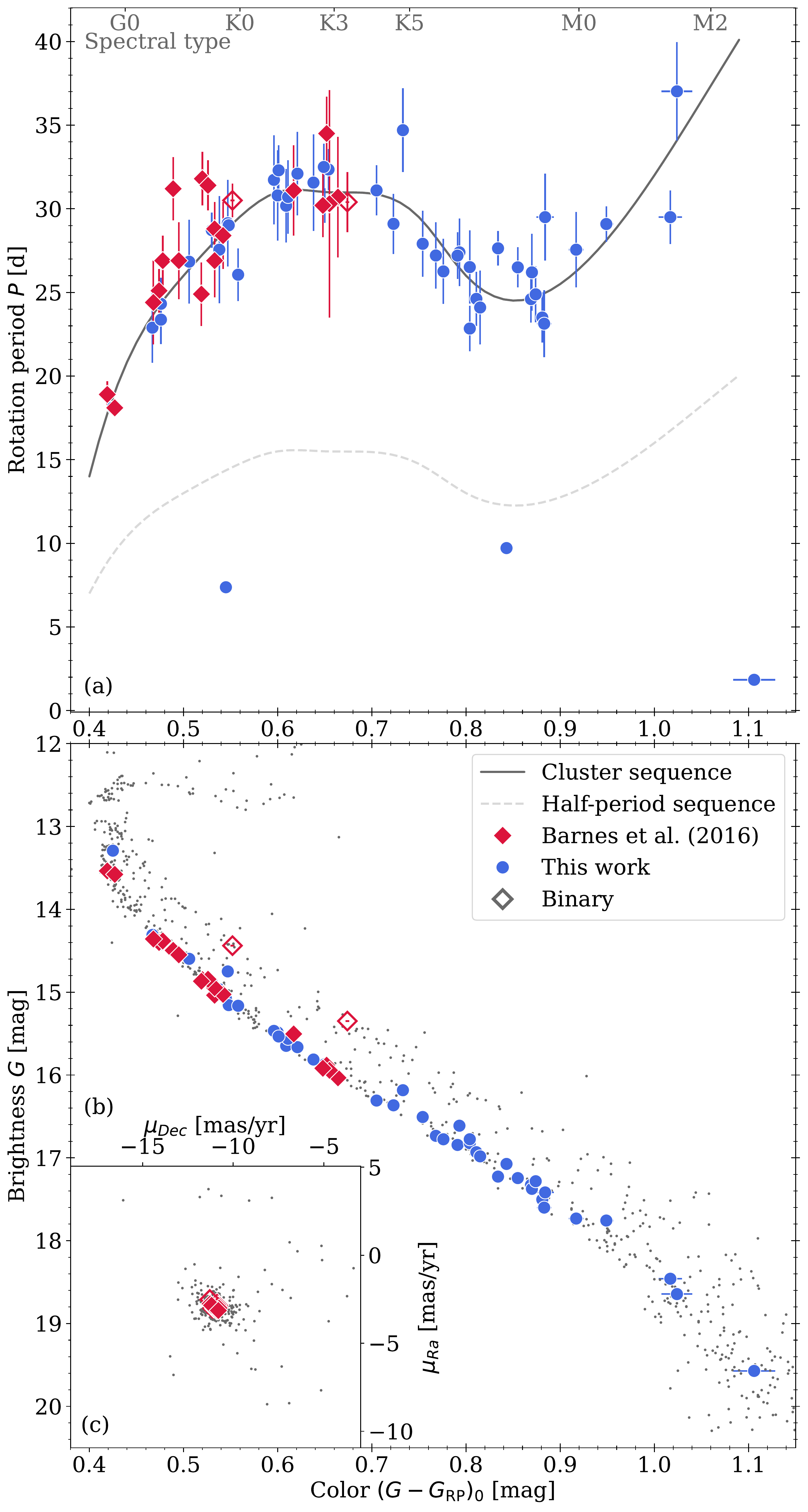}
    \caption{
        Comparison of our results with those of \cite{ApJ...823..2016.16B}. Panel (a) shows a CPD as in Fig.\,\ref{fig_cpd_final} and overplotted in red are the measurements by \cite{ApJ...823..2016.16B}. The gray lines, solid and dashed, are our indication of the cluster and half-period sequences, respectively. Panel (b) shows a CMD of the same stars and the inset (panel c) the proper motions centered on M\,67. Gray dots in panels (b) and (c) are \emph{members} without period determinations.
    }
    \label{fig_cpd_barnes_1}
\end{figure}

\subsubsection{\cite{2016MNRAS.459.1060G}}

    \cite{2016MNRAS.459.1060G} has carried out an extensive analysis on de-trended \ktwo{} C05 light curves as provided in the \ktwo{} archives, subsequently refined in \cite{2016MNRAS.463.3513G} with a larger sample but with a more restrictive selection of rotational signals. Similar to \cite{ApJ...823..2016.16B}, they used the PDC light curves for their analysis.

    The sample includes 98 stars between early G and early M spectral types, spanning periods from 12\,d to 40\,d. The sample distribution found by \cite{2016MNRAS.459.1060G} is atypical as compared with the sequences of known clusters (cf. panel (a) in Fig.\,\ref{fig_cpd_gonzalez_1}). The distribution resembles a double wedge shape with a pinch point around mid-K and large period spreads for both warmer (early-G) and cooler (early-M) stars. As such, it spans an area rather than a sequence in color-period space, unlike any other measured cluster. The lower boundary of this area overlaps with the \cite{ApJ...823..2016.16B} results and with our data in the respective color ranges. The region of lowest period spread in the \cite{2016MNRAS.463.3513G} data (around mid-K) coincides with the red limit of the \cite{ApJ...823..2016.16B} data, and where our present distribution slopes down toward shorter periods. Additionally, \cite{2016MNRAS.463.3513G} has a group of faster ($P\approx14\,$d) late-K stars that fall on our identified half-period sequence.
    
\begin{figure}
    \centering
    \includegraphics[width=\linewidth]{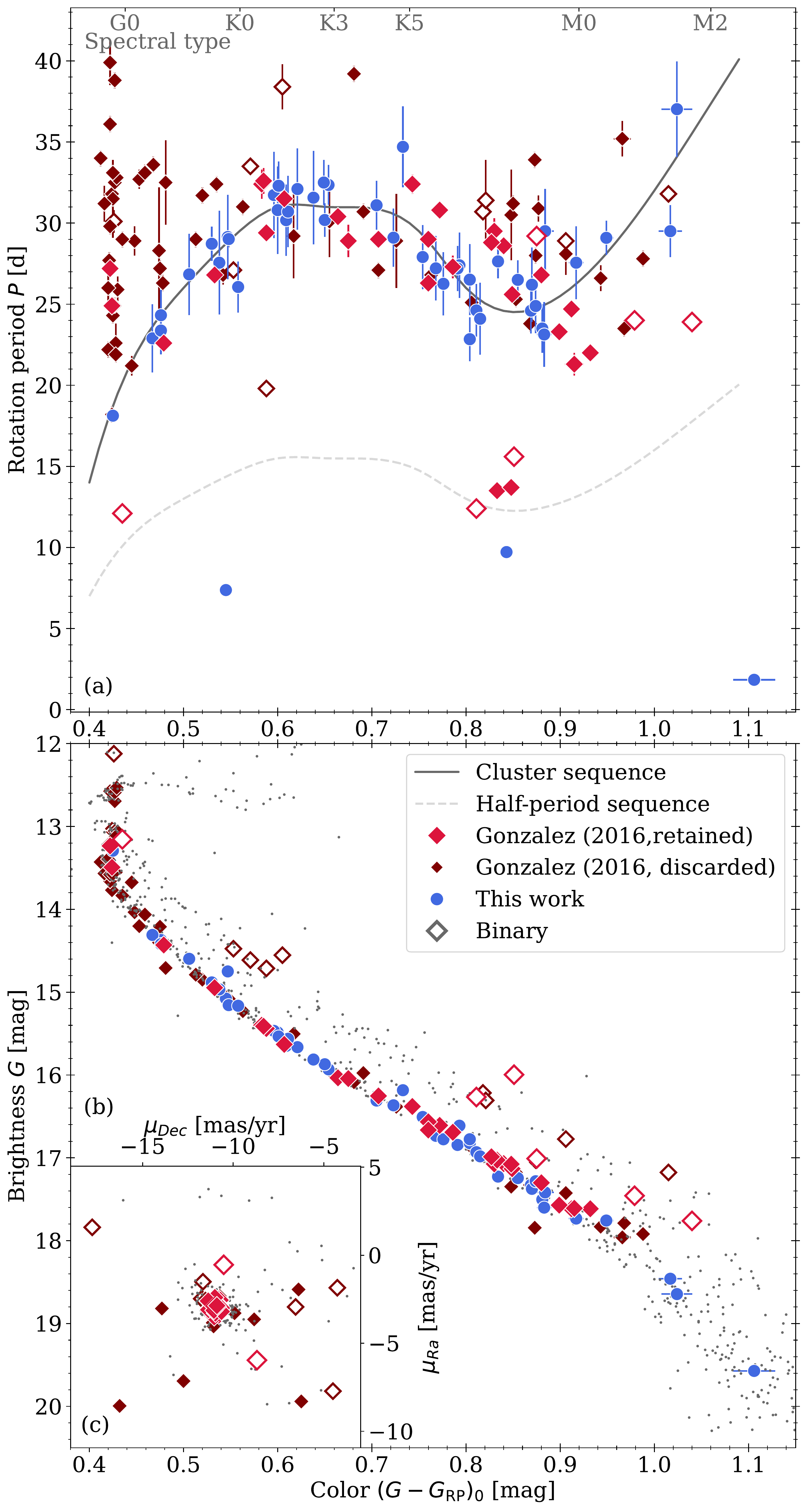}
    \caption{
        Same as Fig.\,\ref{fig_cpd_barnes_1} but with a comparison to results of \cite{2016MNRAS.463.3513G} (red), showing both retained and discarded stars from that sample.
    }
    \label{fig_cpd_gonzalez_1}
\end{figure}    

    If the rotation periods found by \cite{2016MNRAS.463.3513G} are to be truly representative for M\,67, they, and the stars that they have been derived from, must satisfy certain criteria. Firstly, they need to be bona-fide cluster members. \cite{2016MNRAS.463.3513G} worked in the pre-\gaia{} era, but we now have the advantage of \gaia{} astrometry. We match their targets to their GDR3 counterparts and find that that the member selection is good, although certain stars have deviating proper motions. It includes some subgiants and, like \cite{ApJ...823..2016.16B}, a few photometric binaries. Neither of these are responsible for the deviations observed.

    We then inspect the light curves used. \cite{2016MNRAS.463.3513G} used the PDC light curves from the \ktwo{} archive, subjected them to removal of flux outliers and used the result for the period analysis. And here we find the source of the problem. \cite{2016MNRAS.463.3513G} assumes the PDC light curves to be free of systematics and trends. This is emphatically not true. Although apparently corrected for these effects, the light curves are riddled with them. We have taken the light curves from the archives and followed their steps \citep[see Sect.\,2 in][]{2016MNRAS.463.3513G} to recreate the same input data. Their period analysis goes on to identify periodic signals in the light curves where there are clearly (from a manual inspection) none present. Those misidentified signals are generally a combination of low level, somewhat random, intrinsic variation paired with remnants of systematics and trends. Neither of those can be used for the reliable derivation of a rotation period. The maxima in a PDM power spectrum of those light curves is generally low, which supports the picture of misidentifications. 
    
    We therefore manually inspect the light curves of all 98 stars of their sample and select those stars for which a periodicity is obvious in the light curve. For this selection we uses the PDC and \texttt{k2sc} light curves and prepared them in the same way as \cite{2016MNRAS.463.3513G}.In our selection, we err on the side of caution, similar to the selection of our own sample, and identify 33 stars whose periods we adopt. We do not derive a period ourselves for those stars but adopt the ones reported by \cite{2016MNRAS.463.3513G}, even though we may disagree slightly on the assigned period for individual stars. Our selection is purely based on the shape of the light curve. Those 33 light curves are generally those with the greatest variability amplitude, because they dominate the light curve over the systematics and trends and are easy to pick out. We list these 33 stars in Table\,\ref{tab_gonzales_selection}.

    We now compare this subset with our findings. This new subset reduces the \cite{2016MNRAS.463.3513G} sample from an \emph{area} to a \emph{sequence} in the CPD. A particular improvement is seen for the bluest stars of the G16 sample, where we now exclude almost all of the long period G-type stars. This improvement is also visible in the comparison with our data shown in Fig.\,\ref{fig_cpd_gonzalez_1}. The agreement is now much better than with the entire sample. The largest part of the subset occupies the downward sloping section of our sequence, starting at the period maximum at early-K stars and going toward the minimum at early-M, thereby matching our distribution well. However, the downward sloping section appears slightly red-shifted ($\Delta (G-G_\mathrm{RP})_0\approx0.1$\,mag). It also connects well to the \cite{ApJ...823..2016.16B} sample and extends it toward redder stars with a small range of overlap. Interestingly, all five stars of the \cite{2016MNRAS.463.3513G} sample that lie on the half-period sequence made it into our vetted subset.
    
    Finally, we note that G16 adopted 998 light curves and identified rotational signals in 441 ($\approx44$\,\%) for their M\,67 study. This is a very high fraction, even for a much younger, more active cluster, where stellar rotation periods are far easier to identify. For comparison, we identified rotation signals in $<10$\,\% of the sample and \cite{2020A&A...644A..16G} identified rotation periods in 21 out of 102 ($\approx20$\,\%) stars for the 2.7\,Gyr-old open cluster Ruprecht\,147. We therefore conclude that the results of \cite{2016MNRAS.459.1060G} represent an overly inclusive set of stars (unfortunately including misidentifications) in addition to those with true rotational periodicity. Our sample is far more exclusive in comparison.
    When systematics and instrumental trends are excluded, the two distributions can be seen to be substantially similar. 
    
\subsubsection{\cite{2018PhDT........63E}}

    The study of \citet[][E18 hereafter, see also \cite{2018ApJ...859..167E}]{2018PhDT........63E} had a slightly different focus than the ones before. Their main objective was to assess the detectability of rotation signals from light curves of stars as old and inactive as in M\,67. They devised a sophisticated methodology around the artificial injection of variability signals into real \kepler{} data. Mostly as a byproduct, they identified rotation periods from archived light curves (SAP) and self-extracted data.
 
    E18 provides rotation periods for 30 stars (their Table\,A.13) that span from the turn-off point to early K-stars (cf. panel (a) in Fig.\,\ref{fig_cpd_esselstein_1}). Their sample includes a large number of fast rotators while the slow rotators show significant scatter. As before, we scrutinize the sample by identifying binaries and also subject their light curves to manual inspection. E18 lists identifications for \emph{binaries} and \emph{probable  binaries} which we complement with a photometric binary designation from \gaia{} photometry. We come up with a list of 11 binaries. Additionally, three of their stars are clearly evolved past the main-sequence. All fast rotators in the E18 data either show signs of binarity or are post-MS.

\begin{figure}
    \centering
    \includegraphics[width=\linewidth]{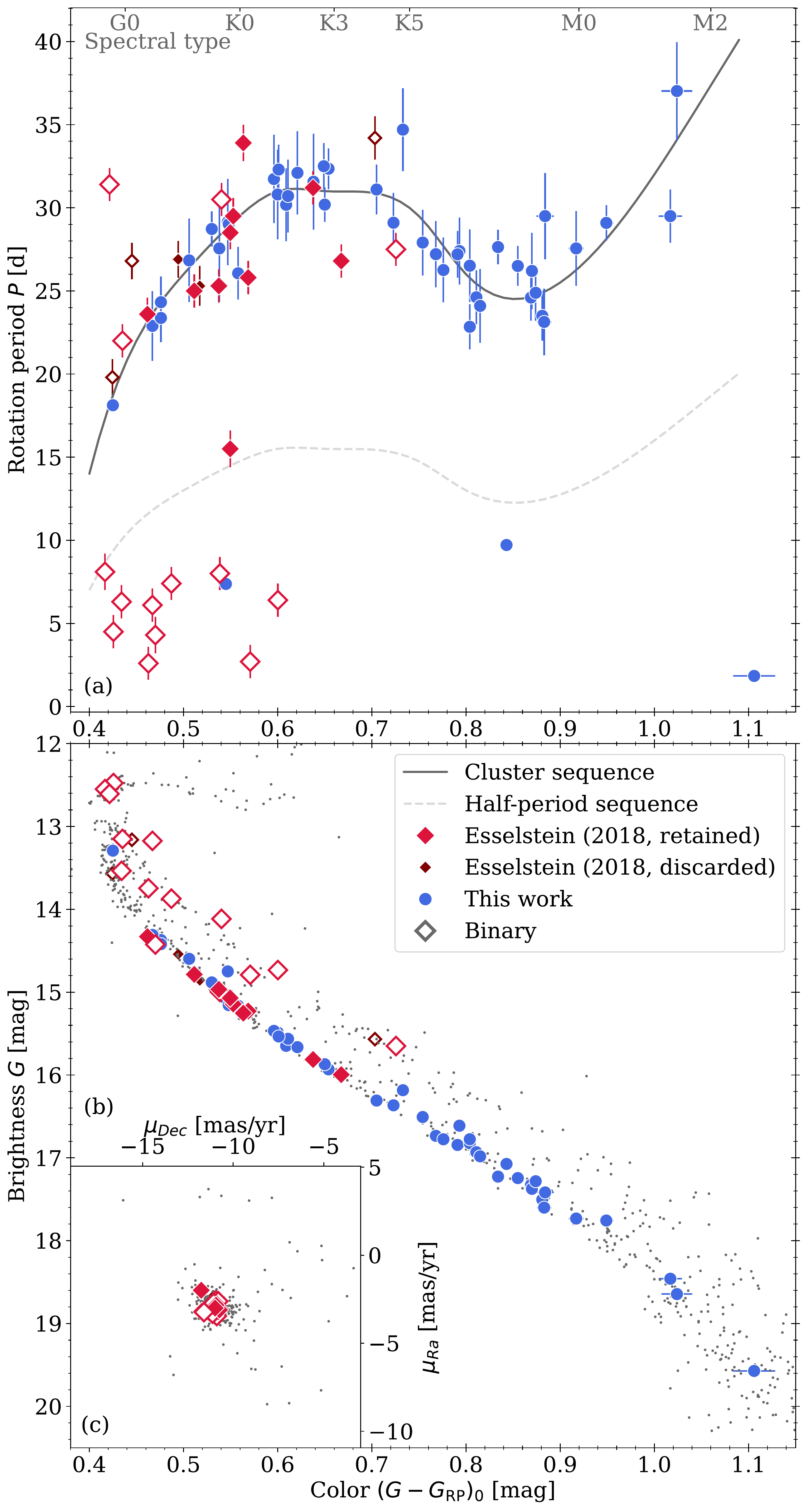}
    \caption{
        Same as Fig.\,\ref{fig_cpd_barnes_1} but with a comparison to results of \cite{2018PhDT........63E} (red), again identifying retained, discarded, and binary stars.
    }
    \label{fig_cpd_esselstein_1}
\end{figure}  
 
    The E18 slow rotators agree reasonably well with our distribution (as verified with B16 and G16). However, theirs shows significantly greater scatter than the other studies. Similar to our investigation of the G16 sample, we inspect the light curves to look for inconsistencies. This, however, is more difficult than before as their light curves are largely only available to us in plotted form (see their Appendix Sect.\,B). We identify five stars for which we cannot verify the rotation signal. However, their elimination from the sample does not change the overall picture. As before, we adopt their period as is even though we may disagree about the derived value for individual stars. One star of the sample (EPIC 211404554) sits arguably on the half-period sequence.
  
    This scrutiny suggests that E18 does agree reasonably well with the overall picture of M\,67 constructed here. Inconsistencies can largely be explained through binarity and post-MS evolution. However, the slow rotator sequence has greater scatter than other results. We argue that this originates in the detection method used. E18 used a Lomb-Scargle analysis. While that is a good choice for fast rotating stars, it ceases to be for slow rotators. This is because the sine wave that is assumed in the LSC becomes less and less suitable for slow rotators due to the occurrence of multiple spots and significant spot evolution between consecutive rotations. Phase dispersion and autocorrelation are much better suited for this period regime. As such, we treat the results of E18 with caution.

\subsubsection{\cite{2022ApJ...938..118D}}

    A recent study by \cite{2022ApJ...938..118D} has used ground-based data acquired over a 3\,yr baseline to investigate the rotation of M dwarfs in M\,67.     They have obtained periods for 383 stars and applied iterative outlier rejection leading them to adopt 64 of the M\,67 M-dwarfs as their cluster sample. The full sample spans a huge range in rotation periods, from a few to several hundred days. After outlier rejection the remaining sample (dubbed as \emph{converged}), a clear sequence appears to emerge, starting around 28\,d periods for late-K stars and showing an increase in period at mid-M to almost 60\,d. Figure\,\ref{fig_cpd_dungee_1} shows their periods together with ours.
    
\begin{figure}
    \centering
    \includegraphics[width=\linewidth]{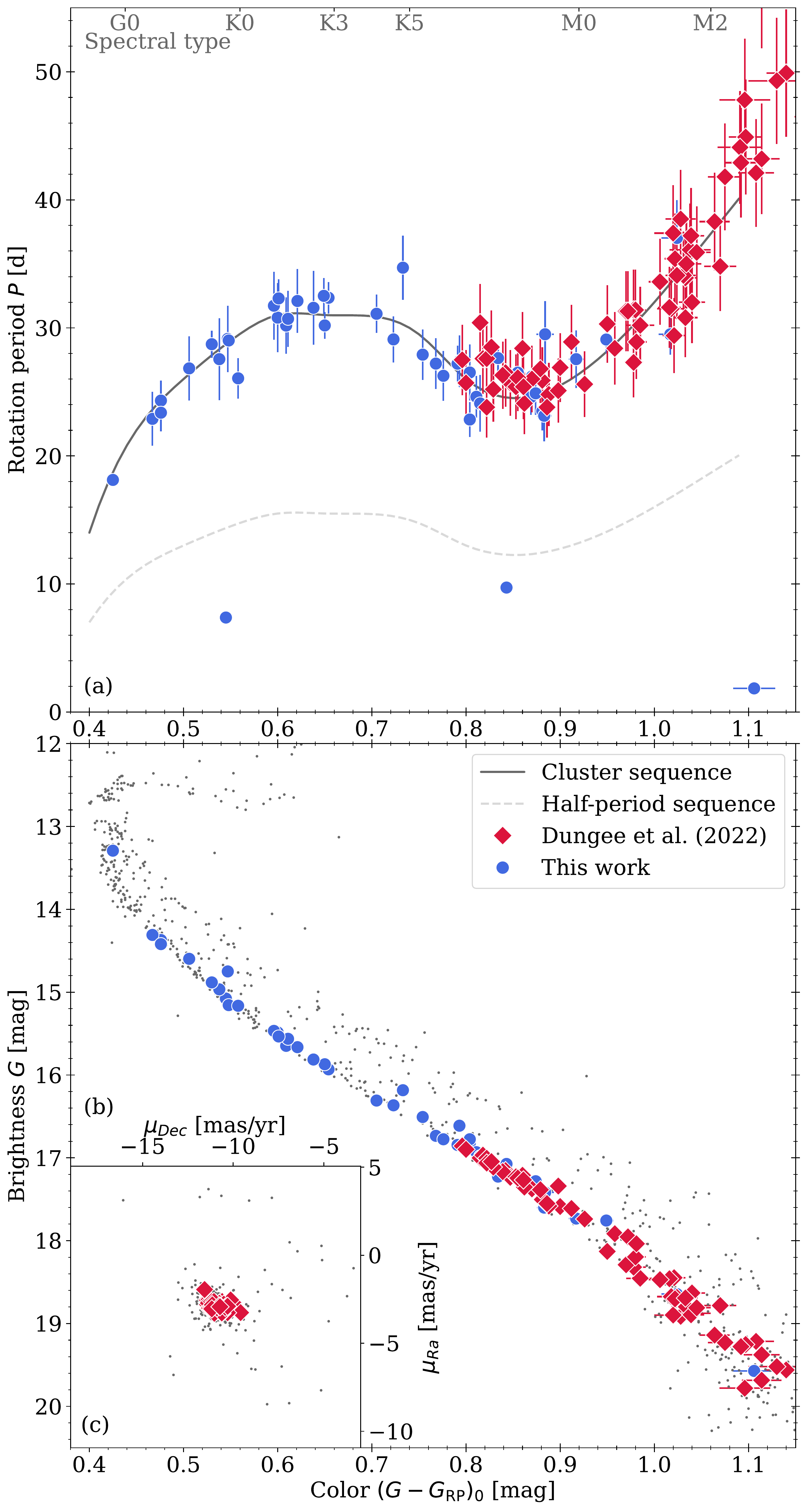}
    \caption{
        Same as Fig.\,\ref{fig_cpd_barnes_1} but comparing with the results of \cite{2022ApJ...938..118D} (red). \cite{2022ApJ...938..118D} assume a flat period error of 10\,\%.
    }
    \label{fig_cpd_dungee_1}
\end{figure}  

    We find very good agreement between our results and those of \cite{2022ApJ...938..118D}. Their sample has its blue boundary at late K-stars and extends toward mid-M. This connects directly to a well populated area in our sample and extends it redward where our data is thinning out but is still in direct agreement with theirs. They have included a photometric cut to their data, eliminating photometric binaries and consequently, we do not find any problematic binary systems in the sample.

    We do not have access to the light curve data from \cite{2022ApJ...938..118D} (only plotted images) to inspect them in similar fashion as we did with the other samples. However, we also see no reason to do so as the agreement between our results and theirs is excellent and their long-baseline ground-based data is not subject to the problems specific to \emph{Kepler} data.

 \subsubsection{A lesson from prior studies}
 
    Based on the above comparisons, we emphasize the need for a very careful approach in the identification of rotation periods from space-based data, especially in the present period regime. Trending and systematics could mask or mimic spot-like variability in a light curve and spot evolution may further complicate period identification. Despite concerns about objectivity, it appears to be essential to supplement all algorithmic period determinations with manual inspection and judgment. We suggest far more exclusivity with similar rotation period derivations generally, but especially in the presence of systematics and when operating with a rather limited observation baseline.

\subsection{Assembly of final sample}\label{sec_max_sample}
  
    Before comparisons with models and to work in other clusters can be performed, we need to assemble a final sample. This consists of the following steps:
    \begin{enumerate}
        \item We accept all stars of our sample that are both on the MS and that are not identified as binaries.
        \item We assume that the stars occupying the half-period sequence are in fact double-spotted stars and as such we double those observed periods. (This assumption is valid to the extent that at the rotation rates we observe, stars tend to be in the double spotted domain. Furthermore, we have several stars among our sample that are clearly double spotted, but with the spot shape and phase shift such that they cannot be mistaken as being single spotted.)
        \item We add the stars from the \cite{ApJ...823..2016.16B} sample, but omit the binaries therein.
        \item We omit the \cite{2016MNRAS.459.1060G} and \cite{2018PhDT........63E} samples entirely.
        \item We include the 64 stars marked as \emph{converged} from the \cite{2022ApJ...938..118D} sample.
        \item We do not remove stars that are common between the samples.
    \end{enumerate}
   The addition of stars from \cite{ApJ...823..2016.16B} significantly strengthens the emergent distribution and bolsters the region around mid-G type stars, while adding the D22 stars completely defines the M-dwarf region.
   
\begin{figure}
    \centering
    \includegraphics[width=\linewidth]{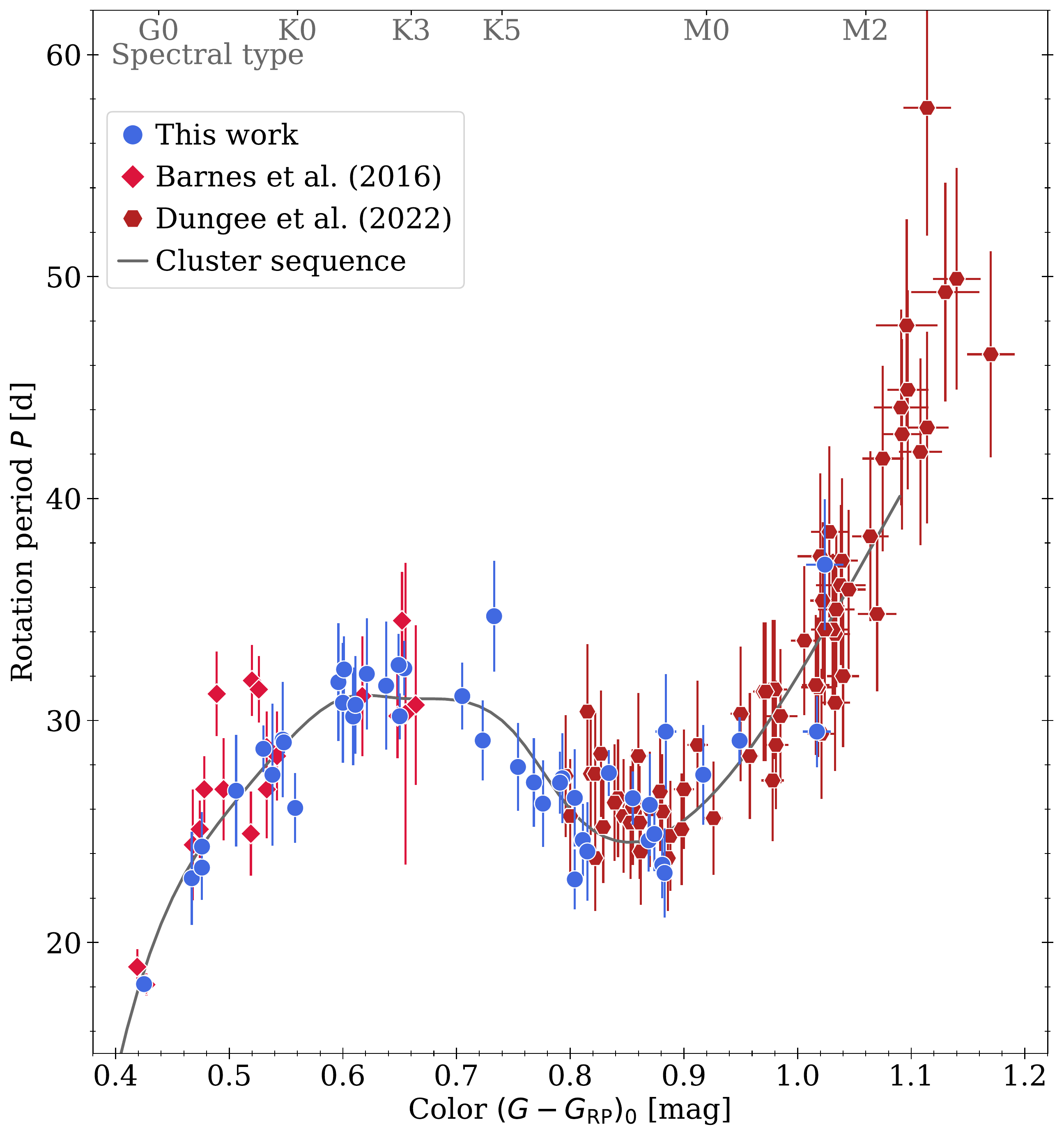}
    \caption{
        Color-period diagram for our final assembled sample \citep[combined with the results of][]{ApJ...823..2016.16B}, now including only main sequence stars with no indications of binarity. Only the slow rotator sequence is shown.
    }
    \label{fig_cpd_m67_full}
\end{figure}
   
    We match the sample to the \tmass{}, \usno{}, and \gsc{} catalogs to obtain a large set of measured stellar colors for each star. The availability of particularly blue bands is understandably sparse toward fainter stars. We also extend this search to non-M\,67 stars, namely those included in rotational studies for other clusters (cf. overview in Sect.\,\ref{sec_intro}). Those will allow a more reliable comparison between the different cluster sequences that emerge later. The final assembled sample, offering as complete a view of the rotational distribution of M\,67 is shown in Fig.\,\ref{fig_cpd_m67_full}.

\subsection{Empirical comparison with observations of other clusters}\label{sec_cluster_work}
    
    We now compare our rotational distribution for M\,67 with those of certain other open clusters. For this comparison we use data for NGC\,6819 \citep[][2.5\,Gyr]{2015Natur.517..589M}, Ruprecht\,147 \citep[][2.7\,Gyr]{2020A&A...644A..16G,2020ApJ...904..140C}, and NGC\,6811 \citep[][1.0\,Gyr]{2011ApJ...733L...9M,2019ApJ...879...49C} as they are the only open clusters equal or older than 1\,Gyr for which rotation periods have been determined.    We crossmatch their individual catalogs analogously to what was done in Sect.\,\ref{sec_max_sample} to obtain a broader and more equally described sample. This requires accounting for the individual reddenings. We obtain $E(B-V)$ for each cluster and calculate $E(G_\mathrm{BP}-G_\mathrm{RP})$ from that. 
    We use the following prescription to obtain $E(G-G_\mathrm{RP})$: As the transformation from $E(B-V)$ to $E(G_\mathrm{BP}-G_\mathrm{RP})$ can be obtained with a simple multiplicative factor (as an approximation) it is reasonable to assume the same works for the transition from $E(G_\mathrm{BP}-G_\mathrm{RP})$ to $E(G-G_\mathrm{RP})$. To determine the relevant factor we use our empirical result for $E(G-G_\mathrm{RP})$ in Eq.\,\eqref{eq_m67_grp} (namely $E(G-G_\mathrm{RP}) = 0.03 \pm 0.005$) for M\,67 and the calculated $E(G_\mathrm{BP}-G_\mathrm{RP})$. We find
    \begin{equation}
        E(G-G_\mathrm{RP}) = 0.556 \cdot E(G_\mathrm{BP}-G_\mathrm{RP}). \label{eq_ebprp2egrp}
    \end{equation}
    We use this to calculate $E(G-G_\mathrm{RP})$ for all clusters.
    
\begin{table}
    \centering
    \caption{
        Reddening parameters used for the individual clusters in Fig.\,\ref{fig_cpd_m67_cluster}. 
    }
    \label{tab_reddening_refs}\small
    \begin{tabular}{rcccc}
            \hline\hline
        Cluster & $E(B-V)$ &  $E(G_\mathrm{BP}-G_\mathrm{RP})$ & $E(G-G_\mathrm{RP})$ & Ref  \\
            \hline
        \object{NGC 6811}       & $0.048$ & $0.065$ & $0.035$ & 1  \\ 
        \object{NGC 6819}       & $0.15$  & $0.201$ & $0.11$  & 2  \\ 
        \object{Ruprecht 147}   & $0.075$ & $0.1$   & $0.055$ & 3  \\ 
        \object{M 67}           & $0.04$  & $0.054$ & $0.03$  & 4  \\ 
            \hline
    \end{tabular}
    \tablefoot{Calculated reddenings use Eqs.\,\eqref{eq_m67_gaia_extinction} and \eqref{eq_ebprp2egrp} for $E(G_\mathrm{BP}-G_\mathrm{RP})$ and $E(G-G_\mathrm{RP})$, respectively. The Ref. column refers to source of $E(B-V)$.}
    \tablebib{
        (1) \cite{2019ApJ...879...49C}; 
        (2) \cite{2015Natur.517..589M};
        (3) \cite{2020A&A...644A..16G}; 
        (4) \cite{2007AJ....133..370T}
    }
\end{table}

\begin{figure*}
    \centering
    \includegraphics[width=\linewidth]{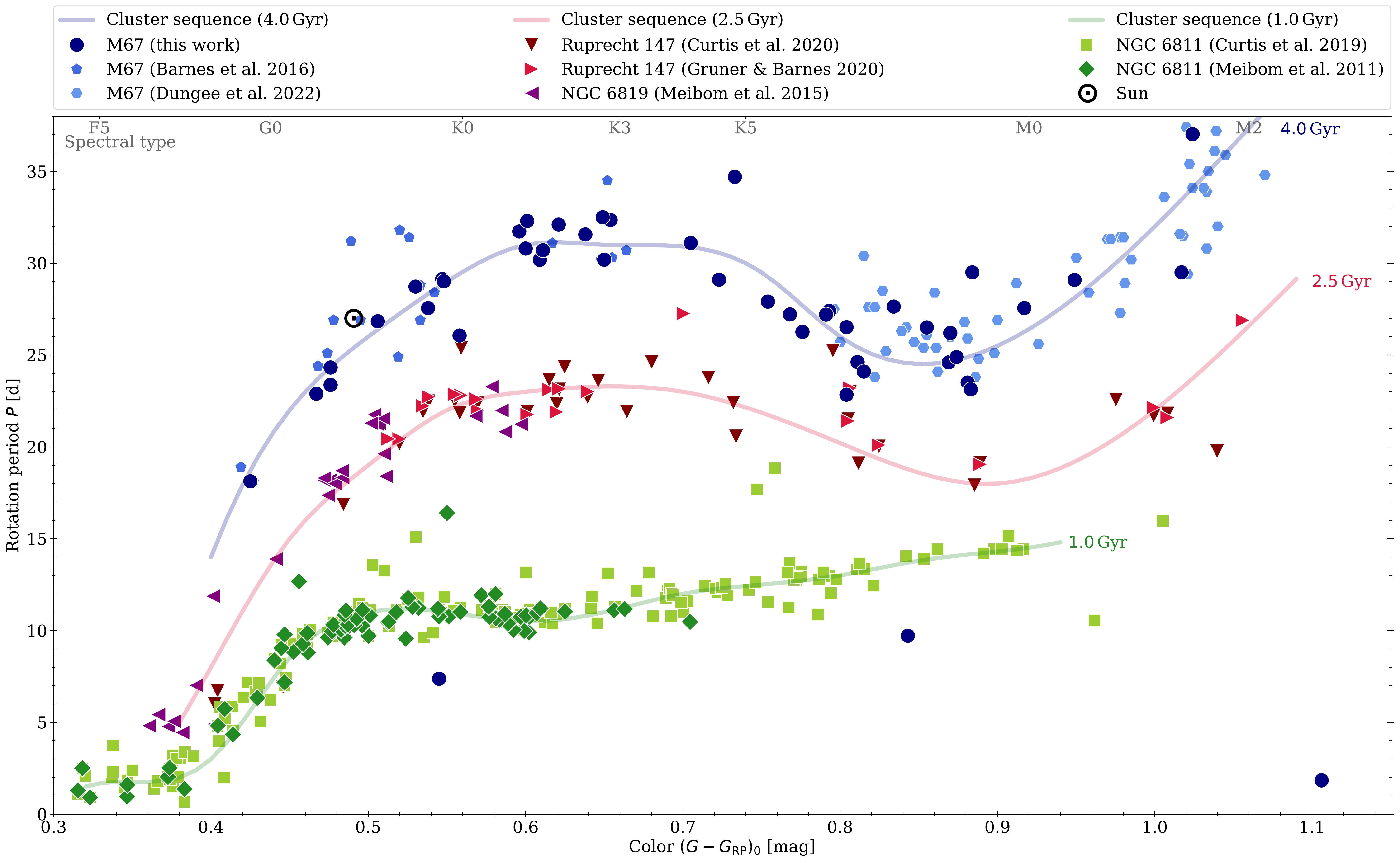}
    \caption{
        CPD of the three open clusters 1\,Gyr or older. Shown are NGC\,6811 \citep[][]{2011ApJ...733L...9M,2019ApJ...879...49C} at 1\,Gyr, NGC\,6819 \citep{2015Natur.517..589M} and Ruprecht\,147 \citep{2020A&A...644A..16G,2020ApJ...904..140C} at 2.5\,Gyr, and M\,67 \citep[][and this work]{ApJ...823..2016.16B,2022ApJ...938..118D} at 4\,Gyr. The color coding groups cluster data by age. Overplotted is a simplified representation of the emerging cluster sequences by age (solid lines color coded by age). The sun is shown with its usual symbol. Figure\,\ref{fig_cluster_cpd_colors} shows equivalent plots in the colors $(G_\mathrm{BP}-G_\mathrm{RP})_0$, $(B-V)_0$, and $(V-K)_0$.
    }
    \label{fig_cpd_m67_cluster}
\end{figure*}

    M\,67 is the oldest open cluster with reliable rotation periods available. It is therefore unsurprising, indeed expected, that its sequence lies above all other clusters in a color dependent sense (cf. Fig.\,\ref{fig_cpd_m67_cluster}). Similar to Ruprecht\,147 and NGC\,6811, M\,67 shows a flattening of its sequence around early-K spectral type, although this feature appears to move to the red with increasing age. The cooler stars (later than mid-K) in Ruprecht\,147 indicated a slight downturn of the sequence, with a minimum around late-K/early-M and a steep increase in period with redder color. However, given the small sample size in that range, the existence of the downturn was uncertain. Our new data for M\,67 now confirms the existence of this downturn and validates the distribution seen for Ruprecht\,147. 
    
    It is visually apparent that M\,67 displays the greatest scatter in rotation period. This fact may be surprising in that the stars are expected to have all converged to the slow rotator sequence. The scatter is, however, a consequence of the difficulty of the period measurement and not an intrinsic property of the distribution. Additionally, the relatively short baseline of the space-based observations, combined with differential rotation results in significant uncertainties on the periods. To help visualize the cluster distributions and unclutter subsequent plots, we have created rough approximations of the cluster sequences (based on cubic interpolations to the points listed in Table\,\ref{tab_cluster_sequences}) as shown with colored lines in Fig\,\ref{fig_cpd_m67_cluster}.
        
\subsection{Comparison with rotation models}\label{sec_spindown_modeling}

    The long-term goal of studies like ours is to understand the rotational evolution of stars and the physics of magnetic braking, to describe the evolution in an empirically constrained model, and to use it to derive stellar ages via gyrochronology. Numerous stellar spindown models have been created over the years with varying levels of detail in their physical descriptions. As our new data extends the knowledge of stellar rotation rates into a parameter space that has not been explored before (lower mass and higher ages), we go on to examine how well the models perform against our new findings.

    We know from prior work in Ruprecht\,147 \citep[e.g.,][]{2020ApJ...904..140C,2020A&A...644A..16G} that models of rotational evolution do not reproduce well the observed distributions of cluster stars at higher ages. Any lingering doubts about discrepancies between the models and the data are removed now. The deviating shape shown by Ruprecht\,147 is also present in M\,67, and is more pronounced. We know now that rotational evolution deviates significantly from model predictions. As such, we limit this section to comparison up to this conclusion and omit a longer discussion on the intricate model details. For a more detailed discussion see the comparison to Ruprecht\,147 in Sect.\,6 of \cite{2020A&A...644A..16G}.
    
    We also limit ourselves to a comparison with the spindown descriptions of \cite{2010ApJ...722..222B} and \cite{2020A&A...636A..76S}. These are chosen among the numerous models available as they are related in their descriptions and exemplify the difference based on the inclusion of an additional parameter. This comparison is essentially similar to the one performed in \cite{2020A&A...644A..16G}, but the observed deviations there are even more pronounced here, and the larger sample allows a stronger conclusion. We note that the models are only available in certain colors and for particular ages. \cite{2020A&A...636A..76S} published a few time steps (notably 2.5, 4.0, and 4.57\,Gyr in the relevant age range) and are available only in $(B-V)_0$ color. \cite{2010ApJ...722..222B} offers a mathematical prescription that we use similarly to \cite{2020A&A...644A..16G} (see their Sect.\,6.1) which intrinsically allows any given age and a variety of Johnson and \tmass{} colors. $B-V$ is especially problematic with respect to reliability and availability for stars as red and faint as we deal with here. As such, an examination based on a more reliable and consistent set of colors is preferred. We opt for \gaia{} $G-G_\mathrm{RP}$ as it omits the problems of \gaia{} $G_\mathrm{BP}-G_\mathrm{RP}$ in the blue and is accessible from one consistent source (unlike e.g., $V-K$). However, we do include those colors below to offer a complete picture.
    
\begin{figure}
    \centering
    \includegraphics[width=\linewidth]{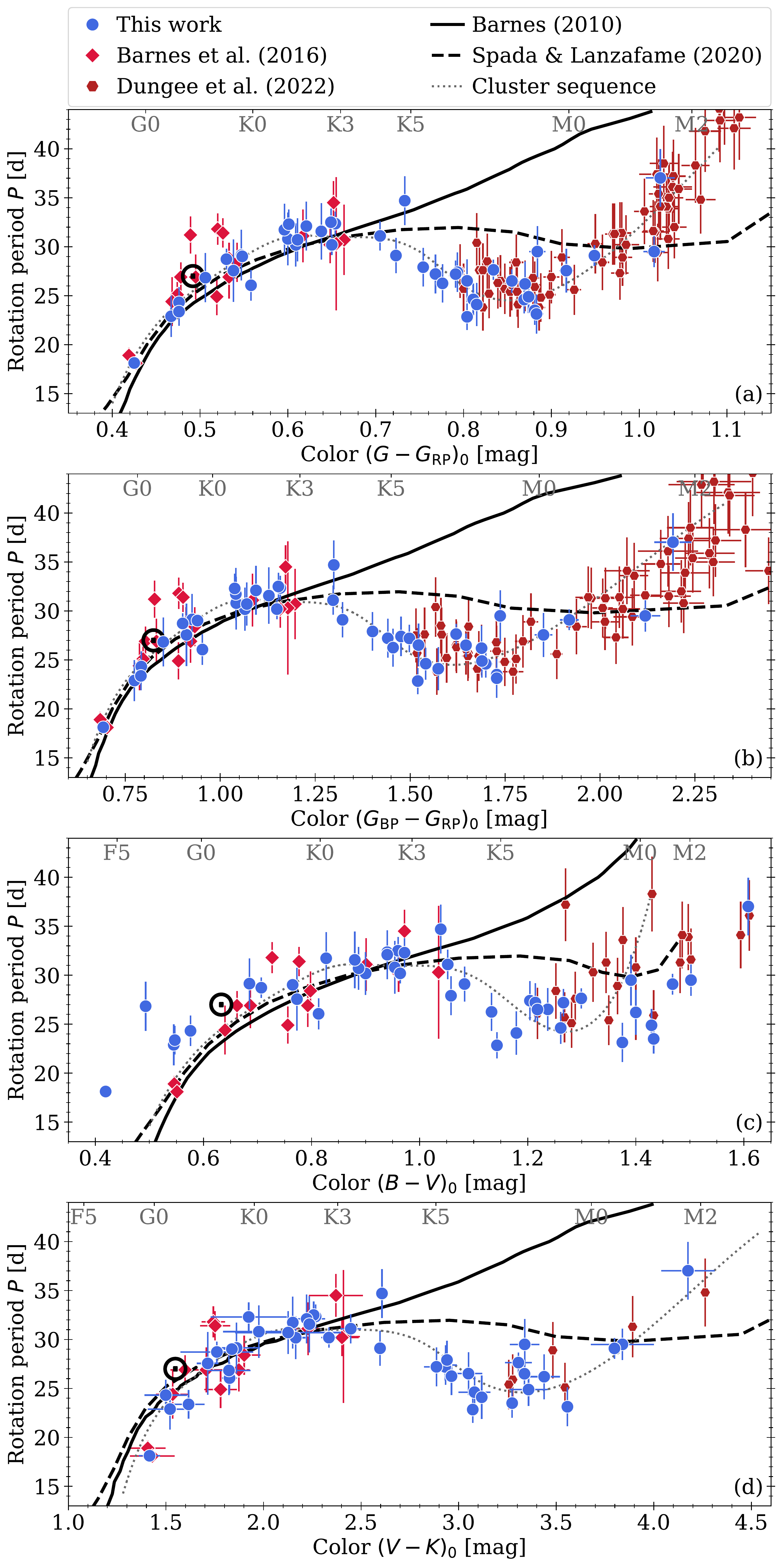}
    \caption{
        Comparison between our final sample of measured rotation periods (this work combined with the results of \cite{ApJ...823..2016.16B} and \cite{2022ApJ...938..118D}) and the stellar rotation models of \cite{2010ApJ...722..222B} and \cite{2020A&A...636A..76S}. Each panel performs the comparison in a different color system: (a) $G-G_\mathrm{RP}$, (b) $G_\mathrm{BP}-G_\mathrm{RP}$, (c) $B-V$, and (d) $V-K$. The errors on $B-V$ are suppressed for visibility reasons.
    }
    \label{fig_cpd_m67_models}
\end{figure}

    To perform the comparison in $G-G_\mathrm{RP}$, we do need the models in said color. We employ a color transformation that is constructed from stellar models in the {\tt PARSEC} isochrones by comparing the different synthetic colors for stars of the same mass, age, and metallicity. This approach has its weaknesses, for example it is problematic toward the very red, but it is sufficient for our comparison. The comparison that is shown in Fig\,\ref{fig_cpd_m67_models} is now based on the following: (1) all stellar colors are measured, that is, obtained from one of the above mentioned catalogs and dereddened according to the description in Sect.\,\ref{sec_cluster_reddening}, (2) both models in $B-V$ and \cite{2010ApJ...722..222B} in $V-K$ are obtained directly from the sources, and (3) $V-K$ for \cite{2020A&A...636A..76S} and both models in \gaia{} colors are obtained via the transformations. We use the same color transformation on our empirical cluster sequence which was originally constructed in $G-G_\mathrm{RP}$ (see Table\,\ref{tab_cluster_sequences}).
    
\subsubsection{The \cite{2010ApJ...722..222B} model}
    
    We find good agreement between our results and the \cite{2010ApJ...722..222B} model prediction for stars earlier than K5 (in the overlap region with the \cite{ApJ...823..2016.16B} data). Redward of K5, the model predicts a steady increase in rotation period with color which is not seen in the data. Instead, the observed stars turn significantly downward toward shorter rotation periods. After this point, the model diverges away from the observed stellar distribution. Similar behavior was observed in Ruprecht\,147 by \cite{2020A&A...644A..16G} but the effect is stronger here, and thanks to the larger sample, can now be stated with much more certainty. It is to be noted that the period increase for the reddest stars in the sample is much steeper than predicted in the model. This leads to an apparent reconvergence between model and observations for mid-M stars. As has now been seen with multiple clusters, the \cite{2010ApJ...722..222B} model fails to predict the stellar rotation rates for older stars later than mid-K. Additional physics is apparently required to describe the spindown adequately.
    
\subsubsection{The \cite{2020A&A...636A..76S} model}
    
    The model of \cite{2020A&A...636A..76S} incorporates the \cite{2010ApJ...722..222B} description for magnetic braking and adds a two-zone model where an additional parameter describes the coupling between those two zones and the angular momentum exchange between them. Generally speaking, angular momentum from the radiative core is transferred to the convective envelope, reducing the stellar spindown to the point where a star may even spin up slightly. This effect is mass-dependent. It should be noted that the \cite{2020A&A...636A..76S} model only describes stars of the slow rotator sequence, in contrast to the \cite{2010ApJ...722..222B} model. However, this is not directly relevant here, since at the age of M\,67, all stars have converged to the slow rotator sequence.
    
    Similar to the \cite{2010ApJ...722..222B} model, the \cite{2020A&A...636A..76S} model provides good agreement for stars bluer than mid-K. For later spectral types, it is closer to the data than the \cite{2010ApJ...722..222B} model, but still does not appear to be able to reproduce the observations in detail. Both the model and observed cluster sequence show first a flattening of the distribution followed by a subsequent downturn in the CPD and eventually a steep increase from blue to red. However, those features appear not to be aligned. The \cite{2020A&A...636A..76S} model predicts a flattening of the distribution and a potential downturn around M0, whereas the observations show said downturn already at mid-K. Furthermore, the amplitude of this downturn in the model is much shallower than observed. Given that the model incorporates a mass-dependent coupling that is responsible for the observed downturn, it is possible that a recalibrated model that incorporates our findings and the recent results for Ruprecht\,147 could improve the level of reproduction of the data and increase the predictive power of the model.
    
\subsubsection{Implications for modeling of rotating stars}
    
    The new M\,67 rotation periods thus continue to challenge theoretical rotational evolution models. No rotational evolution model to date is capable of reproducing the observed behavior for old cool stars later than mid-K. This is unsuprising, as the calibration of the models for a long time could only rely on warmer stars. And it is only recently that we have started to see that the behavior for cool stars differs strongly from that anticipated. This discovery began with the work of \cite{2019ApJ...879...49C} on NGC\,6811, whose behavior in the red sparked the idea of a (temporarily) stalled spindown, found further indication in the works of \cite{2020A&A...644A..16G} and \cite{2020ApJ...904..140C} for Ruprecht\,147, and is now confirmed by the new data on M\,67. The best results are achieved by the \cite{2020A&A...636A..76S} parameterized two-zone model and small adjustments to the model could potentially provide a better match to the observations.

\subsection{Stars at and around the turn-off point}\label{sec_mass_proxy}

    In contrast to younger clusters where all stars redward of the Kraft break are still on the main sequence, M\,67 is old enough that this becomes an issue. For those evolved stars, their colors (and temperatures) are no longer a valid proxy for the stellar mass. Consequently, we need to return to the mass as the relevant indicator, and examine the stars omitted above. We estimate the stellar masses of our sample stars from an isochrone fit with a {\tt PARSEC} isochrone of 3.6\,Gyr age (cf. panel (a) of Fig.\,\ref{fig_mass_proxy}). The uncertainty for this mass estimate is typically on the order of $\Delta M\lesssim \pm 0.05\,M_\odot$.

 \begin{figure}
    \centering
    \includegraphics[width=\linewidth]{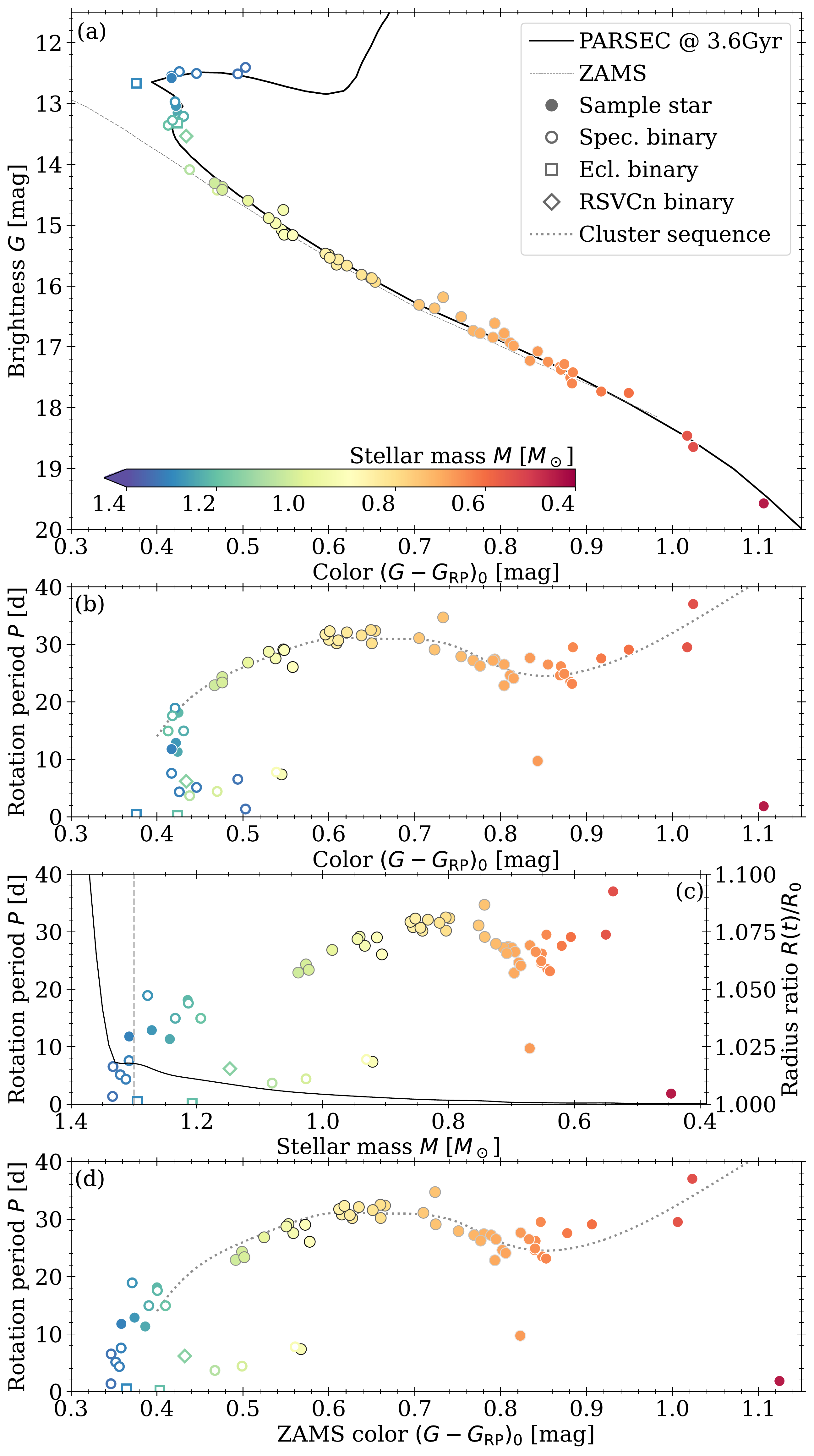}
    \caption{
        CMD and three CPDs highlighting M\,67 sample stars evolved beyond the Main sequence. Panel (a) shows a CMD of our sample with the stars color coded according to their masses. Differing symbols distinguish different types of binaries. Panel (b) shows a corresponding CPD in Gaia colors and identical symbols. Panel (c) plots the rotation periods of the stars against their estimated masses. An approximate indication of the position where stars cease to have an outer convective zones (i.e., the Kraft break) is overplotted with a dashed gray line. The solid black line corresponds to the scaling on the right and denotes the ratio between current ($R(t)$) and ZAMS stellar radius ($R_0$). Panel (d) shows an inferred CPD with stars assigned colors they had when on the ZAMS.
    }
    \label{fig_mass_proxy}
\end{figure}

    Plotting the rotation periods against mass (panel c) instead of color (panel b) unclutters the region around the turn-off point. It also demonstrates that those stars appear to populate a direct extension of the cluster sequence toward the blue. Panel (d) of the same figure projects the stars back to their zero-age main sequence (ZAMS) color, estimated from a 100\,Myr isochrone and the derived stellar masses. For this comparison, we have included binaries whose type does not suggest a tight system,  that means, excluding eclipsing and cataclysmic systems, and those whose CMD position makes a mass estimate difficult, that is, photometric binaries. As expected, eclipsing and cataclysmic systems show much faster rotation rates than comparable stars of the same mass/ZAMS color.

    The impact of post-main sequence expansion of the star's envelope is likely to be a small effect on their current rotation periods as their radii have not increased significantly ($\lesssim 2$\,\%, cf. panel (c) of the figure). This is estimated from a comparison of the radii of ZAMS stars and those at the age of M\,67.
    
\section{Conclusion} \label{sec_conclusion}

    We have studied space-based photometric data from \kepler's \ktwo{} mission for the 4\,Gyr open cluster M\,67 to examine this key object in its role as a gyrochronology calibrator and to extend our knowledge of the evolution of stellar rotation. We used the \ktwo{} superstamp created from Campaign 5 target pixel files together with a stellar sample based on \gaia{} DR3 and performed membership analysis. We constructed light curves of stars in the superstamp region based on aperture photometry and devised a correction algorithm to deal with the well known \ktwo{} systematics. We identified periodic signals in 128 of these light curves, and created a color-period diagram using the 47 stars which are believed to be effectively single main-sequence stars. Those span spectral types from early-G type stars to mid-M and as such reach a region hitherto unexplored.

    Our data connect well to prior studies on M\,67, especially \cite{ApJ...823..2016.16B}. Most of the data of \cite{2022ApJ...938..118D} lies redward of ours but their results agree with ours in the overlap region and extend it smoothly toward the very red. The work of \cite{2016MNRAS.459.1060G} and \cite{2018PhDT........63E} is also shown to be consistent with our work if they are suitably pruned of binaries and light curves with remaining instrumental effects. Despite the current extension toward significantly later-type stars, open cluster rotation period data remains compatible with the idea that effectively single stars populate a unique surface in rotation period-mass-age space. However, the suggestion of a diminishment of spindown for K-type stars by \cite{2019ApJ...879...49C} based on NGC\,6811 data, and subsequently supported by \cite{2020A&A...644A..16G} and \cite{2020ApJ...904..140C} for Ruprecht\,147  now appears to be secure. The shape of the rotation period surface is more complex than originally envisaged, and deviates strongly from the classical Skumanich-style, mass-dependent predictions. A comparison with models of rotational evolution shows that the models appear to be inadequate for stars redder than mid-K. We find that the model of \cite{2020A&A...636A..76S} currently provides the closest description of the spindown by invoking a third mass-dependent timescale that parameterizes internal angular momentum transport.  We conclude that future models likely need to include three distinct physical processes to account for slow, fast, and low-mass rotators if they are to accurately describe stellar spindown.

    One fact about stellar rotation appears to continue to hold in the aftermath of the newly-obtained data. \emph{Single stars continue to occupy a unique surface in rotation period-mass-age space.} However, its shape is now found to be more complex than was indicated schematically in \cite{2015Natur.517..589M}. The warmer part of the sequence appears to behave mostly Skumanich-like, albeit with a strong mass dependence. Current models already reproduce this behavior reasonably, and it is likely that a good description can be found using only one or two mass-dependent timescales. But it is not true for the cooler part of the surface where the behavior becomes more complex. It is likely, and the transition from the \cite{2010ApJ...722..222B} model to the \cite{2020A&A...636A..76S} model strengthens this thought, that the description of the cooler part requires the inclusion of at least a third parameter, an additional distinct mass-dependence of stellar spindown. Finally, it appears that even photometric binary stars of M\,67 age are so diverse in their rotational behavior as compared with single stars that they seem to be unsuitable for gyrochronology at present.

\begin{acknowledgements} 
    We are grateful to an anonymous referee for insightful suggestions that helped to improve the quality of this paper.
    SAB gratefully acknowledges support from the Deutsche Forschungs Gemeinschaft (DFG) through project number STR645/7-1. 
    Some of the data presented in this paper were obtained from the Mikulski Archive for Space Telescopes (MAST). STScI is operated by the Association of Universities for Research in Astronomy, Inc., under NASA contract NAS5-26555. Support for MAST for non-HST data is provided by the NASA Office of Space Science via grant NNX09AF08G and by other grants and contracts. 
    This work has made use of data from the European Space Agency (ESA) mission \gaia{} (\url{cosmos.esa.int/gaia}), processed by the \gaia{} Data Processing and Analysis Consortium (DPAC, \url{cosmos.esa.int/web/gaia/dpac/consortium}). Funding for the DPAC has been provided by national institutions, in particular the institutions participating in the \gaia{} Multilateral Agreement. 
    This paper includes data collected by the \kepler{} mission and obtained from the MAST data archive at the Space Telescope Science Institute (STScI). Funding for the \kepler{} mission is provided by the NASA Science Mission Directorate. STScI is operated by the Association of Universities for Research in Astronomy, Inc., under NASA contract NAS 5–26555.
    This research has made use of the SIMBAD database, operated at CDS, Strasbourg, France. 
    Based on photographic data obtained using The UK Schmidt Telescope. The UK Schmidt Telescope was operated by the Royal Observatory Edinburgh, with funding from the UK Science and Engineering Research Council, until 1988 June, and thereafter by the Anglo-Australian Observatory. Original plate material is copyright (c) of the Royal Observatory Edinburgh and the Anglo-Australian Observatory. The plates were processed into the present compressed digital form with their permission. The Digitized Sky Survey was produced at the Space Telescope Science Institute under US Government grant NAG W-2166.
    This work made extensive use of the \verb+python3+ implementations of the \verb+numpy+ \citep{harris2020array}, \verb+astropy+ \citep{2013A&A...558A..33A,2018AJ....156..123A,2022ApJ...935..167A}, \verb+scipy+ \citep{2020SciPy-NMeth}, \verb+numba+ \citep{10.1145/2833157.2833162}, and \verb+scikit-learn+ \citep{scikit-learn} libraries.
    Indications of dwarf stars' spectral types are based on ``A Modern Mean Dwarf Stellar Color and Effective Temperature Sequence'' \citet[][and continuously updated afterwards, Version 2022.04.16]{2013ApJS..208....9P}\footnote{\url{pas.rochester.edu/~emamajek/EEM_dwarf_UBVIJHK_colors_Teff.txt}}.
\end{acknowledgements}

\bibliographystyle{aa}
\bibliography{paper}

\begin{appendix}
 
\section{M67 cluster membership}\label{appendix_cluster_membership}
 
    In this section we describe the process that we used to determine cluster membership. For this, we use the stellar photometry ($G$ and $G-G_\mathrm{RP}$), proper motions ($\mu_\text{Ra}$ and $\mu_\text{Dec}$), parallax $\varpi$, and radial velocity $v_\text{rad}$ to the extent available in the sample. Based on a comparison with the cluster parameters (cf. Table\,\ref{tab_m67_overview}), each star is then classified as either a \emph{member}, \emph{candidate}, or a \emph{field star} in each of these parameter categories. Stars that lack a particular parameter are designated as \emph{unknown}. Those individual designations are then combined into a single membership assessment. 
 
    For parallax and radial velocity, we define a range around the cluster value in which each star is designated a \emph{member} when the measured value overlaps with that range within its error. A second, larger range is defined in similar fashion for the designation of \emph{candidate} status (see below). For proper motions, we use a circular area around the cluster that is compared with the error ellipse spanned by the proper motion errors in the same way. The ranges around the cluster are defined as 
    \begin{align}
        \mu_\text{M67} &\pm 2.5\,\text{mas/yr for \emph{member}, } \label{eq_member_1} \\
        \mu_\text{M67} &\pm 3.5\,\text{mas/yr for \emph{candidate},}  \\
        v_\text{rad,M67} &\pm 10\,\text{km/s for \emph{member},}  \\
        v_\text{rad,M67} &\pm 15\,\text{km/s for \emph{candidate},} \\
        \varpi_\text{M67} &\pm 0.2\,\text{mas for \emph{member}, and} \\
        \varpi_\text{M67} &\pm 0.3\,\text{mas for \emph{candidate}.} \label{eq_member_2}
    \end{align} 
    An additional constraint for the parallax criterion is that the parallax needs to exceed its error by a factor $>3$. Furthermore, we limit the proper motion offset to $\pm 8$\,mas/yr. Table\,\ref{tab_membership_tab} provides an overview of the sample sizes in each category determined by this process.

\begin{table}[ht!]
    \caption{Membership overview.}
    \centering
    \label{tab_membership_tab}
         \begin{tabular}{lcccccc}
     \hline\hline
        Status   &  phot.  & PM & plx & $v_\text{rad}$ & & M\,67 \\
     \hline
          Member & 1226 & 913  & 856 & 1226 & &   873 \\
       Candidate & 186  & 18   &  63 &  186 & &    -- \\
      Field star & 549  & 1082 & 877 &  549 & &  1140\\
         Unknown & 52   & 0    & 217 &   52 & &    --\\
     \hline
     \end{tabular}

    \tablefoot{For each parameter involved in the membership designation and merged result, we list the number of stars falling into the categories described in the text.}
\end{table}

    Because a cluster follows a complex sequence in the color-magnitude diagram (instead of being approximately represented by a single point), we need a more sophisticated approach. Instead, for each star we calculate the $\chi^2$ distance to a cluster isochrone. We use a {\tt PARSEC} isochrone corresponding to the age and metallicity of M\,67 and apply dilution, extinction, and reddening according to Table\,\ref{tab_m67_overview}. Similar to all other parameters, this $\chi^2$ is then used to evaluate the photometric cluster membership.

\begin{figure}
    \centering
    \includegraphics[width=\linewidth]{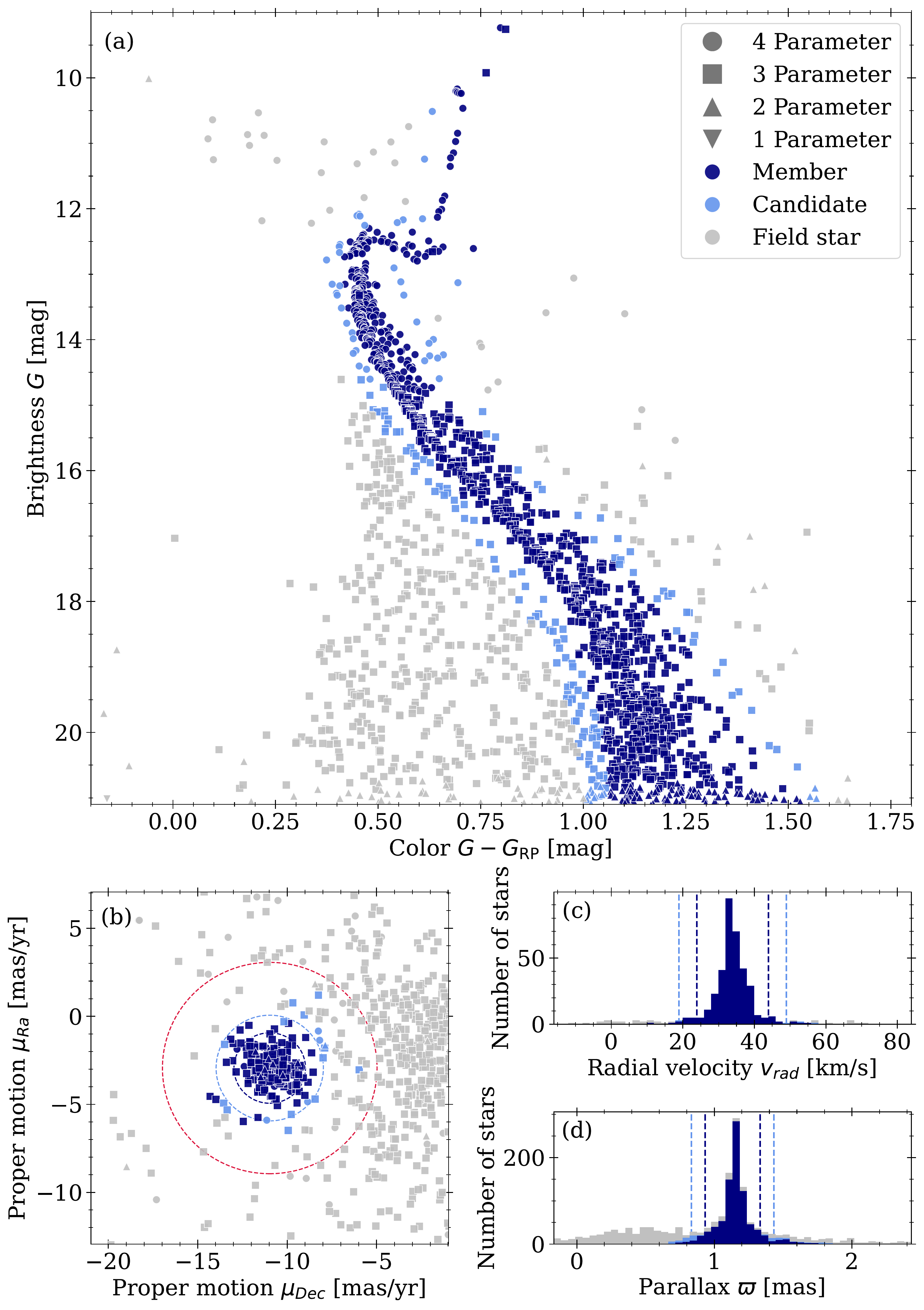}
    \caption{
        M\,67 cluster membership evaluation. The panels (a) to (d) show the individual parameters used for our membership analysis: photometry, proper motions, radial velocity and parallax, respectively. Each point in panels (a) and (b) is a \gaia{} DR3 target in the field of view. Panels (c) and (d) show histograms (gray) of the sample. The parameter numbering in the legend refers to the number of parameters that are available per star. The color coding indicates the assigned membership status. We note that this refers solely to the membership on the respective parameter of panel, not on the total membership. For an overview of the total membership, see Fig.\,\ref{fig_sample_overview}. The dashed lines in panels (b), (c), and (d) indicate the parameter ranges for \emph{member} and \emph{candidate} designations based on Eqs.\,\eqref{eq_member_1} to \,\,\eqref{eq_member_2} (dark and light blue, respectively). The red circle in panel (b) indicates the 8\,mas/yr cutoff.
    }
    \label{fig_membership_assignment}
\end{figure}

    Following the assessment as described above, we have evaluated four individual membership criteria. Figure\,\ref{fig_membership_assignment} shows the assigned category to the sample for each parameter. To combine them into a single, general membership designation, we assign a numeric value to each category. Here, \emph{member} counts as 1.0, \emph{candidate} as 0.5, and \emph{field star} as 0.0. Those values are then added up for each star and divided by the number of available criteria for each star. If the resulting fraction for a star evaluates to a value $\geq 0.5$, we designate that star as a \emph{member}, wheras stars with lower values are designated as \emph{field stars}. The result of this combination is displayed in the final column named \emph{M67} in Table\,\ref{tab_membership_tab} and plotted accordingly in Fig.\,\ref{fig_sample_overview}. Stars with fewer than two criteria available are treated as field stars. This is only relevant for stars with $G>20$\,mag, for which photometry is often the only available parameter.

\section{K2 systematics correction}\label{appendix_correction_main}
 
    In this section, we describe the steps that were taken to extract and correct light curves from the \ktwo{} superstamp. In Sect.\,\ref{sec_lightcurves} have we already introduced the notation regarding \emph{intrinsic variability}, \emph{instrumental systematics}, and \emph{trending} which we will continue to use here. The method for correcting the instrumental systematics was iterative, following the understanding of their nature and an exploration of different ways of approaching the problems present. Below we describe the approach that provided the best results. The method involves a detailed understanding of the systematics present in the data. We illustrate the process on the example of the sample star \emph{Gaia\,DR3\,604895948360165888}\footnote{$G=16.3$\,mag and $G_\mathrm{BP}-G_\mathrm{RP}=1.58$\,mag}.

\subsection{General description of the systematics}\label{appendix_correction_general}
 
    We begin with a description of the systematics because their precise characteristics arise from multiple sources, not apparent immediately. However, those details are crucial to obtaining good results.
 
    As noted before, during the \ktwo{} mission several parts of the \kepler{} telescope ceased to function. Among those were the reaction wheels, crucial for the stability of the telescope's pointing. Their malfunction caused the telescope to drift and to lose its pointing on the sky. This drift was then regularly corrected with the telescope thruster, pointing the telescope back to its intended location. The drift was small and firings were frequent to prevent targets from moving out of the field of view (aside from objects on the edges).
 
    However, this constant motion away and back caused an object to meander across the detector. While the drift itself is slow and the stars do not leave visible trails on the images, they still fall on slightly different clusters of pixels on the detector. Figure\,\ref{fig_image_drift} shows the visible difference in the stellar positions on the detector between two different cadences. This shift is small; a little more than two pixels in a diagonal motion (at least for the \ktwo{} superstamp data). With the large pixels of \kepler{} ($4\,\arcsec$), however, this motion spans $\approx10\,\arcsec$ a significant amount in crowded regions, especially when one considers the overlap of the individual point spread functions (PSF).
 
\begin{figure}[h!]
    \centering
    \includegraphics[width=\linewidth]{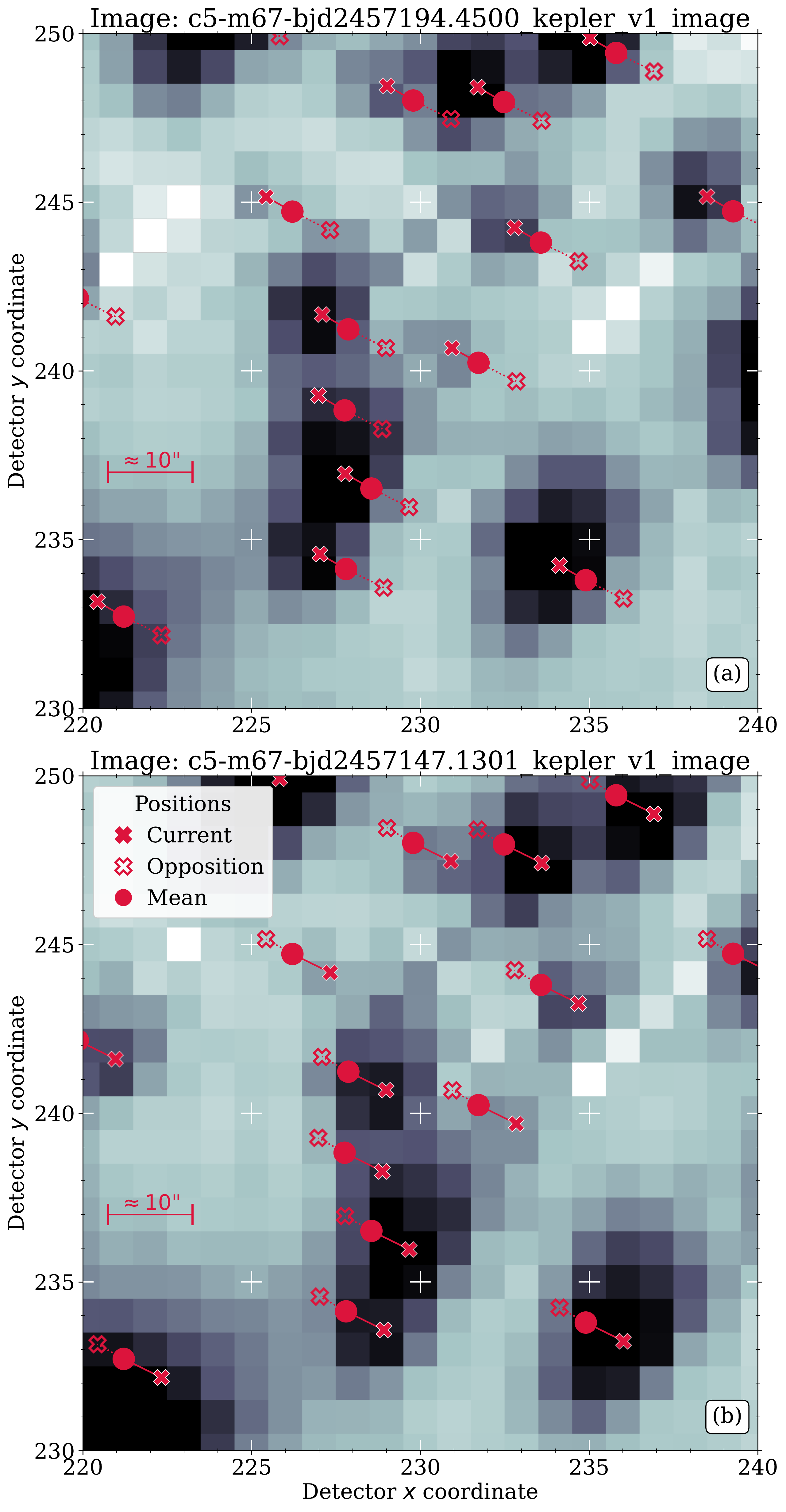}
    \caption{
        Visualization of the two most extreme positions of the telescope drift during C05. Upper and lower panel both show superstamp images ($\Delta t\approx47$\,d) indicating the peak-to-peak positional variation, that is, the minimal and maximal $x$ coordinate of a star. The CCD pixel range is identical in both images. The red dots represent the mean position of a star throughout the observing run, and the filled red crosses connected to them mark the position at the time of the image. Red outlines crosses indicate the extreme opposite positions of a star. The white crosses indicate identical pixel coordinates in both panels to help to visualize the magnitude of the positional changes.
    }
    \label{fig_image_drift}
\end{figure}
 
    While motion alone could probably be corrected relatively easily, an additional effect introduces further complexity; the sensitivity of the detector is not constant across its surface. To be more precise, even on a subpixel scale, two areas of the CCD record different fluxes despite being illuminated by the same amount.

    Considering both the constant drift and varying sensitivity, this means that the recorded flux of an intrinsically constant star is different depending on the epoch of observation. And since the individual regions of the detector are largely independent of each other, those changes in flux are different from position to position (and with that from star to star) on the CCD. Their systematic changes can be small and barely noticeable in a particular star while causing flux differences exceeding 15\,\% for others. Applications that seek to identify variations in the mmag range are therefore difficult.
 
    We are fortunately seeking stars that are not very variable and which change only over the course of tens of days, whereas the systematics occur of time scales of a few hours. As such, we can assume that our targets of interest are constant during a single drift, with only a small change in flux occurring from one drift to another.
 
    However, the motion is more complicated than the description so far; it changes over the course of the campaign, drifts are unequal in length and amplitude, and there are jumps in the pointing. Therefore, we cannot adopt a mean flux from each drift; more precautions have to be taken to obtain reasonable light curves.  Fortunately, the motions do allow us to identify common patterns which are then transferred into chunk-wise processing of the light curves. Images taken during the thruster firing (where the motion during the exposure is significant) exist, but those are fortunately rare (and only one or two at a time) and are consequently simply omitted by us. We also note that the sensitivity changes across the CCD are sufficiently well behaved to allow us to model them with a polynomial function, simplifying the problem.
  
    We use aperture photometry for this work. Photometry that models the point spread function (PSF) generally allows disentangling stars in crowded regions to a certain extent, but there are two problems with the application of PSF photomtery to \ktwo{} data.  The low spatial sampling and large FOV introduce complicated PSF shapes that vary across the detector, and even vary strongly depending on where a star is located on a subpixel scale. This alone can be solved with a sufficiently detailed empirical PSF, for example, see the PATHOS project  \citep{2019MNRAS.490.3806N,2020MNRAS.498.5972N,2021MNRAS.505.3767N,2022A&A...657L...3M} that operates this way on \tess{} data.

    However, given the special problems of \ktwo{} as described above we also face the issue that the PSF changes for a star from image to image, while also being essentially unique for each star in a particular image. The nominal strength of PSF fitting, the assumption that stars have similar PSF shapes despite differing in their fluxes, is not valid for these data.

\subsection{Details of telescope drift}\label{appendix_telescope_drift}
 
    To obtain the information about the motion of the star across the detector, we utilize the world coordinate system (WCS) for each image provided by \cite{2018RNAAS...2Q..25C}.  This allows us to calculate the position $p_0(t)=(x_0(t),y_0(t))$ of a star in each image. While we use the GDR3 ICRS coordinates for this, it would, however, not be a problem to use J2000 (or other) coordinates instead since we are only interested in relative changes.
 
    For simplicity going forward, we will always refer to the telescope drift from the viewpoint of the star moving across the detector. As already noted, the motion of stars is not as simple as a straightforward back-and-forth motion. It is shown in Fig.\,\ref{fig_detector_motion} for one example. The majority of the motion is in a diagonal direction, with an additional slow drift perpendicular to that, the latter jumping with some regularity. These jumps force us to divide the light curve into individual chunks of similar behavior. Since these are all on the same detector, the emerging patterns arethe same for each star, allowing us to construct one mask that works for all.
 
\begin{figure}
    \centering
    \includegraphics[width=\linewidth]{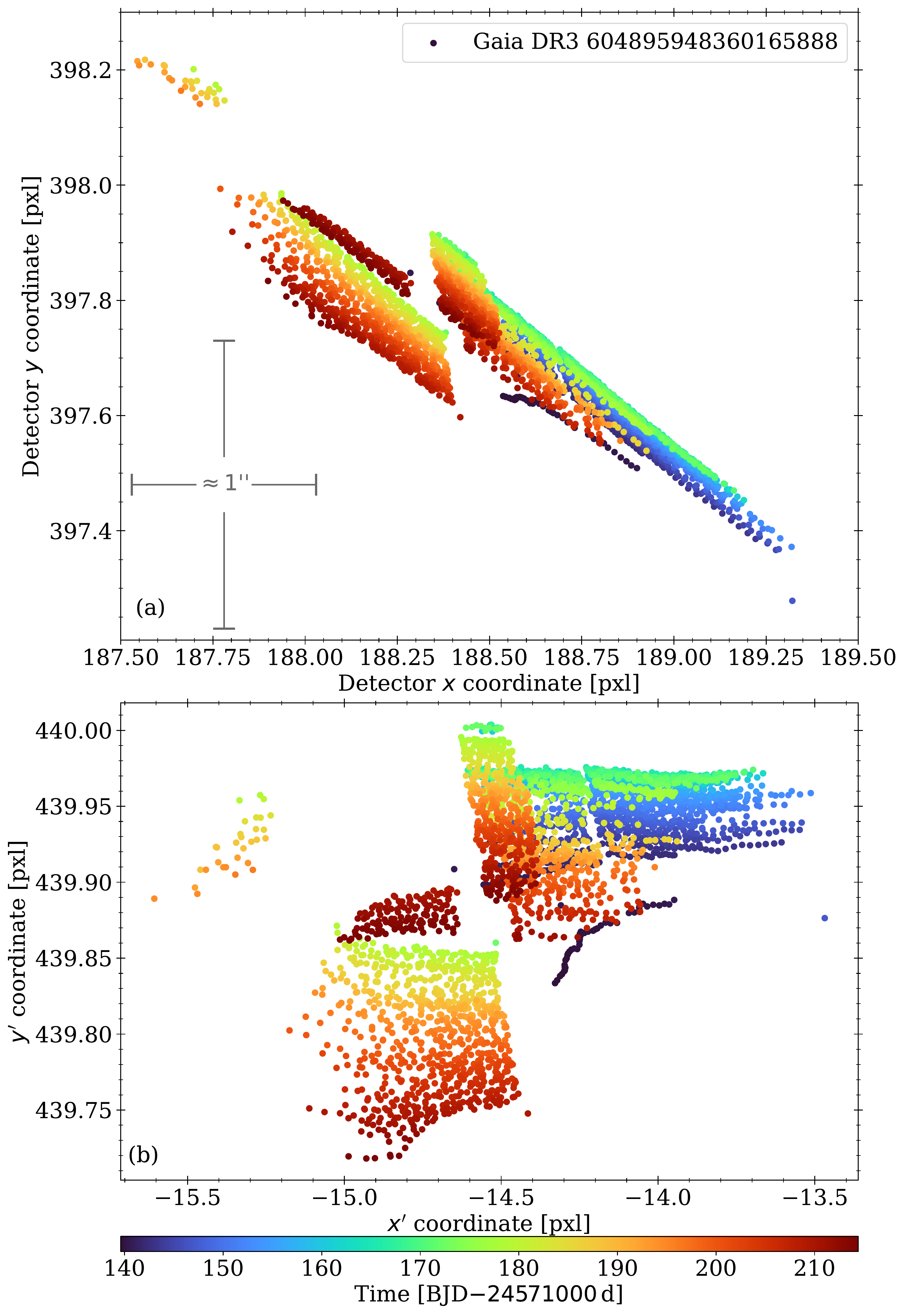}
    \caption{
        Motion of the sample star \emph{Gaia\,DR3\,604895948360165888} across the detector during C05. Each point represents the central position in one image color coded with the time of observation. Panel (a) gives the original detector coordinates and panel (b) those after the rotation applied. The extent of 1\arcsec{} on-sky is indicated. We note the difference in scaling between the x and y coordinate axes, and the change from (a) to (b).
    }
    \label{fig_detector_motion}
\end{figure}
 
    The first step is to simplify this motion by transforming the underlying coordinate system $(x,y)$ to one where the majority of the motion is along one coordinate axis. For that we determine the gradients $m_i$ between two consecutive cadences ($p(t_{i-1})$ and $p(t_{i})$) as
    \[ m_i = \frac{y(t_i) - y(t_{i-1})}{x(t_i) - x(t_{i-1})} \qquad\text{with }i\in\left\lbrace 1,\ldots,3620 \right\rbrace \]
    and take the respective median $\bar{m}$. We note that we purposefully do not take the mean value or use a linear fit to the data.  Both of those approaches fail to provide good approximations of the motion due to the frequent jumps in the data. We then rotate the coordinate system around the angle $\theta = \text{arctan}\,\bar{m}$ such that
    \[ 
        \left(\begin{array}{c} x'(t) \\ y'(t) \end{array}\right) 
        = 
        \Psi \left(\begin{array}{c} x_0(t) \\ y_0(t) \end{array}\right)\quad\text{with}\quad \Psi
        = 
        \left(\begin{array}{cc}
        \text{cos}\theta & \text{sin}\theta \\
        -\text{sin}\theta & \text{cos}\theta \\
        \end{array}\right) 
    \]
    creates a new set of coordinates $p'=(x'(t),y'(t))$. Figure\,\ref{fig_detector_motion} shows the new coordinates for same target. Note the differing scaling for $y'$ and $x'$ in the plot. The motions clearly follow patterns within certain segments of the data in both coordinates. This becomes even more obvious when we plot $x'$ and $y'$ over time as shown in Fig.\,\ref{fig_detector_motion3}.
    
\begin{figure}
    \centering
    \includegraphics[width=\linewidth]{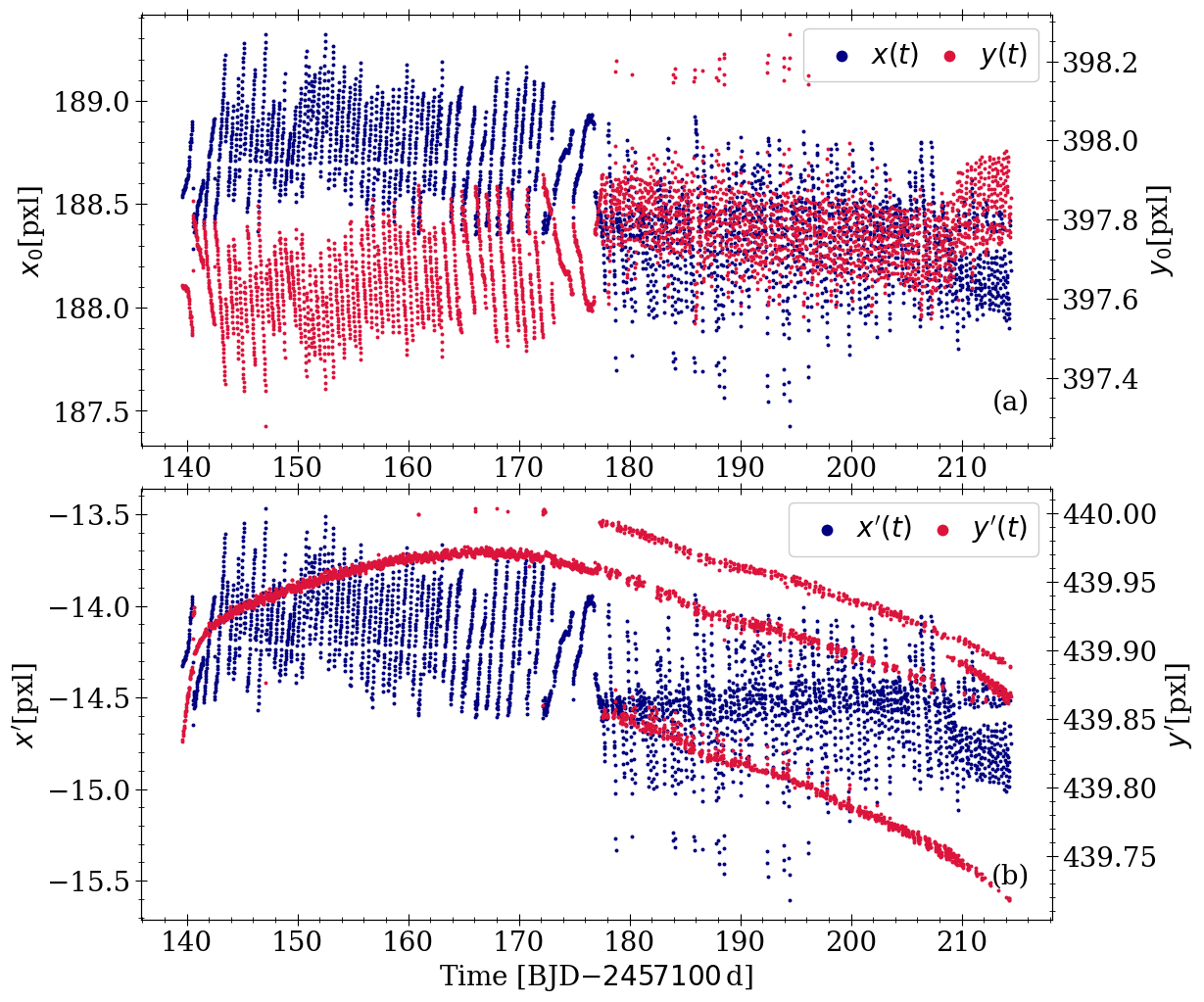}
    \caption{
        Detector coordinates of the sample star \emph{Gaia\,DR3\,604895948360165888} over the course of C05. Panel (a) shows the $x$ and $y$ coordinates while panel (b) shows $x'$ and $y'$ after the rotation. Note that significantly smaller scale for $y'$ compared with $x'$. 
    }
    \label{fig_detector_motion3}
\end{figure}
  
    It turns out that these individual segments of similar behavior are impossible to process together because their individual motion patterns are still too distinctive. Following this, we mask the segments and identify regions of similar behavior for collective processing. Based on the patterns, we mark 12 different groups of pixel behavior which we will henceforth refer to as \emph{segments}. As can be seen in Fig.\,\ref{fig_detector_motion4}, the individual segments are neither identical in size nor are they distinct in time. The latter fact will be advantageous for us below.
  
\begin{figure}[ht!]
    \centering
    \includegraphics[width=\linewidth]{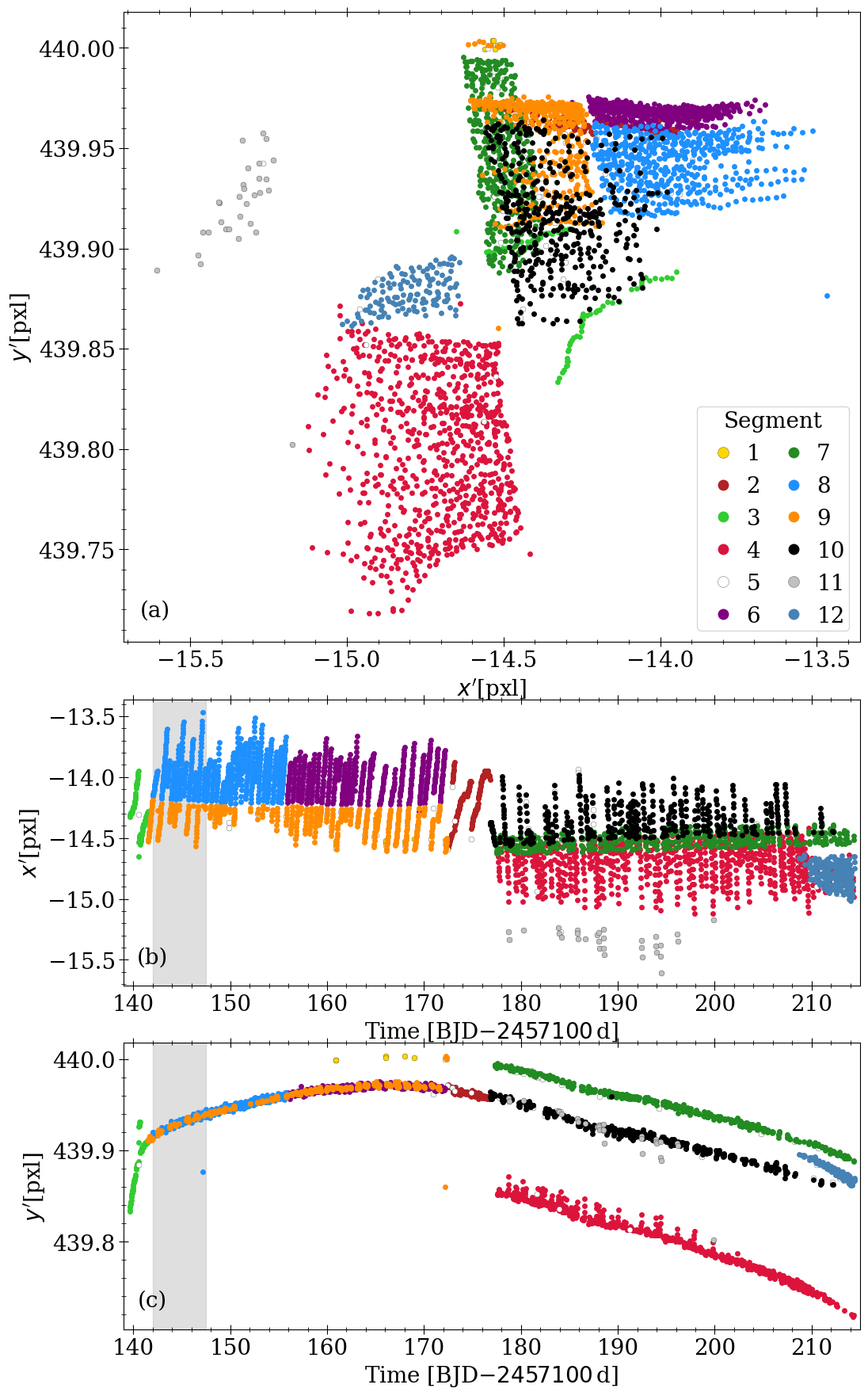}
    \caption{
        Motion of the sample star \emph{Gaia\,DR3\,604895948360165888} across the detector during C05. The correspondence of each point with the segments is color coded. The gray shaded area is inspected more closely in Fig\,\ref{fig_lc_slices} below. 
    }
    \label{fig_detector_motion4}
\end{figure}
  
    This pattern of behavior is indeed real and traceable in the image and not, for example, an artifact of the creation of the WCS. One can easily show this by combining the superstamp images into a movie, with the position of a star according to the WCS indicated and then observe the changes in the observable PSF of the star. These match one another, indicating that the WCS solution is an accurate representation of the positions (and therefore of the motion) of the stars.
  
    There is one potential assumption we have not talked about yet which, if valid, would allow for an additional simplification or validation of our results. It is reasonable to assume that when a star falls two times on a very similar position, the detector response should be identical. With that, all observed flux variations between those two individual cadences should be free of the instrumental systematics,  that means only composed of trending and the intrinsic signal. A closer inspection of Fig.\,\ref{fig_detector_motion4} shows that the distribution of positions of a star would allow this for at least some cadences (upper right region of the figure). However, upon further investigation, we notice that this assumption is not valid in most cases. This can be shown by selecting a star where the photometric noise is negligible and that exhibits a clear signal of variation (e.g., a M67 giant branch star). When we now compare the fluxes of two similar locations and compare those to the overall behavior of the raw light curve, we find that they do not match. We can identify the reason for that. Figure\,\ref{fig_detector_motion4} shows that most the of cadence pairs that we can use for this are made of cadences from different segments. If we limit ourselves to cadence pairs that are only from one segment, the assumption does indeed hold. However, there are too few of those available to be a significant help in the reduction process. At this point we have to look back at the origin of the detector coordinates. They come from the superstamp, which was created from the individual TPFs. Thus they are not necessarily connected to the actual physical pixel coordinates of the detector and may be shifted (by integer multiples of pixels). Judging from what we see, it is reasonable to assume that the distinction into the segments we see finds its origin in a jump in actual pixel coordinates and is not reflected in the superstamp coordinates. It may be possible to verify this assumption from the telescope telemetry itself or the individual TPFs. However, this verification (or rebuttal thereof) does not provide any additional value given that the segments are obvious in the data and therefore, we do not pursue this line of inquiry any further.

    Table\,\ref{tab_segment_overview} shows the extent of the individual segments. Within those segments, we can now express the flux $f$ as a function of the coordinate $x'$ short of one more technical detail we will address below. However, we first need to describe the flux extraction from the superstamp images, because it also needs to be performed segment-wise.

\begin{table}[ht!]
    \centering
    \caption{Different masks used on the light curves.}
    \begin{tabular}{ccccl}
   \hline\hline
   Segment  & Cadences & Block & Slices & Comment \\
   \hline
          1 &     6 &     E &      3 & cut\\
          2 &   211 &     A &      8 & take\\
          3 &    92 &     E &      4 & cut\\
          4 &   796 &     C &    110 & take\\
          5 &    98 &     E &     19 & cut\\
          6 &   505 &     B &     29 & take\\
          7 &   390 &     A &    103 & take\\
          8 &   519 &     B &     32 & take\\
          9 &   414 &     A &     49 & take\\
         10 &   432 &     A &     68 & take\\
         11 &    30 &     D &     13 & cut\\
         12 &   127 &     C &     19 & take\\
   \hline
\end{tabular}

    \label{tab_segment_overview}
    \tablefoot{\emph{Segments} refers to identification used in the text. \emph{Candences} and \emph{Slices} list the number of each included in each segment. The \emph{comment} denotes whether a segment is adopted for the final light curve.}
\end{table}

\subsection{Flux extraction from the full frame images}\label{appendix_aperture_photometry}
 
    The FFI underwent background subtraction in the \ktwo{} pipeline. We see no reason to revisit this, and accept it as is. However, we will introduce an additional background subtraction to reduce the impact of the light from bright stars surrounding a particular target (as detailed below).
    
    As described in Sect.\,\ref{sec_lightcurves}, we use aperture photometry with an individually defined pixelmask rather than a fixed aperture or PSF photometry. A fixed circular or elliptical aperture differs too much from the very non-Gaussian PSF present in the FFI. Furthermore, taking only fractions of the flux stored in a given pixel based on the degree of aperture coverage introduces additional artifacts in the recorded flux.
 
    To include all the relevant flux of the star, we need to account for the motion of the star as well. The peak to peak motion covers more than two pixels. In a situation where the PSF of a star of intermediate brightness roughly covers a cluster of $5\times5$ pixel, this is a significant effect and cannot be neglected. The brightest stars have PSFs about three times this size, whereas the faintest stars can be adequately covered by $2\times2$ pixel. Thus we need to account for the motion in the design of the aperture mask. For stars that are isolated, this does not present a problem as the aperture can just be designed large enough to cover all eventualities, though that would accumulate unnecessary noise. For stars in crowded regions, this is not a valid approach since the apertures begin to overlap and flux from neighboring stars is recorded.
 
    Thus, we need an aperture that moves with the star. As already stated above, taking only fractions of the flux from a pixel introduces additional unwanted effects. Thus, we cannot have the pixelmask gradually moving with the star. Another approach that adds and removes entire pixel to and from the mask based on the current position of the star fails due to similar problems.  Given the originally diagonal motion of the star and the fact that our determined segments separate mostly perpendicular to this motion, we use the following approach: we design different pixelmasks for different segments. It is, however, not necessary to design different pixelmasks for every segment. The coordinates allow different segments to share the same pixelmask. We refer to this grouping as a \emph{block}, denoted with the letters A to E, for the five groups we found. Table\,\ref{tab_segment_overview} includes the block designation in the column of the same name. We note that the blocks D and E only contain regions that will be cut due to the peculiar motion during the involved segments. Therefore, we focus on the blocks A\,--\,C from here on. The raw flux is extracted in D and E with a copy of the pixelmask for A.
 
    We go on to design a pixelmask for each star, based on its visible shape in an image that averages all superstamp FFIs belonging to a certain \emph{block}. This design is performed manually and iteratively. We endeavor to create a pixelmask for each star in each block that includes most of the flux of a star while simultaneously excluding flux from surrounding stars. This is, however, not always possible and so there are stars for which it is impossible to define a pixelmask that allows good extraction of the flux. We discard those from the sample. The design is iterative because the work showed that an incomplete pixelmask \emph{may} leave strong imprints on the extracted flux that cannot be corrected. Thus, we iterate between pixelmask design and flux correction until we achieve light curves for the individual stars that are appropriate for our science case. In Figure\,\ref{fig_pixelmask}, we show the pixelmask for the example star in block A together with a comparison to a high resolution image from the \emph{Digitized Sky Survey 2} (DSS2\footnote{\url{archive.eso.org/dss/dss}}).
 
    Given the extent of manual work on each target in the FOV, we limit ourselves to targets of interest and a certain number of additional stars to verify the method. At this point we also discard all stars which cannot be separated well-enough from neighboring stars.
 
\begin{figure}[ht!]
     \centering
     \includegraphics[width=\linewidth]{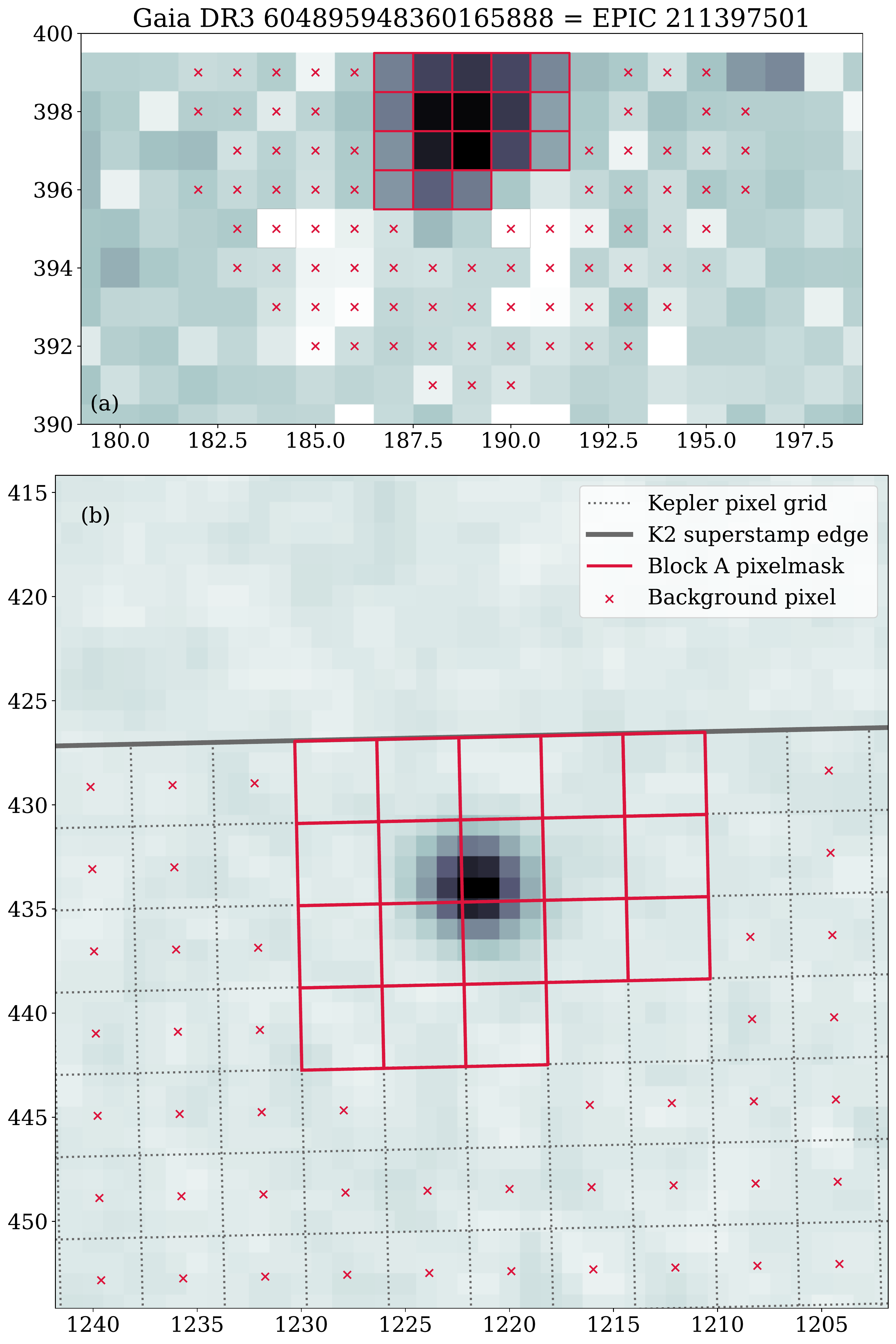}
     \caption{
        Pixelmask and background selection for \emph{Gaia\,DR3\,604895948360165888} in block A. Panel (a) shows an averaged \ktwo{} superstamp image around the target. Panel (b) shows a comparison with a DSS2 (red channel) image of the same region. \ktwo{} pixels outlined in red mark the chosen pixelmasks pixels selected for the background estimate are marked with an X. The DSS2 mage has the \ktwo{} pixelgrid indicated by the dotted lines and the solid line marks the edge of the superstamp.
    }
    \label{fig_pixelmask}
\end{figure}
 
\begin{figure*}[ht!]
    \centering
    \includegraphics[width=\linewidth]{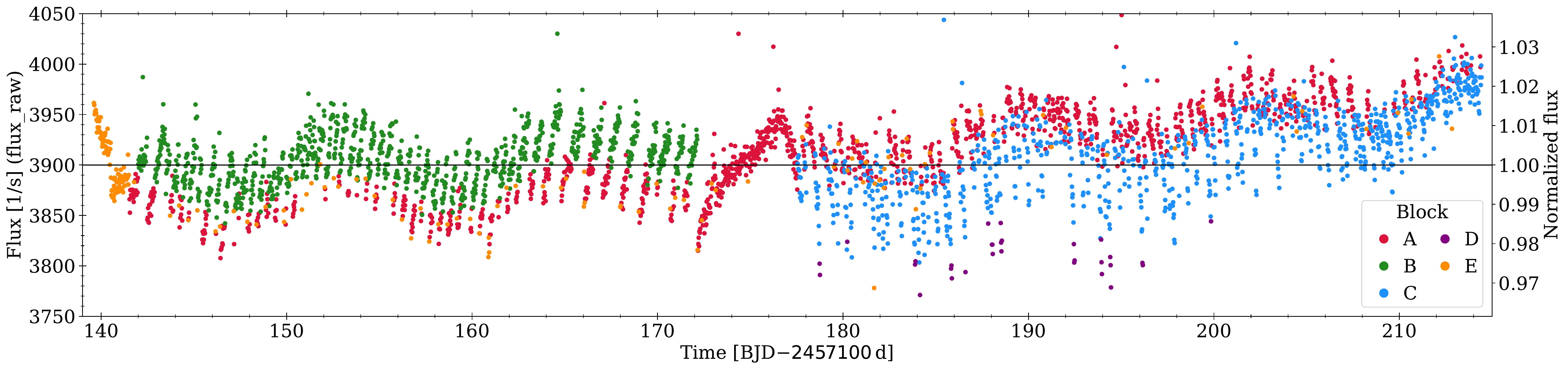}
    \caption{
        Raw light curve $f_\text{raw}(t)$ of \emph{Gaia\,DR3\,604895948360165888}. 
    }
    \label{fig_lightcurve_raw}
\end{figure*}
 
\subsection{Extraction of the light curve and background subtraction}\label{appendix_correction_extraction}
 
    We can now extract the raw flux $f_\text{raw}(t)$ based on the pixelmasks found by simply adding the flux from all included pixels for each cadence. Before doing so, we create a mean image for each block and determine all pixels fainter than the median in each. We assume that those pixels constitute the background. However, we do not use all those pixels for each star. From this selection, we use the ones closest to the extracted star. Here we sort the pixels by their distances to the mean position of the star in the block, exclude a potential overlap with the pixelmask itself, and include all pixels until we have at least three times as many background pixels as we have in the pixelmask. We then use the median of those pixels as the background and subtract it, weighted by the number of pixels included. This procedure allows the removal of stray light from stars in the vicinity. The number of pixels included and the use of the median are based on extensive testing of different variations and the final procedure presented appears to provide the best result overall. This provides us with the raw light curve $f_\text{raw}(t)$ which is shown in Fig.\,\ref{fig_lightcurve_raw} for our example star. The typical systematic patterns due to the telescope motion are obvious.  We note that due to differently shaped pixelmasks, the individual segments are generally not as well aligned as seen in Fig.\,\ref{fig_lightcurve_raw} but show offsets (cf. Fig.\,\ref{fig_lightcurve_comaparison_pdcsap}). Those are eliminated when the light curves are realigned (see below).
 
 \subsection{Correction of instrumental systematics}\label{appendix_correction_instrumental}
 
    With the raw, but background subtracted light curves $f_\text{raw}(t)$ at hand, we can now proceed to correct the instrumental systematics. This process continues to be on a star-by-star basis and is also carried out segment-by-segment. When we investigate the position dependency of the flux $f_\text{raw}(x')$ (see Fig.\,\ref{fig_lightcurve_slice_correction}) we see that the correlations are well-behaved enough to be modeled with a low order polynomial. To a large degree, the trends are almost linear but even a slight curvature can have a significant impact. We find that a 5th-order polynomial s the best compromise for the fit and use it to reproduce $f_\text{raw}(x')$ for each segment. However, one more difficulty has to be accounted for before doing so. When we directly fit a polynomial to $f_\text{raw}(x')$, it is very dependent on the range covered in $x'$ and the number of points responsible for said coverage.
 
    At this point we introduce the a third (and last) masking variable which we will refer to as a \emph{slice}. A slice is a short section of the light curve which covers one continuous episode of motion between two jumps. In Figure\,\ref{fig_detector_motion3}, each apparent upward (or downward) stripe for $x'$ is distinguished as a slice. (There are jumps and gaps that are not apparent in the plot, causing the number of slices to be higher than obvious.) By virtue of the design of the process, each slice in its entirety is part of only one segment.  Figure\,\ref{fig_lc_slices} provides a zoomed-in view of the light curve in Fig.\,\ref{fig_detector_motion4}, with the slices distinguished by the color coding. Slices vary strongly in their extent for all involved parameters ($x'$, $t$, number of cadences). They typically cover 5 to 15 cadences, with the smallest containing only one individual cadence. Table\,\ref{tab_segment_overview} includes a column that lists the number of slices in each segment.
  
\begin{figure}
    \centering
    \includegraphics[width=\linewidth]{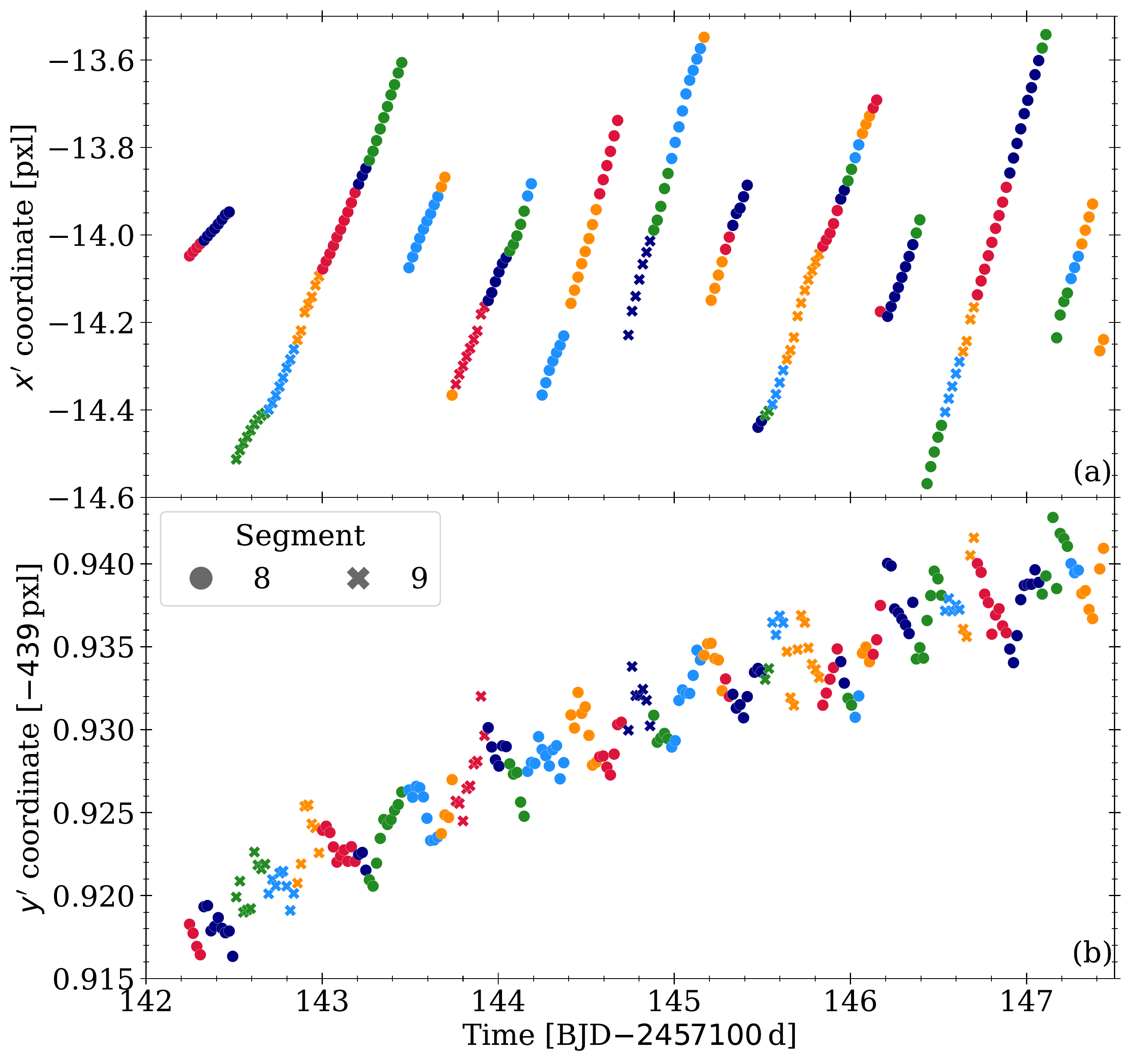}
    \caption{
        Designation of the slices in a short light curve segment. Panels (a) and (b) show the time dependency of $x'(t)$ and $y'(t)$, respectively. The alternating color coding distinguishes individual slices and different symbols refer to different segments as indicated in panel (b). Only a small cutout from the full light curve is shown, corresponding to the shaded area in Fig.\,\ref{fig_detector_motion4}. It contains 40 \emph{slices} for the sample star \emph{Gaia\,DR3\,604895948360165888}.
    }
    \label{fig_lc_slices}
\end{figure}
 
    We now utilize the slices that were introduced above. All slices of one segment exhibit the same systematic behavior (cf. Fig.\,\ref{fig_lightcurve_slice_correction}). However, given the fact that they are spread over several days (the segments cover typically about half the baseline of the light curve), even slow variations in the light curve leave their imprints in the mean flux of a slice. Furthermore, the individual \emph{slices} have a different degree of coverage in $x'$. This complicates a polynomial fit to the data. We need an additional technical step to proceed.
 
\begin{figure}
    \centering
    \includegraphics[width=\linewidth]{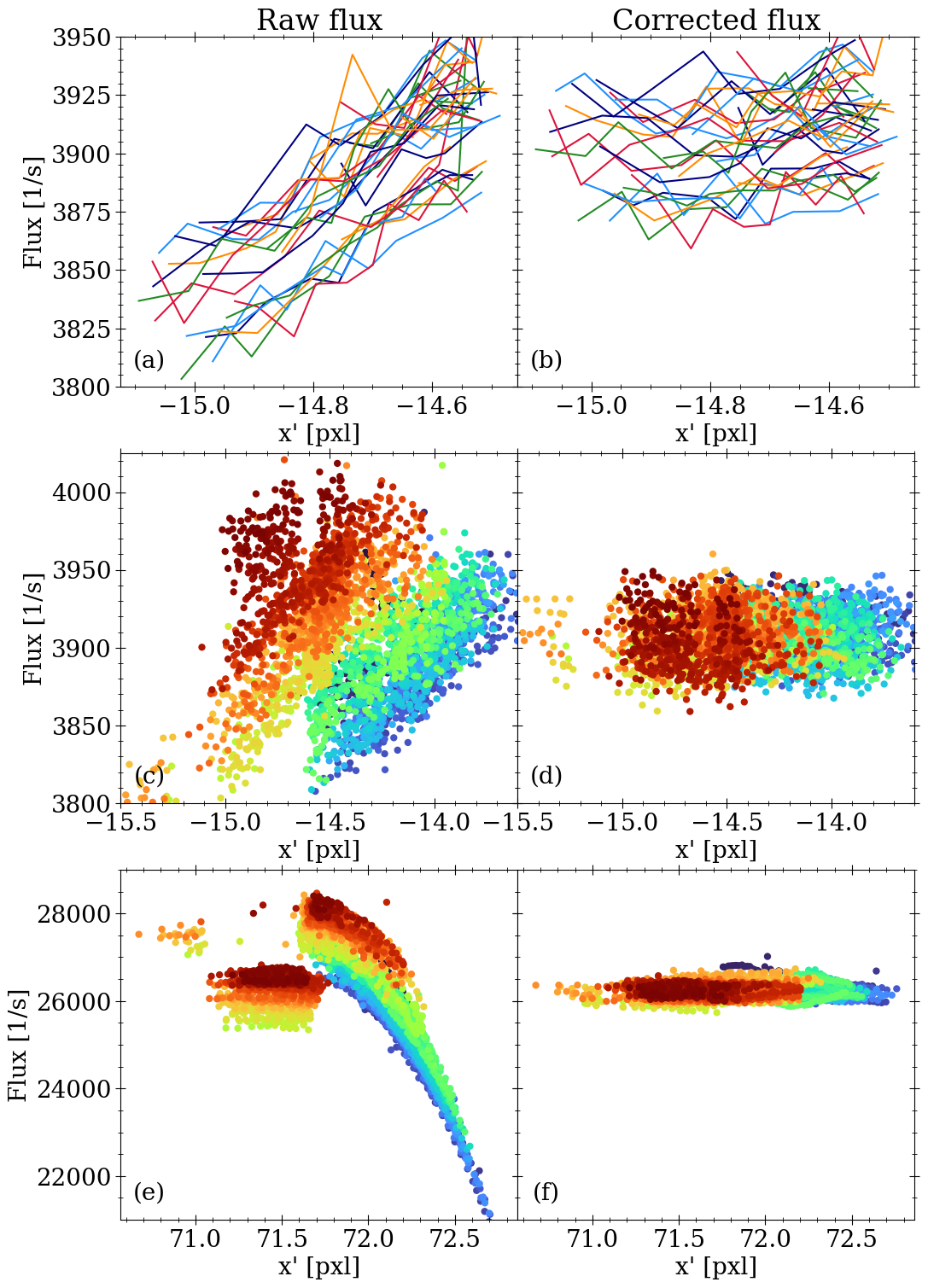}
    \caption{
        Position dependence of the recorded raw flux $f_\text{raw}(x')$ and the correction thereof. Panel (a) shows the raw flux from \emph{Gaia\,DR3\,604895948360165888} as a function of the detector coordinate and clearly shows the linear dependence. Flux values are plotted for the same \emph{slices} as in Fig.\,\ref{fig_lc_slices}. Each line represents one \emph{slice}. Panel (b) shows the same \emph{slices} but with the flux after the correction is applied. Panels (c)\,+\,(d) are similar to (a)\,+\,(b), only for the full light curve of \emph{Gaia\,DR3\,604895948360165888} and the color coding representing the time as in Fig.\,\ref{fig_detector_motion}. Panels (e)\,+\,(f) are the same as (c)\,+\,(d) but for \emph{\object{Gaia DR3 604917629355039360}} for which the position dependency becomes nonlinear.
    }
    \label{fig_lightcurve_slice_correction}
\end{figure}
 
    We designate the one \emph{slice} which has the most extensive coverage in $x'$ in each segment as the \emph{prime slice}. Each \emph{slice} is then rescaled to this prime slice. We proceed to fit a 5th-order polynomial $p_5(x')$ to the resulting distribution of fluxes, and apply it as a correction to the individual \emph{slice} (their unscaled fluxes!) as
    \[ f_\text{cor1,slice}(x') = f_\text{raw,slice}(x')/p_5(x') \cdot \overline{f_\text{raw,seg}(x')} \]
    to create the first step in the correction process $f_\text{cor1}$. We multiply with the average flux of segment $\overline{f_\text{raw,seg}(x')}$ because the polynomial normalizes the fluxes to the \emph{segment} mean. The result can be seen for a few selected slices in Fig.\,\ref{fig_lightcurve_slice_correction}. This processes is carried out for all \emph{slices} in a \emph{segment} and for all \emph{segments}. We note that this process conserves long term flux changes and differences between individual \emph{slices}. The resulting light curve for our example star \emph{Gaia\,DR3\,604895948360165888} is shown in Fig.\,\ref{fig_lightcurve_cor1}.
 
\begin{figure}
    \centering
    \includegraphics[width=\linewidth]{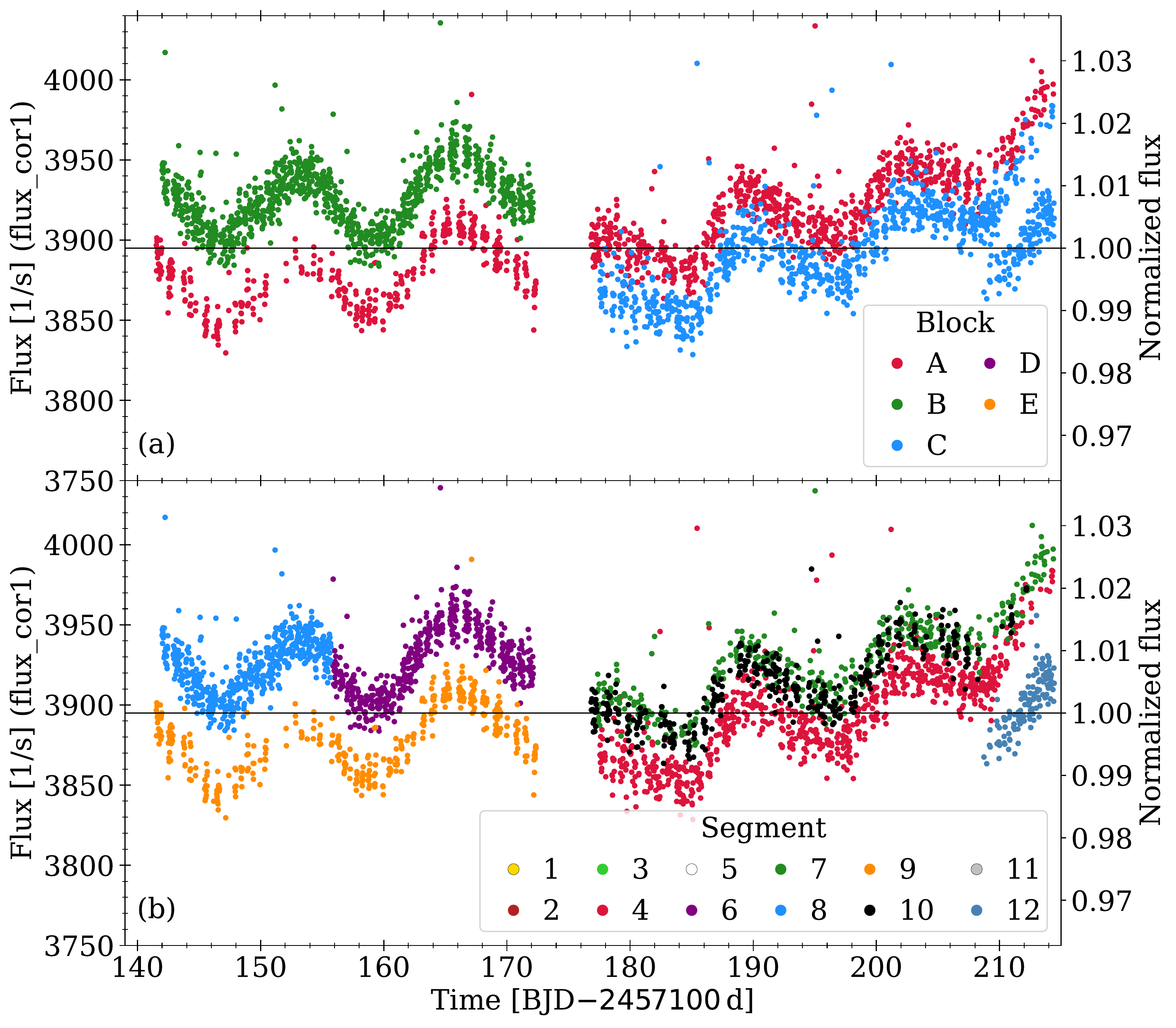}
    \caption{
        Light curve of \emph{Gaia\,DR3\,604895948360165888} after the application of the first correction for the instrumental systematics. Both panels show the same light curve. The color coding in the upper panel indicates the block as in Fig.\,\ref{fig_pixelmask}. In the lower panel, the colors indicate the segments as in Fig\,\ref{fig_detector_motion4}.
    }
    \label{fig_lightcurve_cor1}
\end{figure}
 
    This process reveals that two regions of the light curve cannot be processed in this way. The first $\approx 3\,$d and a central part ($t\approx 172$\,--\,177\,d) correspond to exceedingly long \emph{slices} that exhibit behaviors that cannot be found anywhere else and therefore those regions cannot be corrected this way (cf. the same regions in Fig.\,\ref{fig_detector_motion4}). We mask those regions and cut them from the final light curves. Furthermore, there are individual cadences in the light curve from times of fast telescope motion (thruster firing). Those cannot be processed as well. We mask them in the correction step and later, when all other \emph{segments} are processed, recreate those points from a linear interpolation in the corrected light curve. Those regions are entirely part of the \emph{blocks} D and E (cf. the comment column Tab.\,\ref{tab_segment_overview}).
  
    Thanks to the processing described above, the relative fluxes between the individual slices is conserved and the (fragmented) light curve of a \emph{segment} may act as a valid light curve all by itself. However, this is not true when we compare the fluxes of the individual segments. They are not aligned. This is, however, not a result of our processing but a consequence of the instrumental systematics and different pixel masks that gives each segment a slightly different recorded average flux (cf. Fig.\,\ref{fig_lightcurve_raw}). To align those fluxes, we merge the segments successively with a scaling to the chronologically first sector. The scaling factor is determined by minimizing the average standard deviation in the merged light curve for a series of windows 15 cadences wide along the light curve. This rescaling provides us with the next level of correction $f_\text{cor2}$ which is shown for example star in Fig.\,\ref{fig_lightcurve_cor2}. For the chronologically first segment we set $f_\text{cor2}=f_\text{cor1}$.
 
\begin{figure}[ht!]
    \centering
    \includegraphics[width=\linewidth]{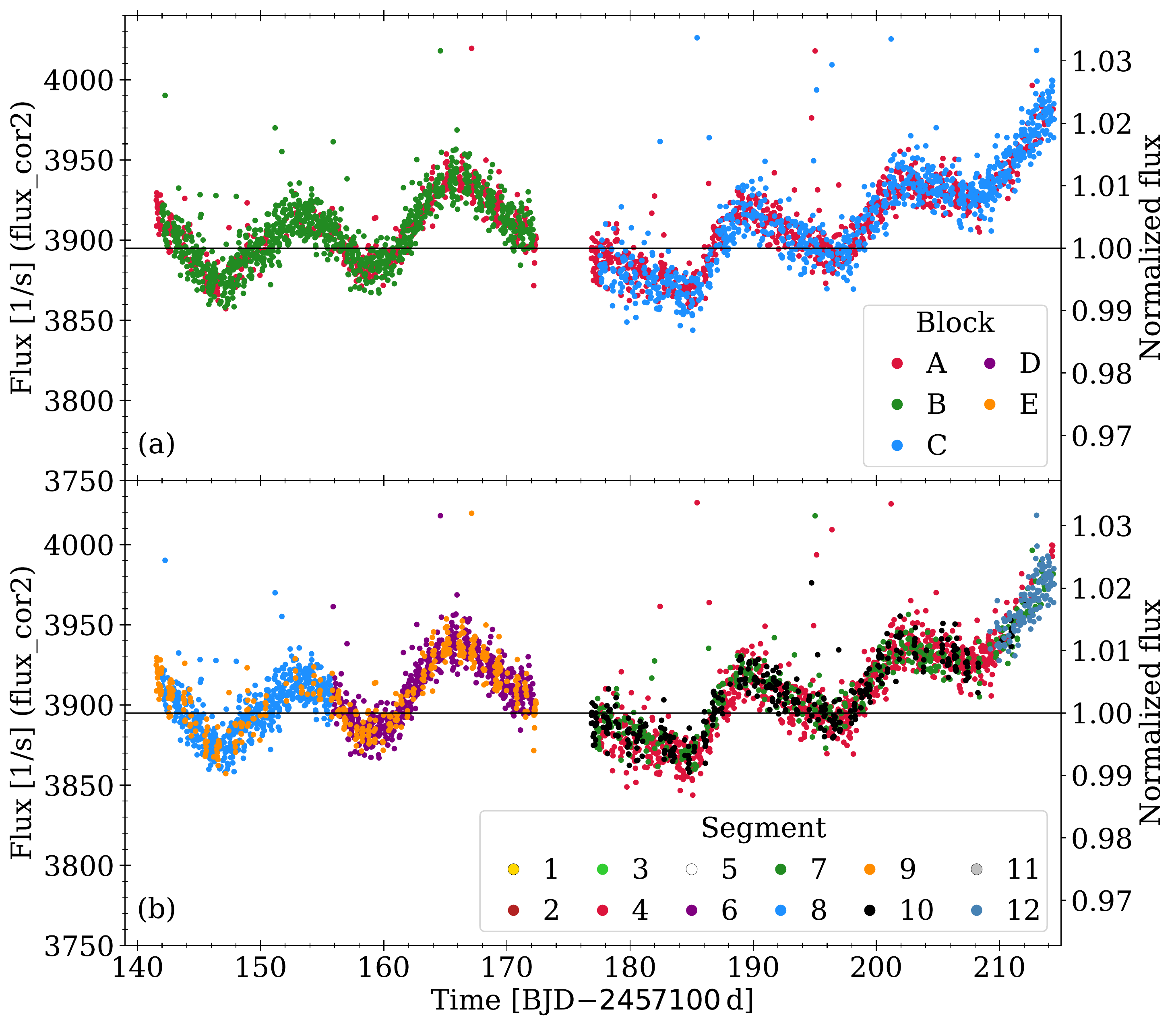}
    \caption{
        Light curve of \emph{\object{Gaia DR3 604895948360165888}} after the application of the second correction, the realignment of the segments. Both panels show the same light curve. The color coding in the upper panel indicates the \emph{block}. In the lower panel, the colors indicate the segments as in Fig\,\ref{fig_detector_motion4}. We emphasize that the flux axis is identical between the Figs.\,\ref{fig_lightcurve_raw}, \ref{fig_lightcurve_cor1}, and this one.
    }
    \label{fig_lightcurve_cor2}
\end{figure}
 
    The corrected and realigned flux $f_\text{cor2}$ now represents the light curve in which we have removed all instrumental systematics. However, this is not fully true for all light curves. A polynomial fit can only do so much and an insufficient pixelmask may introduce effects that cannot be corrected. As such, the process fails for stars in very crowded regions and on the edges of the superstamp. The slices generally cover a few hours. Intrinsic variations that occur on the same timescale may be misidentified by the polynomial. However, such rapidly varying targets are not of interest to us here, allowing us to ignore this problem. Despite this, we note that rapid signals are generally still visible reasonably well in the final light curves, thanks to their large amplitudes.

\subsection{Cleaning the data}\label{appendix_correction_cleaning}

    From this point onward, there is no further separation into blocks, segments, or slices, and the light curves are always treated as a whole. In the next step we apply $\sigma$-clipping to the data to clean it of outliers. This is done only at this stage, because the instrumental systematics may create or hide actual flux outliers and a $\sigma$ estimate is dominated by it. For each point along the light curve we calculate the mean and standard deviation in a 1\,d window around the point and replace it with a linear interpolation between its neighbors if it exceeds a $3\sigma$ deviation from the mean. We note that this may create artifacts for fast transients like transits or flares. However, since we are not interested in such phenomena, we elect to ignore those issues.
 
    At this point we have constructed light curves for target stars that are corrected for instrumental systematics and cleaned of outliers. The light curves exhibit a gap of a few days around the middle and still include long term trending effects. The removal of the latter will be performed using Principal Component Analysis (PCA). 
 
\subsection{Principal Component Analysis}\label{appendix_correction_pca}
 
    PCA is performed in a very similar manner to the light curve processing for Ruprecht 147 \citep[from \ktwo{} C07, see][Appendix B]{2020A&A...644A..16G}. However, we modify the process slightly and describe the deviations below.
 
    Because we do not have a large number of ($\sim 10$k) light curves from the entire campaign for this dataset, we select stars for the PCA basis from our processed sample. We identify 769 stars for the said basis, covering a large range of brightnesses, colors, and locations on the detector (cf. Fig.\,\ref{fig_pca_stats_1}). From this basis, it becomes obvious that the trending signal is not universal for all stars. It shows a strong brightness dependence and a weak location dependence. We therefore limit the PCA basis applied for each target  to stars of similar brightness and location. This means we take only the 125 nearest stars on the FFI within $\Delta G\pm1$\,mag. Both of these choices are the result of testing different parameter combinations; these particular ones provided the best results throughout the sample. We note that, depending on the brightness, there could be fewer than 125 stars overall in the brightness range and the PCA basis for such stars is correspondingly smaller.
 
\begin{figure}
    \centering
    \includegraphics[width=\linewidth]{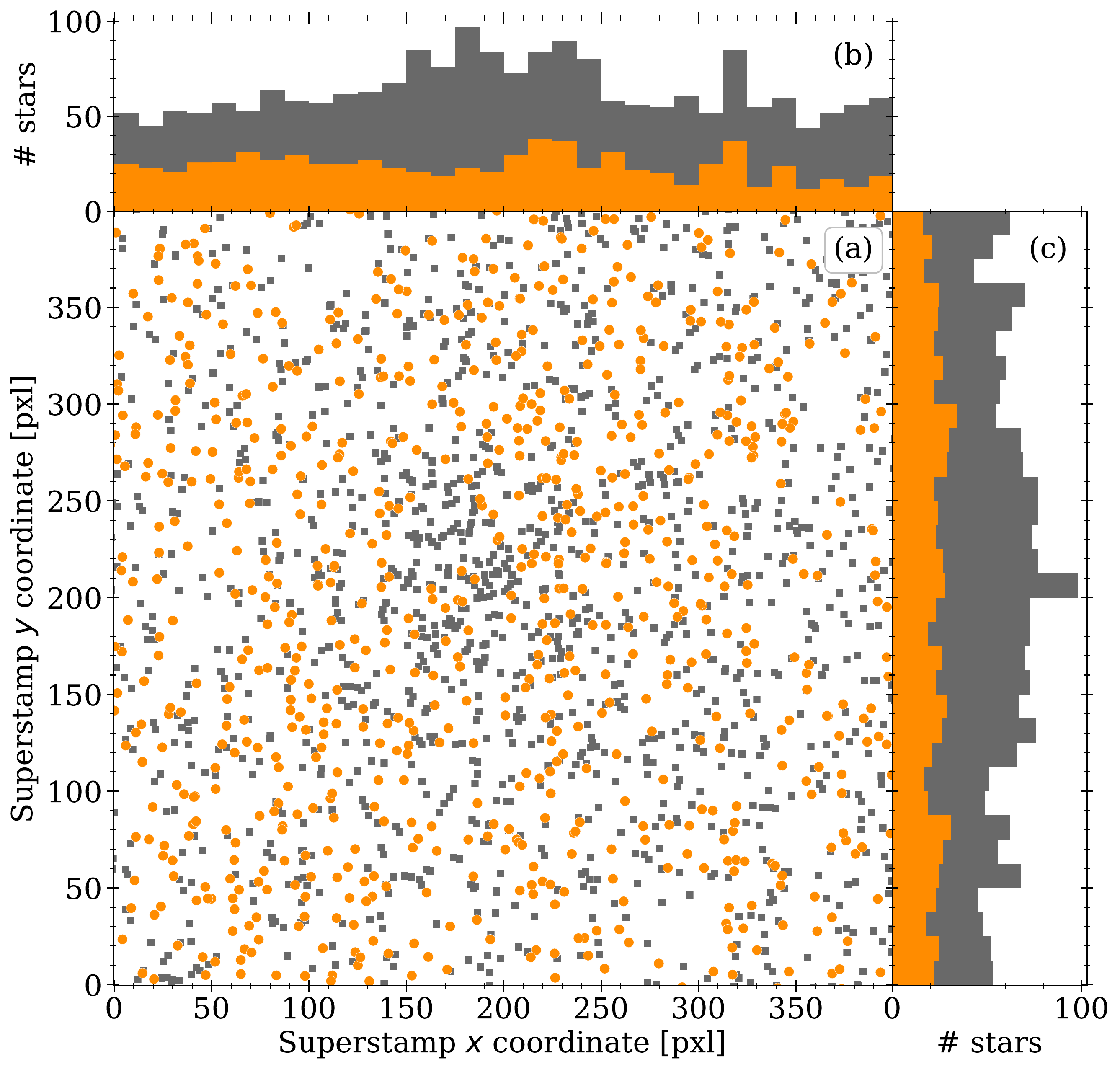}
    \includegraphics[width=\linewidth]{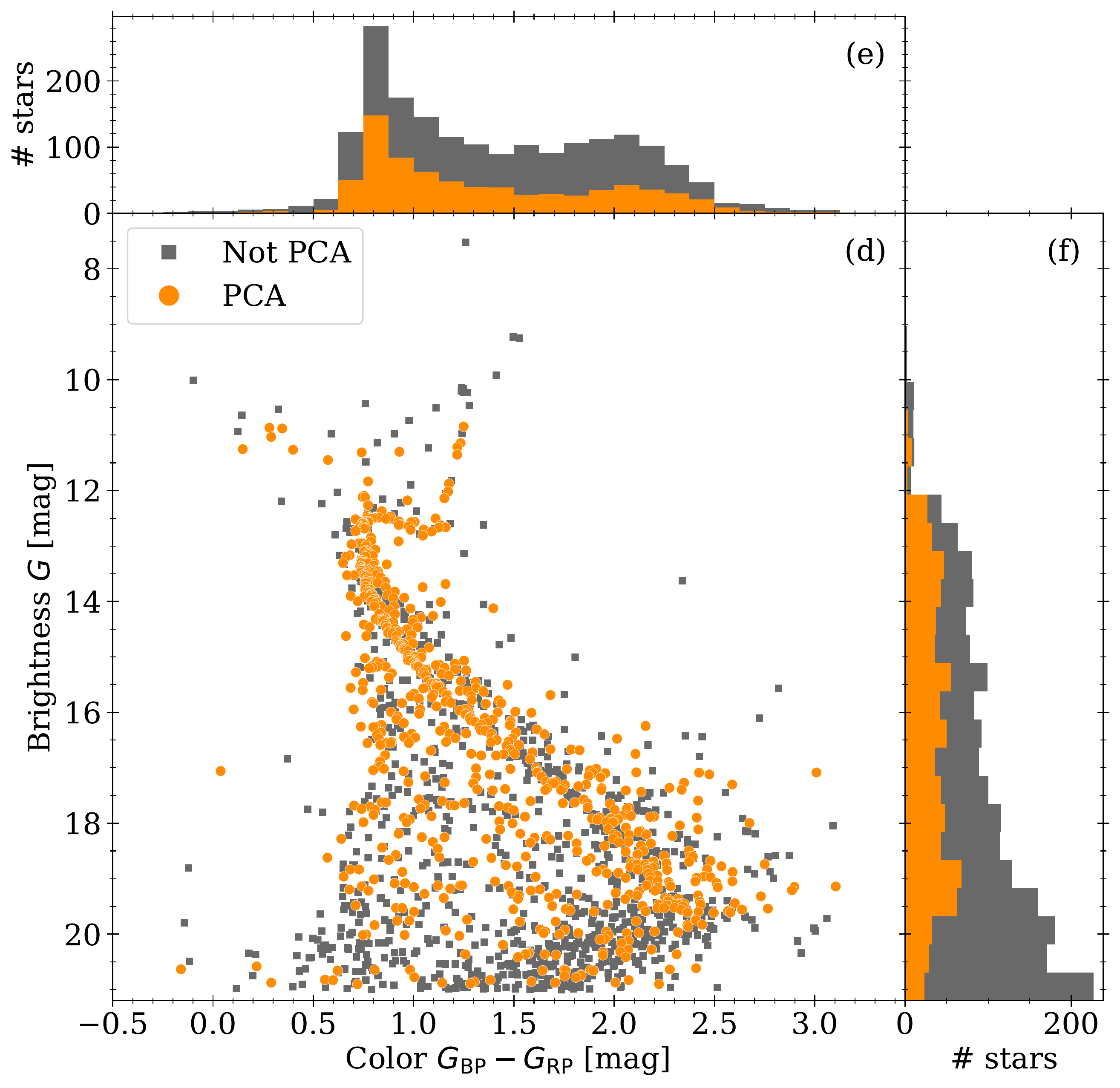}
    \caption{
        Star selection for the Principal Component Analysis, indicating those used and those not used. 
    }
    \label{fig_pca_stats_1}
\end{figure}
 
    We smooth the light curves before entering them into the PCA to remove the impact of the noise from the analysis. This smoothing is done by replacing each flux value by the mean of a 1\,d wide window around this point. We note that this smoothing is only in place to create the PCA correction and is not used beyond that. Furthermore, we normalize all light curves to their respective means before entering them into the PCA. This smoothing removes rapid-trending signals from the PCA input. This means that such signals are not accounted for in the PCA correction and thus remain in the final light curve. However, their impact is small with respect to our purposes and as such we do not address them any further. Traces of this can be seen in the light curve plots in Fig.\,\ref{fig_lightcurve_sample_1} and are generally of two different types: 
    \begin{enumerate}
        \item A very rapid fluctuation, as seen for instance, between $t=150$\,d and $t=160$\,d for \emph{\object{Gaia DR3 604914880575997056}}  and
        \item A distinct short term signal, as seen for instance, between $t=150$\,d and $t=170$\,d for \emph{\object{Gaia DR3 604915773929008384}}.
    \end{enumerate}
 
    The (long-term) trending signals are dominant and the PCA can identify their different shapes relatively easily. We use only two components for the reconstruction; this is enough to remove the trending signals adequately. This is a consequence of the limited PCA basis (brightness and position), which preremoves very different trending signals from the basis. Figure\,\ref{fig_pca_application} shows the application of the PCA correction on the example star and on an essentially constant star.

\begin{figure}
    \centering
    \includegraphics[width=\linewidth]{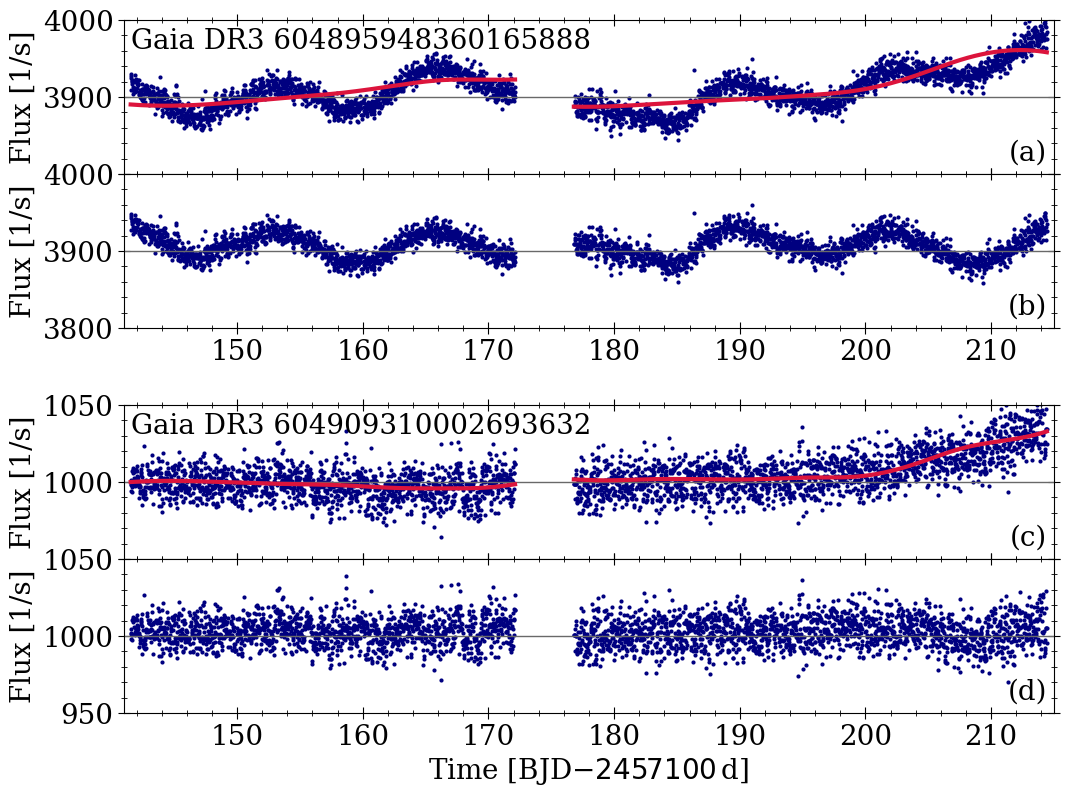}
    \caption{
        Application of the PCA correction to the example star \emph{Gaia\,DR3\,604895948360165888} (panels a and b), and the fainter and essentially constant star \emph{\object{Gaia DR3 604909310002693632}} (panels c and d). Panels (a) and (c) each show the result of the instrumental systematics correction (blue) and the PCA correction (red) whereas panels (b) and (d) show the final light curve after the PCA correction was applied.
    }
    \label{fig_pca_application}
\end{figure}
 
    At this point we are in possession of the final light curves $f_\text{fin}$ which are corrected for instrumental systematics, cleaned for outliers, and are free of trending. Figure\,\ref{fig_full_lightcurve_process} summarizes the individual steps for the example star. This process is by no means perfect, and does not work for all targets. We only use a carefully selected subsample of all potential targets, and remove all light curves where residuals of the systematics or trending cause uncertainties. Most removed light curves suffer from an insufficient pixelmasking (mostly due to being in a crowded area in the superstamp FFI) or because the corresponding star coincides with a region of bad pixels. Both problems make corrections nearly impossible. Only for a few stars does the correction itself directly fail. For those, the flux does not behave well enough to be adequately modeled with a polynomial. However, their numbers are small enough not to merit the effort required for an adjusted correction procedure.

    The above-described process is specifically designed for the \ktwo{} C05 superstamp around M67. However, it should be possible to adapt it for other parts of the \ktwo{} survey as well. The principal idea should hold and, assuming that the detector variability stays similar in its magnitude, provide good results without modifying the core components. The parts that would need to be modified are those that are fine-tuned to the superstamp. The pixelmasks for the individual targets are the first among these. Secondly, the masking (block, segment, slice) would also need to be adjusted to a different campaign. However, we can reasonably assume that our masks work for other regions of C05 without modification. Caveats such as the difficulties with rapidly varying stars and crowded regions do, of course, remain. A precondition for such work is the existence of a reference for the telescope motion, and with that of motion of the stars on the detector. The \ktwo{} TPFs do not provide this, but some of the other superstamps created by \cite{2018RNAAS...2Q..25C} do. We note that the SPICE kernels for the \ktwo{} mission only include position and velocity of the telescope but not its orientation, a necessity to be able to leverage those products. 
 
\subsection{Performance of the reduction}\label{appendix_performance}

    The correction process described above works well for stars that are isolated and whose PSF does not exceed the FFI range, that is, stars for which we can define a pixelmask that includes all of the stellar flux and none from surrounding sources. This condition is not fulfilled on the edges of the superstamp, and especially in the central region of the cluster. Figure\,\ref{fig_other_lightcurve_examples} highlights some details that show the performance and limitations of our corrections.

\begin{figure}[ht!]
    \centering
    \includegraphics[width=\linewidth]{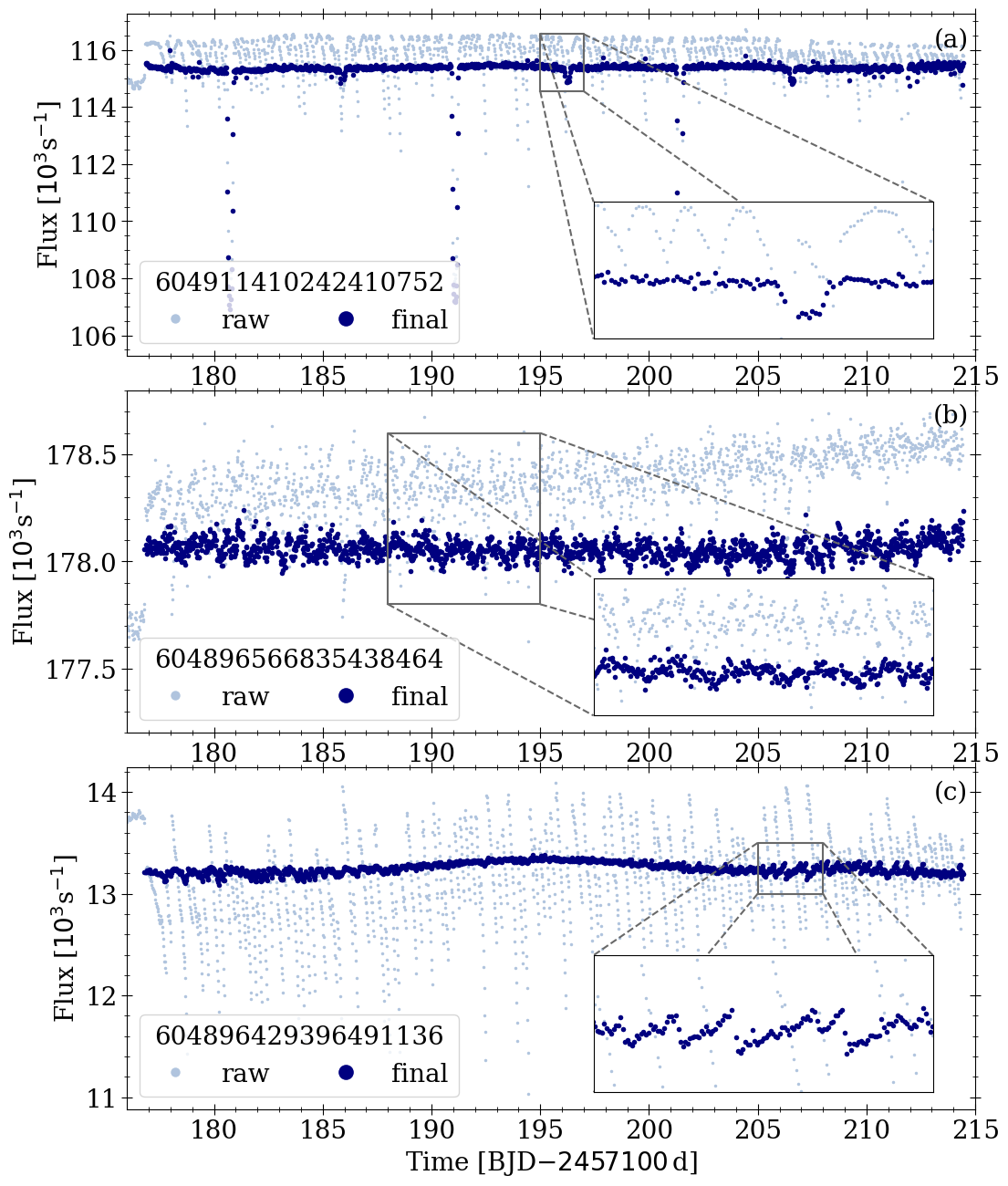}
    \caption{
        Example light curves illustrating the capabilities and limitations of our correction process. Panel (a) shows the light curve of an eclipsing binary with a significant primary eclipse and a secondary eclipse that only becomes really apparent after the correction was carried out. Panel (b) shows a star where the instrumental systematics completely obscure the intrinsic signal despite their being of comparable amplitude. Panel (c) shows the light curve of a star located at the edge of the superstamp and which suffers from artifacts created by the correction process as a consequence. Each panel includes highlighted regions to show the details of the relevant effects.
    }
    \label{fig_other_lightcurve_examples}
\end{figure}

    Whenever the pixelmask cannot, for whatever reason, be created in a way that includes all the stellar flux the recorded raw flux $f_\text{raw}$ includes a second systematic effect. This effect invalidates the original assumption we had to make, that the raw flux only changes because of variations in detector sensitivity, and is otherwise somewhat constant on short time scales. With that, it very much depends on the individual star's location on the detector whether our correction still works or whether we under- or over-correct the apparent systematics. The star we used as an example for the correction process is located on the edge of the superstamp (c.f. Fig.\,\ref{fig_pixelmask}) but is well-behaved enough to allow a good correction process. Meanwhile, \emph{\object{Gaia DR3 604896429396491136}}, whose light curve is depicted in panel (c) of Fig.\,\ref{fig_other_lightcurve_examples}, shows remnants of an over-correction in the highlighted area due to proximity to the superstamp edges and subsequent loss of flux.

\subsection{Comparing our results with K2SC, SAP, EVEREST}\label{appendix_validation}

    We validate our light curves by comparing them with those produced as part of the original survey and from two commonly used works that implemented a systemtatics correction, {\sc k2sc}, and {\sc everest}. This comparison again highlights one important fact that supports our approach with creating our own light curves --  not many light curves are available for targets within the superstamp. Figure\,\ref{fig_lightcurve_distribution} shows the distribution of stars with available light curves.
    \begin{figure}
        \centering
        \includegraphics[width=\linewidth]{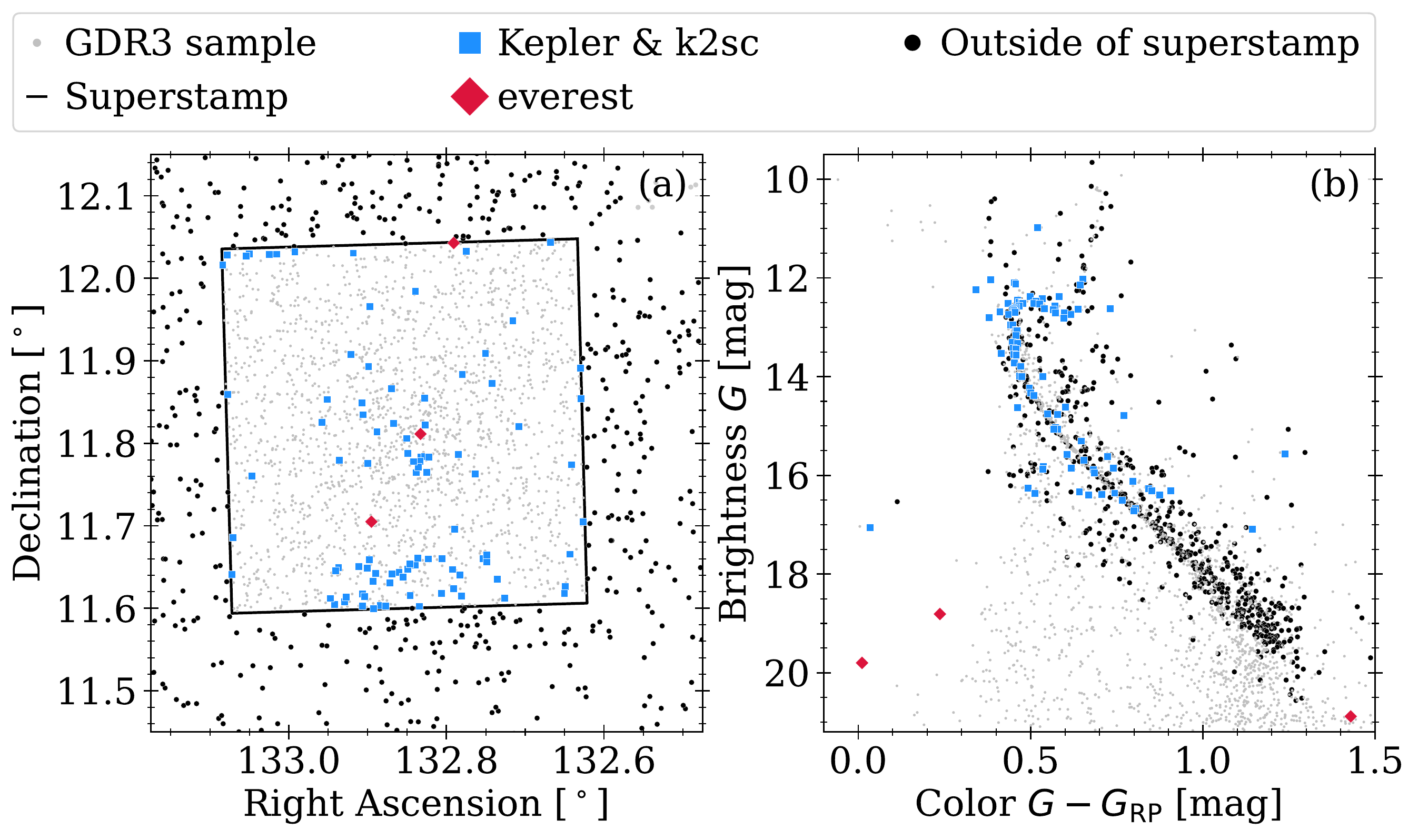}
        \caption{
            Spatial distribution of stars with available light curves in the archives for different data products. Panel (a) shows a map centered on M67 with the extent of the superstamp indicated (black box). Targets with archived light curves within (red and blue) and outside (black) of superstamp are overplotted. Panel (b) shows the same stars in a CMD.
        }
        \label{fig_lightcurve_distribution}
    \end{figure}
    Only 96 are within the field of view of the superstamp, as compared with GDR3 listing $\sim$2000 sources. This most likely originates from the differing nature of the superstamp as compared with the normal TPFs. The latter are typically associated with a proposed star and therefore are automatically processed. This is not true for the superstamp. As far as we are aware, all other projects that dealt with the \ktwo{} systematics only operate on the same sample. Therefore, other extant work does not venture outside this 96-star sample.

    We extracted the 96 light curves in the FOV of the superstamp from the \kepler{} archive\footnote{\url{archive.stsci.edu/k2/data_products.html}}. These include the light curves based on simple aperture photometry (SAP, $f_\text{SAP}$) on the TPFs and corrected light curves that resulted from the Presearch Data Conditioning (PDC, $f_\text{PDCSAP}$) module of the \kepler{} pipeline \citep{2020ksci.rept....8S}. The PDC is employed to remove data systematics and trending while retaining astrophysical signals. This generally works reasonably well but leaves traces of both artifacts. In some cases the trend correction removes astrophysical signal. For our comparison we employ the SAP flux as well as the PDCSAP with the aim of comparing the most similar products from each. For the former we compare with our raw extracted flux. However, we generally do not expect a large agreement in the shape of the raw light curves on short term scales due to the differently shaped aperture mask used. Long term effects are expected to be the same. The PDCSAP flux $f_\text{PDCSAP}$ is compared with our final light curve $f_\text{fin}$ (lower panel of Fig.\,\ref{fig_lightcurve_comaparison_pdcsap}) as both have the same level of processing (systematics and trend removal), whereas the SAP flux $f_\text{SAP}$ is equivalent to our raw flux $f_\text{raw}$ (upper panel of Fig.\,\ref{fig_lightcurve_comaparison_pdcsap}).
    \begin{figure}
        \centering
        \includegraphics[width=\linewidth]{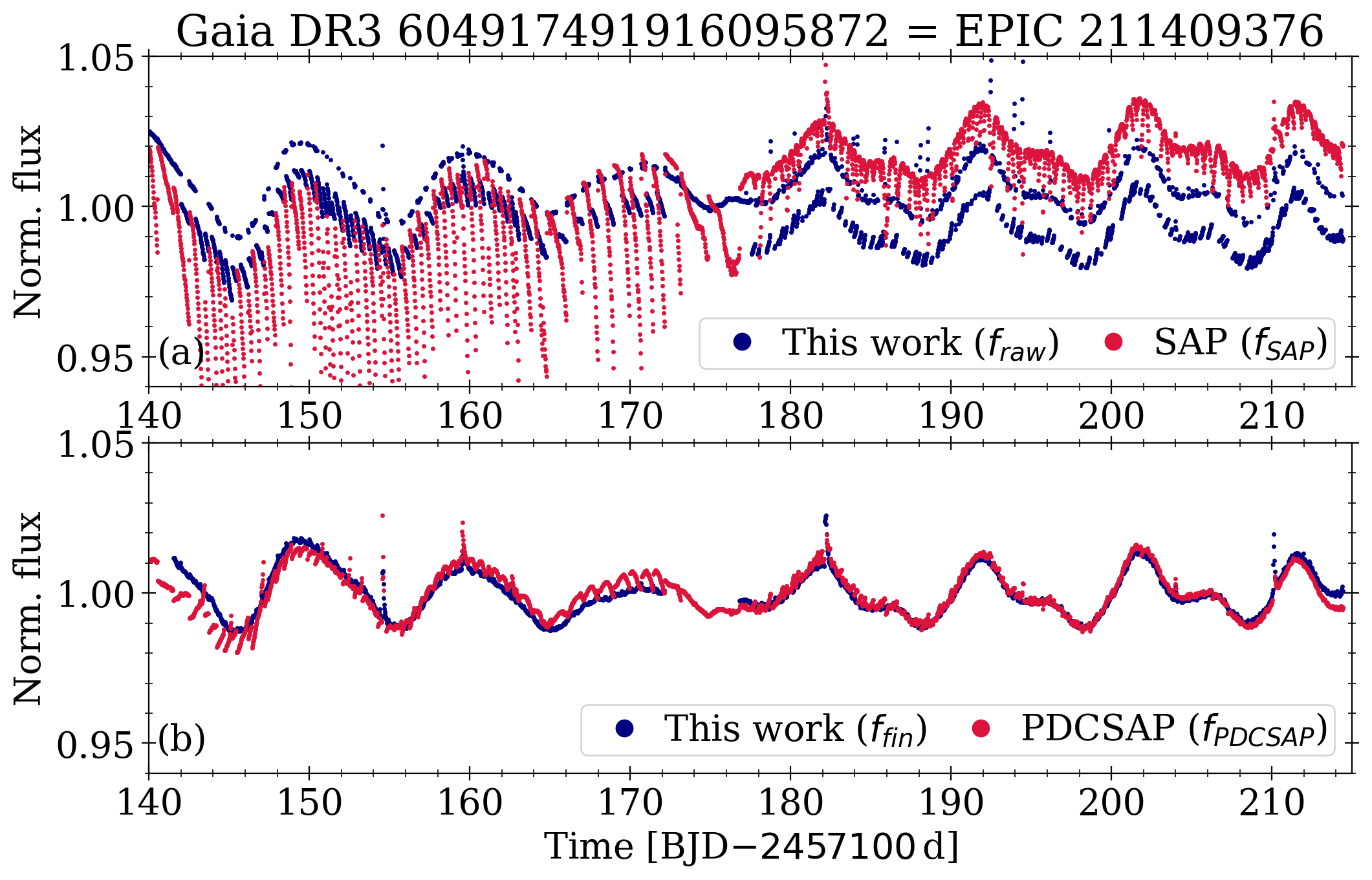}
        \caption{
            Comparison between the \kepler{} PDCSAP light curves and the ones created by us. In both panels, our light curves are shown in blue and the archival light curves are shown in red. Panel (a) shows the uncorrected aperture photometry fluxes ({\tt flux\_raw} from ours and {\tt SAP\_FLUX} for theirs). Panel (b) shows the final light curve product -- {\tt flux\_fin} from ours and {\tt PDCSAP\_FLUX} for theirs.
        }
        \label{fig_lightcurve_comaparison_pdcsap}
    \end{figure}

    We can see that, as expected, the raw light curves generally do not agree with respect to the short-term systematics while having the same long-term behavior. However, after the correction is applied, we find very good agreement between the \kepler{} and our pipeline results. Despite our relatively simple approach for an empirical correction, we not only do match the quality of the official product but partially exceed it. Generally, our approach is slightly better in the removal of the \emph{instrumental systematics}.
 
    The \emph{K2 Systematics Correction} ({\sc k2sc}) pipeline implements Gaussian processes to correct for the telescope jitter  \citep{2015MNRAS.447.2880A,2016MNRAS.459.2408A}. It is geared toward exoplanet detection; as such the pipeline is set up in a way that also removes the astrophysical signal from the final product $f_\text{k2sc}$. However, they provide the removed systematics as part of their data product, differentiated in a position-dependent $f_\text{posi}$ and a time-dependent part $f_\text{time}$. Long term variability is retained in the time-dependent part, together with trending. Recombining their final light curve with both those parts yields back the PDCSAP light curve ($f_\text{PDCSAP} = f_\text{k2sc}\cdot f_\text{time}\cdot f_\text{posi}$) which was the starting point for the {\sc k2sc} pipeline. Thus, the {\sc k2sc} light curves benefit from the long-term trend correction in the \kepler{} PDC pipeline. Technically, with their procedure they also implement a correction for instrumental systematics twice. To obtain a light curve that is corrected to a similar degree as our final product, we recombine the final flux and the time-dependent trending part. Figure\,\ref{fig_lightcurve_comaparison_k2sc} shows the comparison. The additional systematics removal pays off, as the light curves are improved as compared with PDCSAP, and reach a fidelity similar to, or perhaps are even superior to ours. All 96 PDCSAP light curves are processed with {\sc k2sc}. The final {\sc k2sc} light curve ($f_\text{k2sc}$) is not comparable to any other, as it blatantly removes any long term signal, including physical ones (cf. upper panel of Fig.\,\ref{fig_lightcurve_comaparison_k2sc}). A comparison makes sense between our light curves before the PCA ($f_\text{cor2}$) and the {\sc k2sc} one with the time-dependent signal included ($f_\text{k2sc}\cdot f_\text{time}$, cf. lower panel of Fig.\,\ref{fig_lightcurve_comaparison_k2sc}).
 
\begin{figure}
    \centering
    \includegraphics[width=\linewidth]{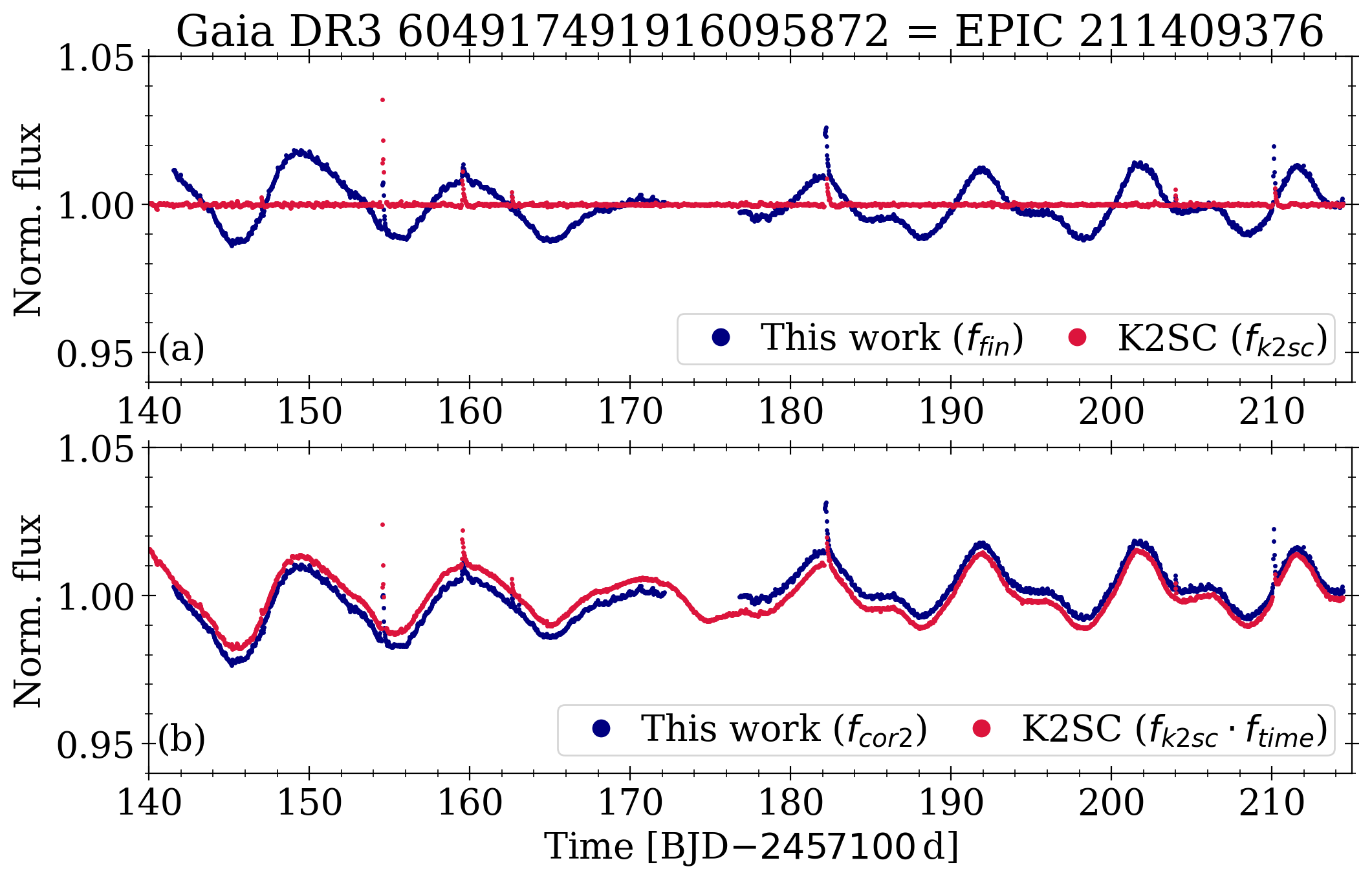}
    \caption{
        Same as Fig.\,\ref{fig_lightcurve_comaparison_pdcsap} but for the comparison between the {\sc k2sc} light curve of \emph{\object{Gaia DR3 604917491916095872}} and the ones created by us.
    }
    \label{fig_lightcurve_comaparison_k2sc}
\end{figure}
 
    The \emph{EPIC Variability Extraction and Removal for Exoplanet Science Targets} ({\sc everest}) light curves \citep{2016AJ....152..100L,2018AJ....156...99L} use a pixel-level decorrelation (PLD) method that operates on pixel level light curves and, since its update to v2.0, includes the telescope motion for the correction. They are, as far as we know, the best available light curves based on \ktwo{} data and the study of variability longer than a few days. Only long-term trending is a remaining issue; such trends are generally more pronounced than in other light curves. However, with those retained, there is also no problem with accidentally removed astrophysical signal. We have worked successfully with them in the past \citep{2020A&A...644A..16G}. However, they are even more limited in their availability regarding the superstamp region. Only three stars in the FOV of the superstamp have an {\sc everest} light curve. And all three of those stars are very faint. Two are white dwarfs, while the third is a very faint red dwarf at the brightness limit of \gaia{} (cf. panel (d) of Fig.\,\ref{fig_lightcurve_distribution}). All three stars are nearly constant in their light curves, and dominated by noise and some weak long-term trending. We use the red dwarf for the comparison in Fig.\,\ref{fig_lightcurve_comaparison_everest}. The extent to which we can correct for instrumental systematics is similar in both products, however, {\sc everest} reaches a slightly higher photometric precision. PDCSAP and {\sc k2sc} fail to fully remove either trends or systematics, and their photometric precision is lower compared with ours or {\sc everest} for the faint stars.
 
\begin{figure}
    \centering
    \includegraphics[width=\linewidth]{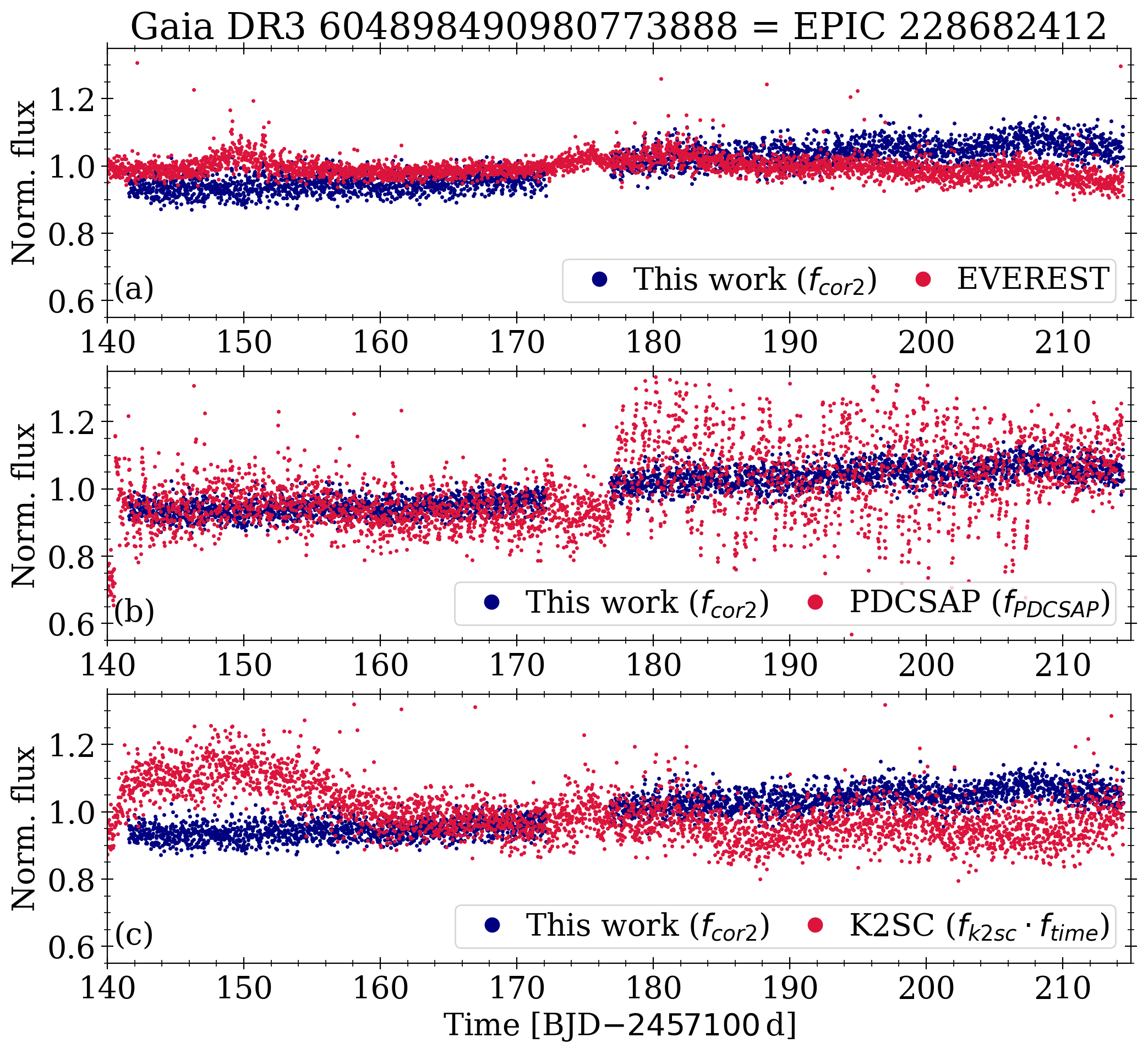}
    \caption{
        Same as Fig.\,\ref{fig_lightcurve_comaparison_pdcsap} but for the comparison between the {\sc everest}, PDCSAP, and {\sc k2sc} light curves of \emph{Gaia\,DR3\,604898490980773888} (panels a, b, and c, respectively) and the ones created by us.
    }
    \label{fig_lightcurve_comaparison_everest}
\end{figure}

% \newpage

\vspace{-1.0em}
\section{Supplements}
  
\subsection{Supplementary tables}

\begin{table}[hb!]
    \centering\small
    \caption{Datapoints for the empirical cluster sequences.\label{tab_cluster_sequences}}\vspace{-1.0em}
    \begin{tabular}{ccccccc}
   \hline\hline
$G-G_\mathrm{RP}$ & $G_\mathrm{BP}-G_\mathrm{RP}$ &           $B-V$ &           $V-K$ &         $P_\text{1}$ &       $P_\text{2.5}$ &         $P_\text{4}$ \\{} 
          [mag] &           [mag] &                  [d] &                  [d] &                  [d] &                  [d] &                  [d] \\ 
   \hline
           0.32 &            0.50 &            0.38 &            0.99 &                  1.5 &                  nan &                  nan \\ 
           0.38 &            0.61 &            0.46 &            1.20 &                  2.0 &                  5.0 &                  nan \\ 
           0.40 &            0.65 &            0.49 &            1.28 &                  3.0 &                  8.0 &                 14.0 \\ 
           0.45 &            0.74 &            0.58 &            1.44 &                  8.5 &                 15.0 &                 22.0 \\ 
           0.50 &            0.83 &            0.67 &            1.64 &                 11.0 &                 19.0 &                 26.0 \\ 
           0.55 &            0.93 &            0.76 &            1.84 &                 11.0 &                 22.0 &                 29.0 \\ 
           0.60 &            1.03 &            0.85 &            2.09 &                 10.5 &                 23.0 &                 31.0 \\ 
           0.65 &            1.14 &            0.93 &            2.32 &                 11.0 &                  nan &                 31.0 \\ 
           0.70 &            1.25 &            1.02 &            2.57 &                 12.0 &                 23.0 &                  nan \\ 
           0.75 &            1.37 &            1.11 &            2.83 &                 12.5 &                  nan &                 29.5 \\ 
           0.80 &            1.48 &            1.19 &            3.08 &                 13.0 &                 20.2 &                 26.0 \\ 
           0.85 &            1.61 &            1.27 &            3.32 &                 13.8 &                  nan &                  nan \\ 
           0.90 &            1.74 &            1.33 &            3.55 &                 14.3 &                 18.0 &                 25.5 \\ 
           0.95 &            1.87 &            1.37 &            3.79 &                 15.0 &                  nan &                  nan \\ 
           1.00 &            2.02 &            1.39 &            4.03 &                  nan &                 22.0 &                 32.0 \\ 
           1.10 &            2.33 &            1.44 &            4.55 &                  nan &                 30.0 &                 41.0 \\ 
   \hline
\end{tabular}

    \vspace{-.7em}
    \tablefoot{The manually drawn sequences used in the CPDs throughout this work are plotted from cubic interpolation between the listed colors and periods. The subscript to the period label indicates the corresponding age in Gyr.}
\end{table}

% \newpage

\begin{table*}[h!]
    \centering
    \caption{Results of the period analysis.}
    \label{tab_results_large}
    \begin{tabular}{lccccccccccc}
   \hline\hline
                 Gaia DR3 &            EPIC &        $P$ &  $P_\text{err}$ &        $G$ & $(G-G_\mathrm{RP})_0$ & $(G_\mathrm{BP}-G_\mathrm{RP})_0$ &      $(B-V)_0$ &      $(V-K)_0$ &        HPS\\
                          &                 &        [d] &             [d] &      [mag] &          [mag] &          [mag] &          [mag] &          [mag] &           \\
   \hline
       598903678707639296 &   \it 211398541 &     $32.4$ &             1.3 &      15.93 &           0.65 &           1.16 &           0.94 &           2.27 &           \\
       604896498115959296 &                 &     $23.5$ &             1.5 &      17.50 &           0.88 &           1.73 &           1.43 &           3.27 &           \\
       604896635554924672 &   \it 211397512 &     $32.5$ &             1.4 &      15.87 &           0.65 &           1.15 &           0.96 &           2.26 &           \\
       604897558972113024 &   \it 211400106 &     $29.1$ &             2.6 &      14.75 &           0.55 &           0.93 &           0.69 &           1.86 &           \\
 \\ [-0.8em] 
       604897833850019328 &   \it 211400500 &     $24.3$ &             1.6 &      14.37 &           0.48 &           0.79 &           0.58 &           1.50 &           \\
       604898490980772352 &                 &     $27.4$ &             2.1 &      16.61 &           0.79 &           1.48 &           1.20 &           2.93 &           \\
       604899831010539904 &                 &     $18.1$ &             0.4 &      13.29 &           0.42 &           0.69 &           0.42 &           1.42 &          x\\
       604900071528704128 &                 &     $34.7$ &             2.5 &      16.18 &           0.73 &           1.30 &           1.04 &           2.61 &           \\
 \\ [-0.8em] 
       604900651348634240 &                 &     $23.1$ &             2.1 &      17.60 &           0.88 &           1.73 &           1.38 &           3.56 &           \\
       604903331408222208 &                 &     $29.5$ &             1.6 &      18.46 &           1.02 &           2.12 &           1.50 &           3.84 &           \\
       604903438783070208 &                 &     $37.0$ &             3.0 &      18.64 &           1.02 &           2.19 &           1.61 &           4.17 &           \\
       604906840397139584 &                 &     $29.1$ &             1.1 &      17.76 &           0.95 &           1.92 &           1.47 &           3.80 &          x\\
 \\ [-0.8em] 
       604907046555568896 &                 &     $27.2$ &             1.4 &      16.84 &           0.79 &           1.50 &           1.27 &           2.93 &           \\
       604908004332577152 &   \it 211400002 &     $22.9$ &             2.2 &      14.31 &           0.47 &           0.77 &           0.55 &           1.52 &           \\
       604909206923484160 &                 &     $27.6$ &             1.1 &      17.23 &           0.83 &           1.62 &           1.30 &           3.31 &          x\\
       604909756679296640 &                 &     $30.2$ &             2.2 &      15.65 &           0.61 &           1.06 &           0.90 &           2.17 &           \\
 \\ [-0.8em] 
       604911204083987584 &                 &     $7.37$ &            0.14 &      15.08 &           0.55 &           0.92 &           0.78 &           1.89 &           \\
       604913540546033024 &                 &     $31.1$ &             1.5 &      16.31 &           0.70 &           1.30 &           1.05 &           2.45 &           \\
       604913952863073920 &                 &     $26.5$ &             2.2 &      16.82 &           0.80 &           1.52 &           1.24 &           3.05 &           \\
       604914880575997056 &                 &     $29.1$ &             1.8 &      16.37 &           0.72 &           1.32 &           1.08 &           2.60 &           \\
 \\ [-0.8em] 
       604915773929008384 &                 &     $29.0$ &             0.5 &      15.16 &           0.55 &           0.94 &           0.77 &           1.84 &           \\
       604916117526551680 &       211405832 &     $23.4$ &             1.5 &      14.42 &           0.48 &           0.79 &           0.55 &           1.62 &           \\
       604917148318674816 &                 &     $30.8$ &             2.7 &      15.49 &           0.60 &           1.04 &           0.95 &           1.98 &           \\
       604919587860083328 &                 &     $29.5$ &             2.6 &      17.42 &           0.88 &           1.74 &           1.39 &           3.34 &           \\
 \\ [-0.8em] 
       604920549932807296 &                 &     $24.6$ &             1.4 &      17.33 &           0.87 &           1.70 &            nan &            nan &           \\
       604920549932809344 &                 &     $26.1$ &             1.6 &      15.16 &           0.56 &           0.95 &           0.81 &           1.83 &           \\
       604922130480588544 &                 &     $27.2$ &             2.0 &      16.73 &           0.77 &           1.44 &           1.21 &           2.89 &           \\
       604922229264424448 &                 &     $9.72$ &            0.25 &      17.07 &           0.84 &           1.61 &           1.27 &           3.19 &           \\
 \\ [-0.8em] 
       604922817675316096 &                 &     $27.6$ &             3.2 &      14.97 &           0.54 &           0.91 &           0.77 &           1.71 &           \\
       604923333071376512 &                 &     $32.1$ &             2.5 &      15.66 &           0.62 &           1.09 &           0.94 &           2.22 &           \\
       604923848467470976 &                 &     $26.2$ &             2.3 &      17.37 &           0.87 &           1.69 &           1.40 &           3.44 &           \\
       604930681760054656 &                 &     $27.6$ &             2.3 &      17.73 &           0.92 &           1.85 &            nan &            nan &           \\
 \\ [-0.8em] 
       604943674036665472 &       211418998 &     $26.8$ &             2.6 &      14.60 &           0.51 &           0.85 &           0.49 &           1.82 &           \\
       604944120713062784 &                 &     $31.7$ &             2.7 &      15.47 &           0.60 &           1.04 &           0.83 &           2.15 &           \\
       604963362166649856 &                 &     $22.8$ &             1.4 &      16.78 &           0.80 &           1.52 &           1.14 &           3.07 &           \\
       604969061592133376 &                 &     $1.84$ &            0.01 &      19.57 &           1.11 &           2.52 &            nan &            nan &           \\
 \\ [-0.8em] 
       604969237681908480 &                 &     $26.3$ &             2.0 &      16.78 &           0.78 &           1.46 &           1.13 &           2.96 &           \\
       604969267746267520 &                 &     $26.5$ &             1.2 &      17.24 &           0.85 &           1.65 &           1.22 &           3.34 &           \\
       604969267746269696 &                 &     $30.7$ &             2.2 &      15.56 &           0.61 &           1.07 &           0.89 &           2.13 &           \\
       604969306401373824 &                 &     $30.2$ &             1.1 &      15.87 &           0.65 &           1.15 &           0.96 &           2.33 &           \\
 \\ [-0.8em] 
       604969306401373952 &                 &     $24.6$ &             1.7 &      16.93 &           0.81 &           1.54 &           1.26 &           3.08 &           \\
       604969336465748352 &                 &     $24.9$ &             1.7 &      17.28 &           0.87 &           1.69 &           1.43 &           3.36 &           \\
       604970131035099008 &                 &     $24.1$ &             2.3 &      16.98 &           0.81 &           1.57 &           1.18 &           3.12 &           \\
       604971466769552128 &                 &     $27.9$ &             2.0 &      16.51 &           0.75 &           1.40 &           1.06 &           2.94 &           \\
 \\ [-0.8em] 
       604972467498966400 &                 &     $32.3$ &             1.5 &      15.53 &           0.60 &           1.04 &           0.97 &           1.92 &           \\
       604973979325779328 &                 &     $28.7$ &             1.1 &      14.88 &           0.53 &           0.90 &           0.71 &           1.76 &          x\\
       604974151124482944 &                 &     $31.6$ &             2.9 &      15.81 &           0.64 &           1.13 &           0.88 &           2.23 &           \\
   \hline
\end{tabular}

    \tablefoot{For convenience, we have limited the sample shown here to data used to create Fig.\,\ref{fig_cpd_final}. \emph{HPS} indicates whether are star is found on the half-period sequence. A star indicated as such is listed here with its period doubled, that being assumed to be the actual period. EPIC ids in italics denote stars for which a PDCSAP light curve is available.}
\end{table*}

\null

\vfill

\null

\begin{table}[h!]
    \centering
    \caption{Selected stars from the \cite{2016MNRAS.459.1060G,2016MNRAS.463.3513G} sample.}
    \label{tab_gonzales_selection}
    \begin{tabular}{lcccc}
   \hline\hline
                 Gaia DR3 &        $P$ &  $P_\text{err}$ & $(G-G_\mathrm{RP})_0$ &     binary\\
                          &        [d] &             [d] &          [mag] &           \\
   \hline
       604907454576711040 &     $27.2$ &             0.5 &           0.42 &           \\
       598689411379091328 &     $24.9$ &             0.5 &           0.42 &           \\
       605000024007420416 &     $12.1$ &             0.5 &           0.43 &         PB\\
 \\ [-0.8em] 
       598692675554477056 &     $22.6$ &             0.5 &           0.48 &           \\
       604960235430488960 &     $26.8$ &             0.5 &           0.53 &           \\
       598901788922041728 &     $32.4$ &             0.9 &           0.58 &           \\
 \\ [-0.8em] 
       604949961868553856 &     $32.6$ &             0.8 &           0.58 &           \\
       598689926775182208 &     $29.4$ &             0.5 &           0.59 &           \\
       604895325589137920 &     $31.5$ &             0.5 &           0.61 &           \\
 \\ [-0.8em] 
       598902716634970240 &     $30.4$ &             0.5 &           0.66 &           \\
       604987139105593344 &     $28.9$ &             1.0 &           0.68 &           \\
       604898731498904704 &     $29.0$ &             0.5 &           0.71 &           \\
 \\ [-0.8em] 
       604964010706281856 &     $32.4$ &             0.6 &           0.74 &           \\
       604901239759778176 &     $26.3$ &             0.5 &           0.76 &           \\
       604965178937376512 &     $29.0$ &             0.5 &           0.76 &           \\
 \\ [-0.8em] 
       604704633336193152 &     $30.8$ &             0.5 &           0.77 &           \\
       598955729416267264 &     $27.3$ &             0.7 &           0.79 &           \\
       604895948360165888 &     $12.4$ &             0.5 &           0.81 &         PB\\
 \\ [-0.8em] 
       604946693397490816 &     $28.8$ &             0.5 &           0.83 &           \\
       598889213257785984 &     $29.5$ &             0.8 &           0.83 &           \\
       604899448757770624 &     $13.5$ &             0.5 &           0.83 &           \\
 \\ [-0.8em] 
       604894535315152512 &     $28.6$ &             0.5 &           0.84 &           \\
       605000058367154176 &     $13.7$ &             0.5 &           0.85 &           \\
       604901686436367232 &     $25.6$ &             0.5 &           0.85 &           \\
 \\ [-0.8em] 
       604980713834545664 &     $15.6$ &             0.5 &           0.85 &         PB\\
       598903128951834496 &     $29.2$ &             0.5 &           0.88 &         PB\\
       604930509961364352 &     $26.8$ &             0.5 &           0.88 &           \\
 \\ [-0.8em] 
       604930445537379712 &     $23.3$ &             0.5 &           0.90 &           \\
       598956485330513536 &     $24.7$ &             0.5 &           0.91 &           \\
       604966518966784256 &     $21.3$ &             0.7 &           0.92 &           \\
 \\ [-0.8em] 
       604713669947377024 &     $22.0$ &             0.5 &           0.93 &           \\
       598901995079871744 &     $24.0$ &             0.5 &           0.98 &         PB\\
       604926043198690176 &     $23.9$ &             0.5 &           1.04 &         PB\\
 \\ [-0.8em] 
   \hline
\end{tabular}

    \tablefoot{Only those stars from the \cite{2016MNRAS.459.1060G,2016MNRAS.463.3513G} sample that have been classified by us as having reliable periods and subsequently adopted in this study.}
\end{table}

\null

\vfill

\null

% \columnbreak

% \clearpage

\FloatBarrier
 
% \newpage

% 
 
\subsection{The data provided}\label{appendix_data_product}
 
    Table\,\ref{tab_results_large} gives an overview of the most relevant sample of the stars for which we found periodic signals. As a supplement to this work we provide the extend version containing all 136 stars with additional columns in machine-readable form. The columns included are described in Table\,\ref{tab_aux_overview}. Additionally, we provide the light curves, including, raw fluxes, the intermediate steps, the PCA, and all masking.
 
\begin{table}[h!]
    \centering
    \caption{Sample table overview.}
    \label{tab_aux_overview}
    \begin{tabular}{rcl}
        \hline\hline
        Column             & Unit   & Description  \\
        \hline
        \verb+Star+        & & Running catalog number  \\ 
        \verb+Sample+      & & =\verb+Y+ if in M\,67 single MS star sample  \\ 
        \verb+Gaia DR3+    & & Gaia DR3 identifier  \\  
        \verb+EPIC+        & & EPIC identifier  \\    
        \verb+2MASS+       & & 2MASS identifier  \\    
        \\ [-0.8em] 
        \verb+period+      & [d]    & derived period  \\    
        \\ [-0.8em]
        \verb+Xmag+        & [mag]  & $X$-band magnitude for \\
                           &        &$X\in\left\lbrace B,V,R,I,J,H,K,G,G_\mathrm{BP},G_\mathrm{RP} \right\rbrace$ \\    
        \\ [-0.8em]
        \verb+M67+         & & member \verb+M+, candidate \verb+C+, field star \verb+F+ \\
        \verb+M67_pmu+     & & proper motions cluster member \\
        \verb+M67_plx+     & & parallax cluster member \\
        \verb+M67_vrad+    & & radial velocity cluster member \\
        \verb+M67_phot+    & & photometric cluster member \\
        \\ [-0.8em]
        \verb+G_RP+         & [mag]  & $G-G_\mathrm{RP}$ color \\
        \verb+G_RP_0+       & [mag]  & $(G-G_\mathrm{RP})_0$ color \\
        \\ [-0.8em]
        \verb+BP_RP+         & [mag]  & $G_\mathrm{BP}-G_\mathrm{RP}$ color \\
        \verb+BP_RP_0+       & [mag]  & $(G_\mathrm{BP}-G_\mathrm{RP})_0$ color \\
        \\ [-0.8em]
        \verb+B_V+          & [mag]  & $B-V$ color \\
        \verb+B_V_0+        & [mag]  & $(B-V)_0$ color \\
        \\ [-0.8em]
        \verb+V_K+          & [mag]  & $V-K$ color \\
        \verb+V_K_0+        & [mag]  & $(V-K)_0$ color \\
        \\ [-0.8em]
        \verb+ra_icrs+      & [$^\circ$]  & GDR3 Right Ascension $\alpha$ \\
        \verb+de_icrs+      & [$^\circ$]  & GDR3 Declination $\delta$ \\
        \verb+plx+          & [mas]       & GDR3 parallax $\varpi$ \\
        \verb+pmra+         & [mas/yr]    & GDR3 proper motion $\mu_\text{RA}$ \\
        \verb+pmde+         & [mas/yr]    & GDR3 proper motion $\mu_\text{Dec}$ \\
        \verb+vrad+         & [km/s]      & radial velocity $v_\text{rad}$ \\
        \\ [-0.8em]
        \verb+otype+         & & Simbad object type \\
        \verb+sptype+        & & Simbad spectral type \\
        \verb+binary+        & & =\verb+Y+ if binary/multiple \\
        \verb+phot_bin+      & & =\verb+Y+ if photometric binary \\
        \verb+MS+            & & =\verb+Y+ if star is main sequence \\
        \verb+HPS+           & & =\verb+Y+ if star is half-period sequence \\
        \verb+DD+            & & =\verb+Y+ if star shows >1 spot \\
        \verb+x_stamp+       & [pxl]  & mean $x$ coordinate on the superstamp \\
        \verb+y_stamp+       & [pxl]  & mean $y$ coordinate on the superstamp \\
        \\ [-0.8em]
        \verb+kepler+        & & =\verb+Y+ if star has Kepler light curve \\
        \verb+k2sc+          & & =\verb+Y+ if star has {\sc k2sc} light curve \\
        \verb+everest+       & & =\verb+Y+ if star has {\sc everest} light curve \\
        \hline
    \end{tabular}
    \tablefoot{Measured and derived values such as photometry, astrometry, and the rotation period have an additional column with a \texttt{e\_{}} prefix which denote the associated errors. Those are not listed here separately.}
\end{table}

\FloatBarrier

\null

\newpage

\subsection{Supplementary figures}\label{appendix_supp_figures}
 
\begin{figure}[h!]
    \centering
    \includegraphics[width=\linewidth]{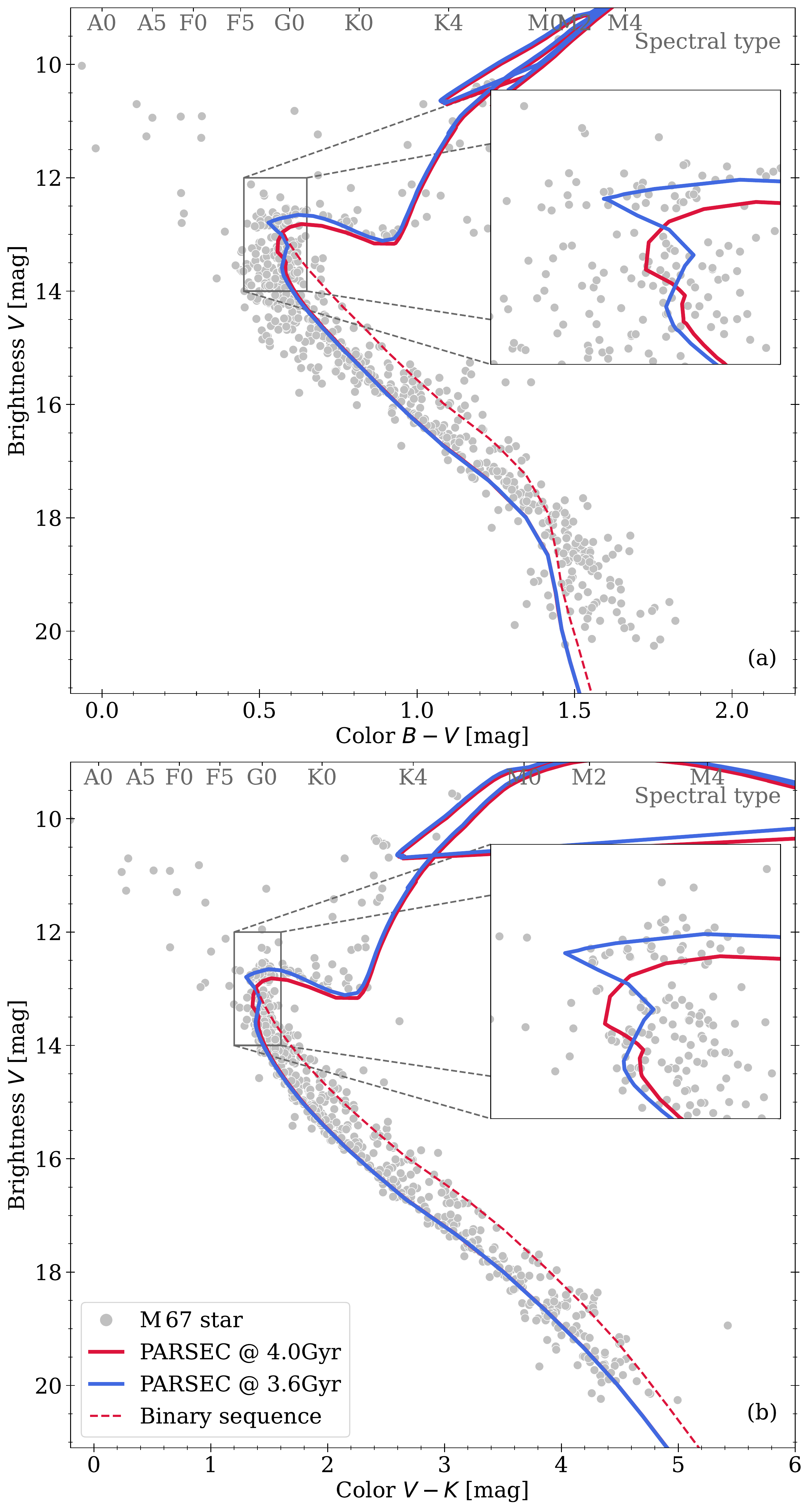}
    \caption{
        Same as Fig.\,\ref{fig_m67_membership} but for $B-V$ and $V-K$ colors.
    }
    \label{fig_m67_johnson_cmd}
\end{figure}
  
\begin{figure}[h!]
    \centering
    \includegraphics[width=\linewidth]{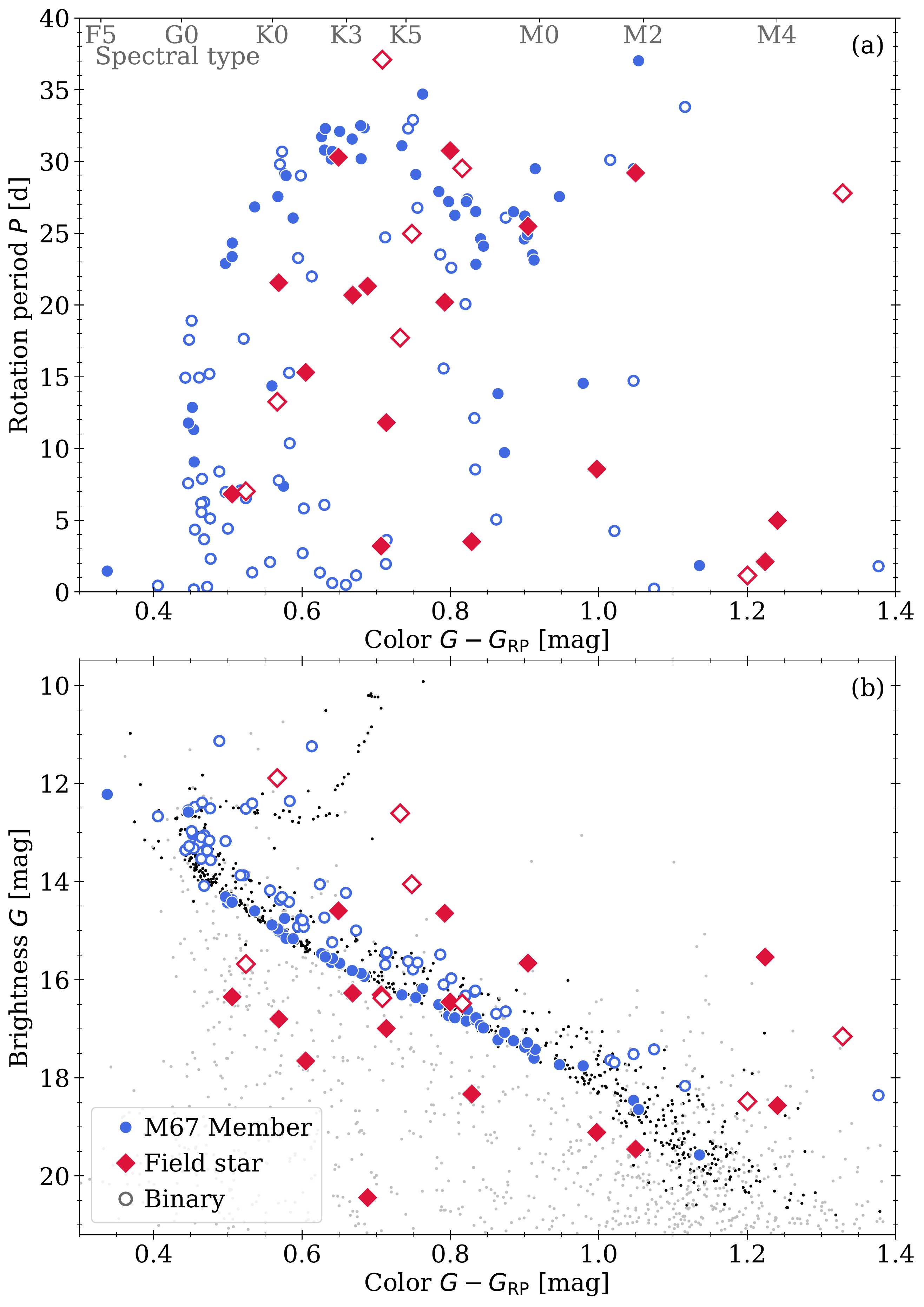}
    \caption{
        Color-period diagram (CPD, panel a) and color-magnitude diagram (CMD, panel b) of the complete superstamp sample. Blue symbols indicate the M\,67 members as discussed above. Overplotted in red are the field stars for which we found a periodic signal. Note that the colors here are not reddening corrected. Open symbols denote binaries.
    }
    \label{fig_field_star_cpd}
\end{figure}

\begin{figure}[ht]
    \centering
    \includegraphics[width=\linewidth]{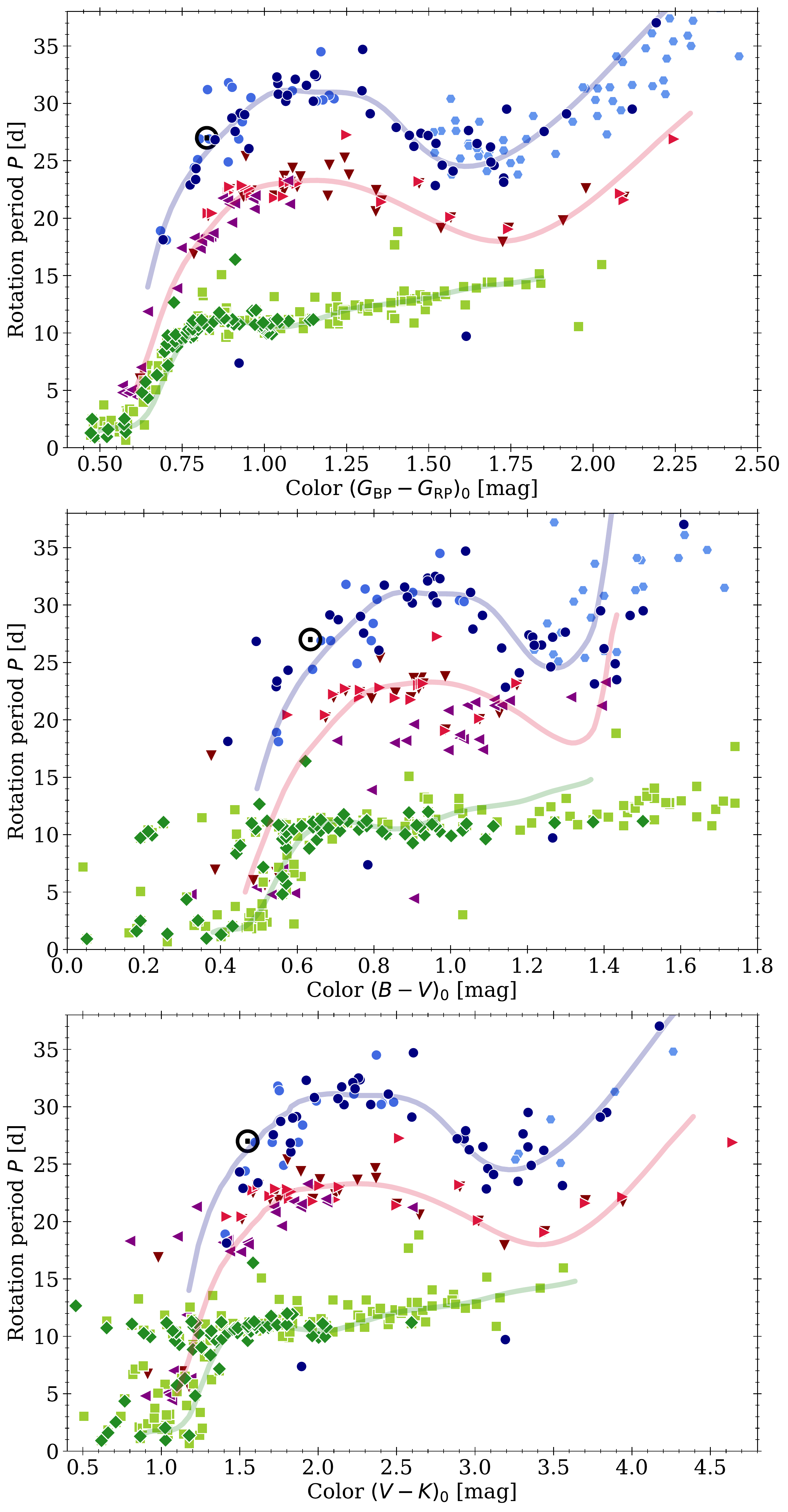}
    \caption{
        Same as Fig.\,\ref{fig_cpd_m67_cluster} but for $(G_\mathrm{BP}-G_\mathrm{RP})_0$, $(B-V)_0$, and $(V-K)_0$ colors.
    }
    \label{fig_cluster_cpd_colors}
\end{figure}

\FloatBarrier

\null

\vfill

\newpage

\subsection{Sample light curves}\label{appendix_lightcurve_plots}

\begin{figure}[h!]
    \centering
    \includegraphics[width=\linewidth]{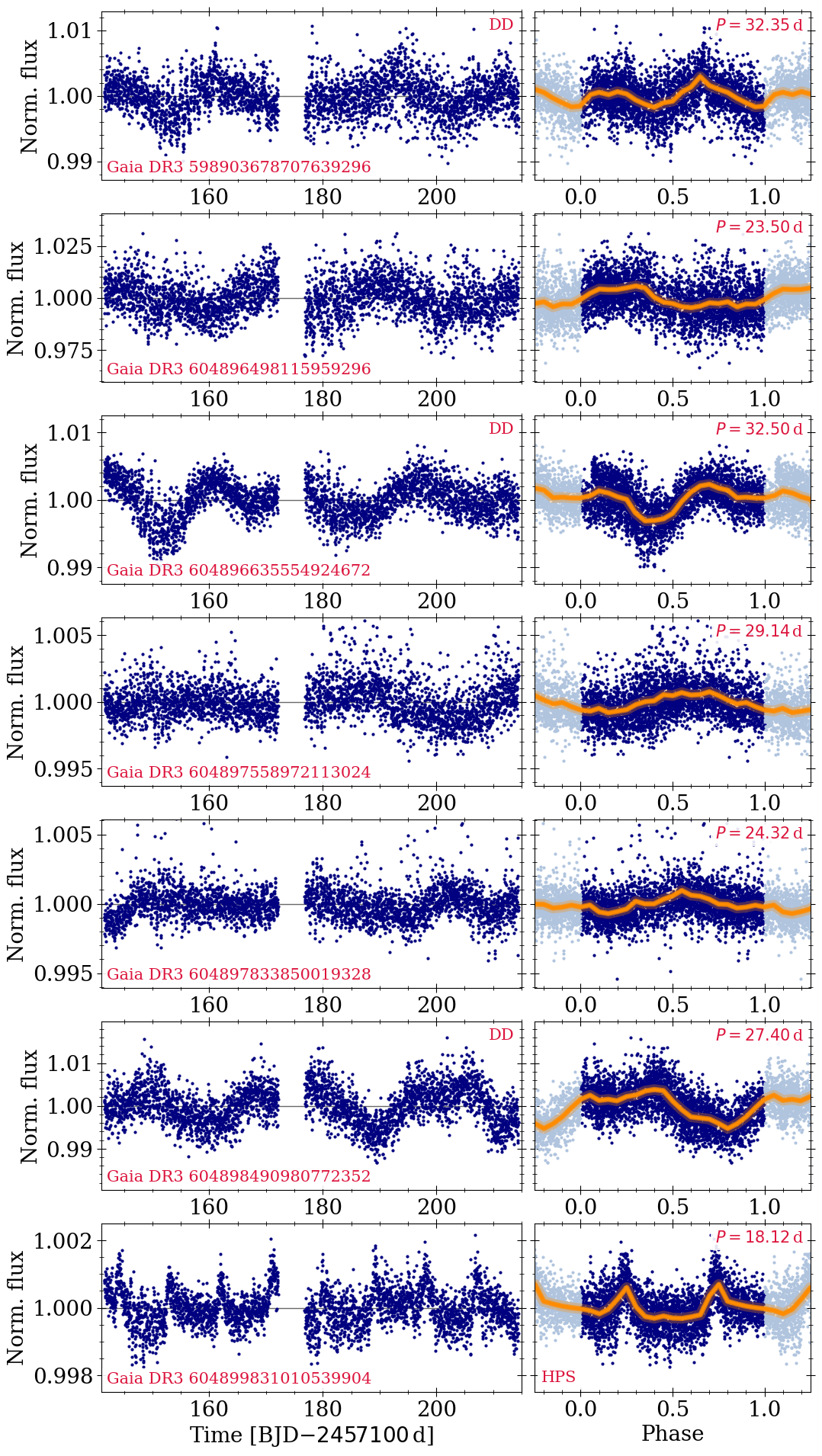}
    \caption{
        Light curves for our sample stars. Each row in the figure is a star, with the name indicated. The left panel shows the light curve, the right panel their phase folded counterparts. The period used for the phase folding is shown in the upper right. The light curve panel displays the \gaia{} DR3 id in the lower left corner. Additional labels indicate whether a star's light curve exhibits multiple spot features, i.e., is \emph{double dipping} (DD, upper right corner in the left panel) or is a member of the \emph{half-period sequence} (HPS, lower left corner of the right panel). The orange line indicates a 0.01 binning. Note that double dipping refers to all stars whose light curves exhibit signs of more than one spot. Stars are sorted as in Table\,\ref{tab_results_large}, by their \emph{Gaia} Id.
    }
    \label{fig_lightcurve_sample_1}
\end{figure}
 
\begin{figure}\ContinuedFloat
    \centering
    \includegraphics[width=\linewidth]{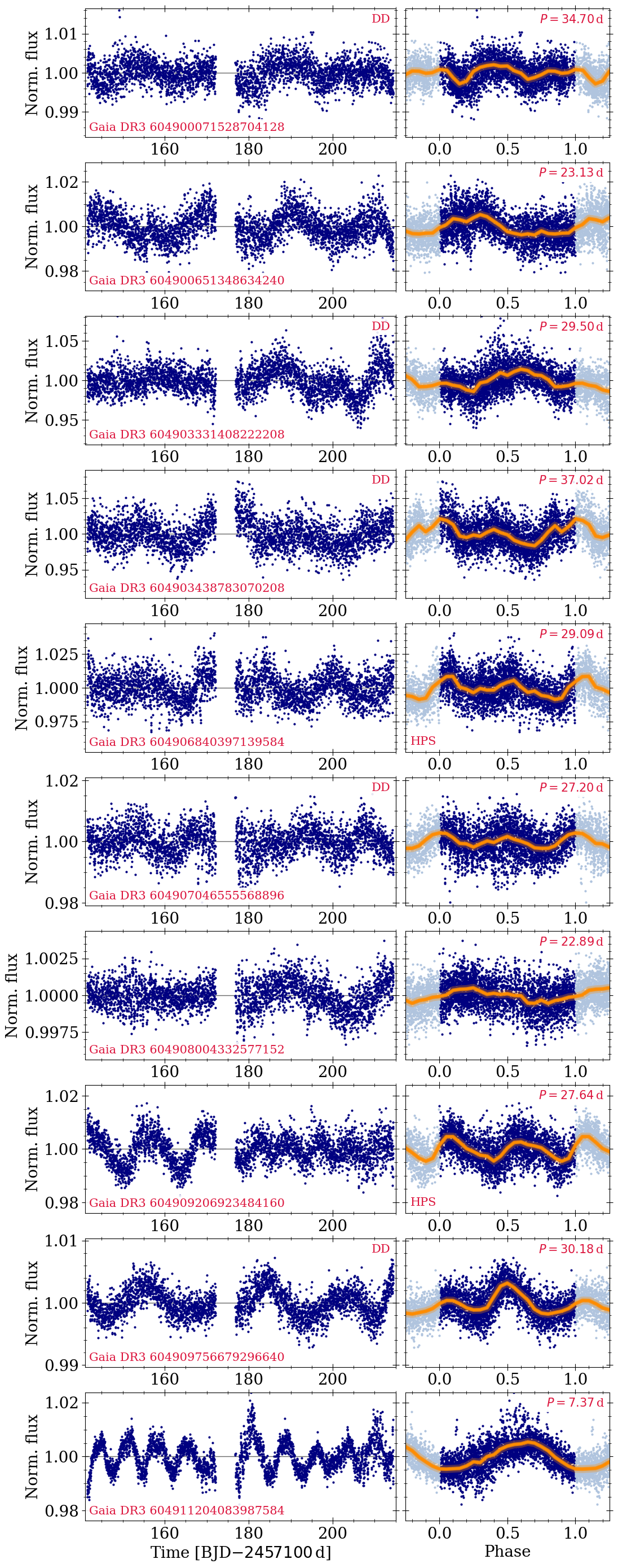}
    \caption{continued.}
\end{figure}

\begin{figure}\ContinuedFloat
    \centering
    \includegraphics[width=\linewidth]{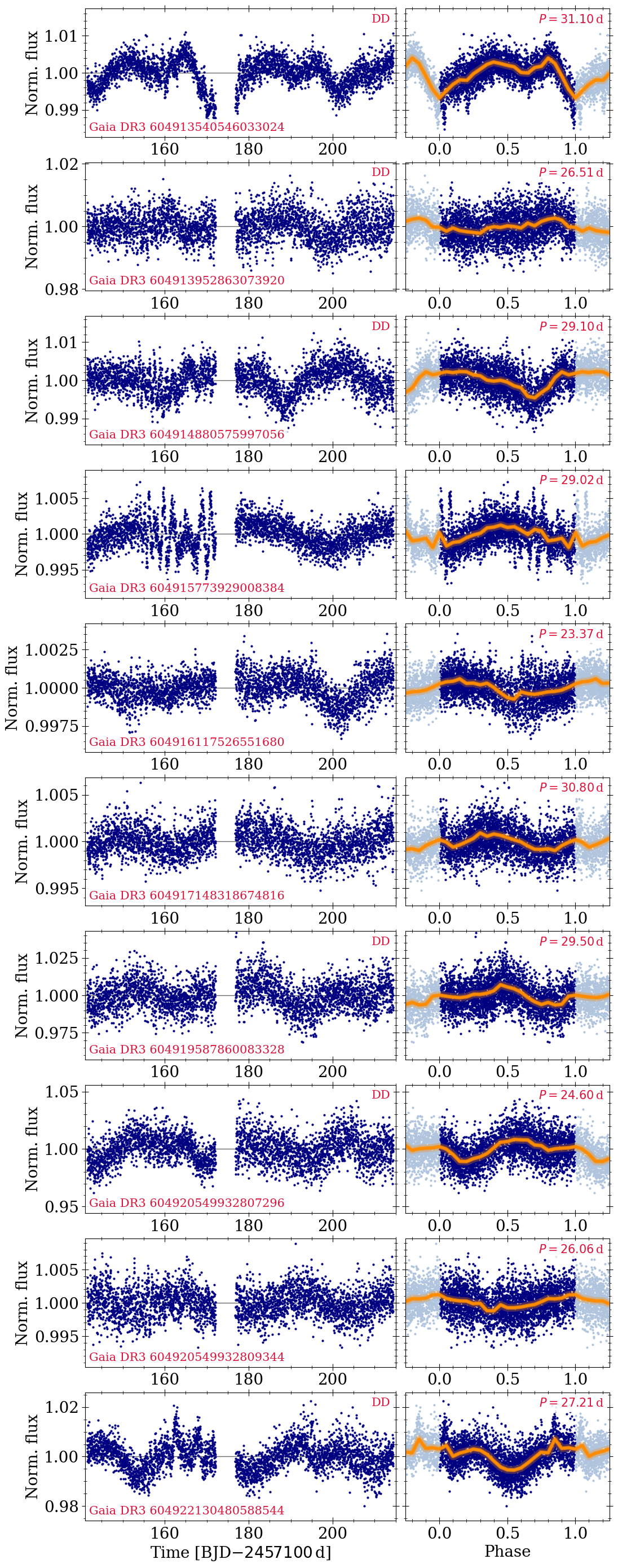}
    \caption{continued.}
\end{figure}

\begin{figure}\ContinuedFloat
    \centering
    \includegraphics[width=\linewidth]{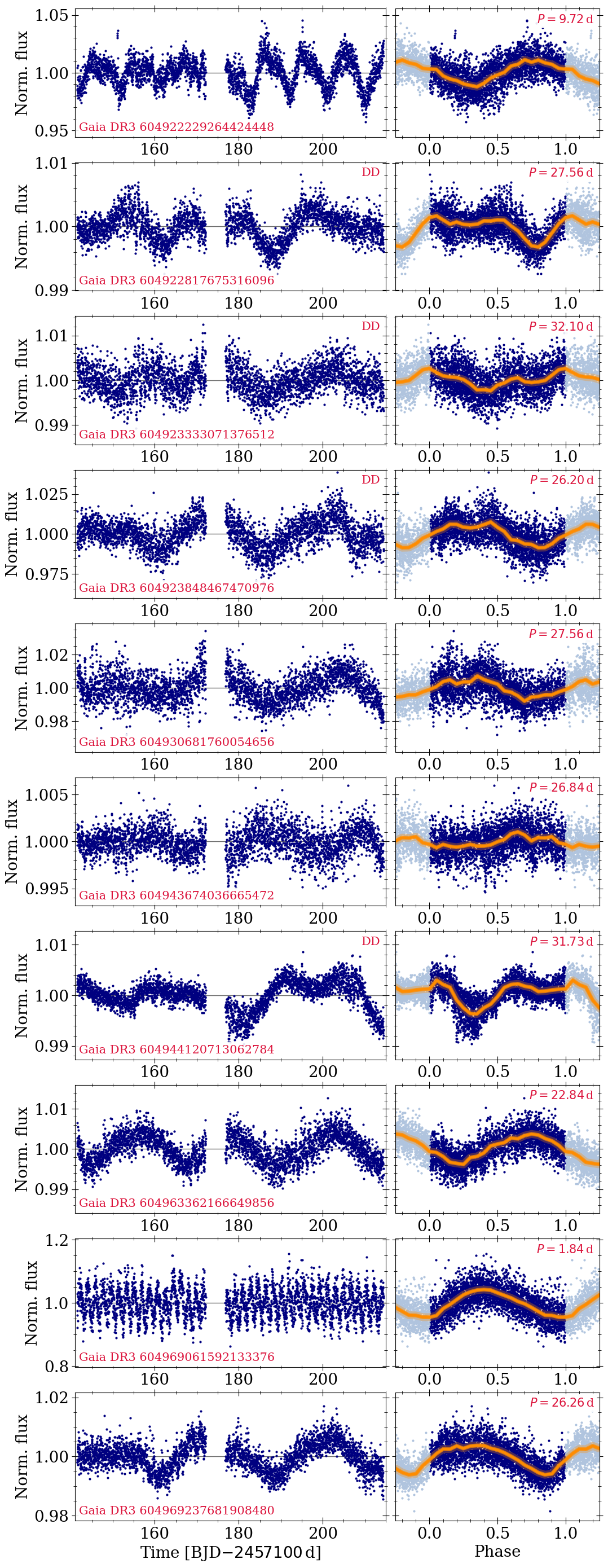}
    \caption{continued.}
\end{figure}

\begin{figure}\ContinuedFloat
    \centering
    \includegraphics[width=\linewidth]{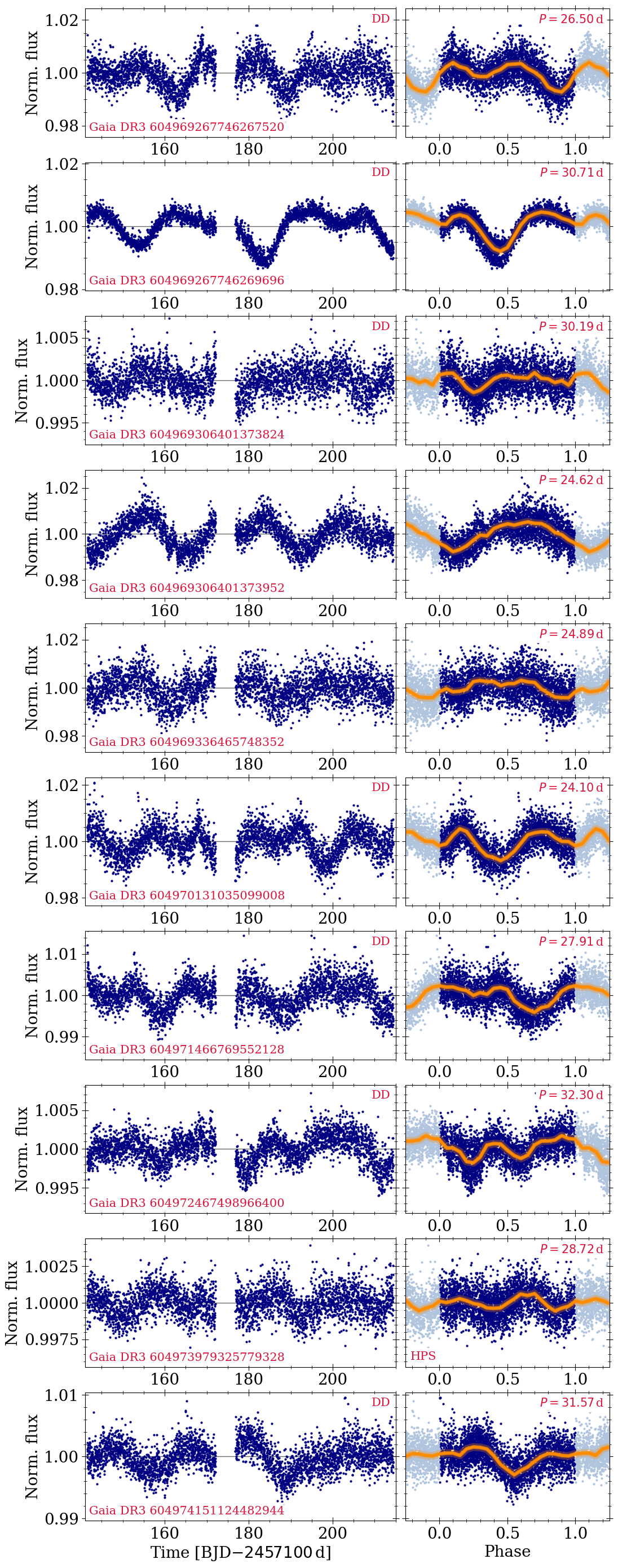}
    \caption{continued.}
\end{figure}

\end{appendix}

\end{document}